\documentclass[%
 reprint,
 amsmath,amssymb,
 aps,
prb,
 superscriptaddress
]{revtex4-2}

\usepackage{graphicx}% Include figure files
\usepackage{dcolumn}% Align table columns on decimal point
\usepackage{bm}% bold math
\usepackage{color}
\usepackage{times}%Font for better visibility
\usepackage[colorlinks=true,linkcolor=blue,urlcolor=blue,citecolor=blue]{hyperref}

\usepackage{booktabs}
\usepackage{multirow}

\begin{document}

\preprint{APS/123-QED}

\title{Asymptotic Scaling of Precision Limits in Continuous Gaussian Quantum Metrology}% Force line breaks with \\

\author{Kazuki Yokomizo}
\affiliation{Department of Physics, The University of Tokyo, 7-3-1 Hongo, Bunkyo-ku, Tokyo, 113-0033, Japan}
\author{Aashish A. Clerk}
\affiliation{Pritzker School of Molecular Engineering, The University of Chicago, Chicago, Illinois 60637, USA}
\author{Yuto Ashida}%
\affiliation{Department of Physics, The University of Tokyo, 7-3-1 Hongo, Bunkyo-ku, Tokyo, 113-0033, Japan}
\affiliation{Institute for Physics of Intelligence, The University of Tokyo, 7-3-1 Hongo, Tokyo 113-0033, Japan}


%\affiliation{
%\affiliation{
%\affiliation{%


\begin{abstract}
Continuous quantum metrology holds promise for realizing high-precision sensing by harnessing information progressively carried away by the radiation quanta emitted into the environment. Despite recent progress, a comprehensive understanding of the precision limits of continuous metrology with bosonic systems is currently lacking. We develop a general theoretical framework for quantum metrology with multimode free bosons under continuous Gaussian measurements. We derive analytical expressions for the asymptotic growth rates of the global quantum Fisher information (QFI) and the environmental QFI, which quantify the total information encoded in the joint system-environment state and the information accessible from the emitted radiation, respectively. We show that the asymptotic growth rates of the global and environmental QFIs coincide in a class of continuous sensing protocols with dissipative system-environment couplings, while they are in general qualitatively distinct when a system-environment coupling exhibits no damping. We further derive bounds on these quantities, showing that while a quadratic scaling with the number of modes is attainable, the precision scales at most linearly with time and a meaningful energy resource. To illustrate our findings, we analyze several concrete setups, including coupled cavity arrays and trapped particle arrays. While a local setup yields a linear scaling with resources, a globally coupled setup can achieve a quadratic scaling in terms of the mode number. Furthermore, we demonstrate that a nonreciprocal setup can leverage the non-Hermitian skin effect to realize an exponentially enhanced global QFI. Notably, however, this enhancement cannot be reflected in the environmental QFI, highlighting a fundamental distinction between the information stored within the joint state and the information radiated into the environment. These findings establish an understanding of the resource trade-offs and scaling behaviors in continuous bosonic sensing.
%\begin{description}
%\end{description}
\end{abstract}
\pacs{Valid PACS appear here}% PACS, the Physics and Astronomy
\maketitle
%
%\tableofcontents
%

\section{\label{sec1}Introduction}
The pursuit of high-precision measurements is a cornerstone of modern science and technology, driving progress across diverse fields such as quantum optics~\cite{Giovannetti2011,Barbieri2022,Huang2024}, condensed matter physics~\cite{Degen2017,Hsieh2019,Tsukamoto2022,Vaidya2023,Sasaki2023,Montenegro2025}, and bio-imaging~\cite{Aslam2023}. Quantum metrology seeks to harness nonclassical phenomena, such as entanglement and squeezing, to develop a measurement protocol whose sensitivity surpasses the limits achievable with purely classical resources~\cite{Giovannetti2004,Huang2024}. The theoretical foundation of quantum metrology is built upon the quantum Cram\'{e}r-Rao inequality, which provides a lower bound on the variance of any unbiased estimator for a given parameter~\cite{Braunstein1994,Paris2009}. This ultimate precision limit is quantified by the quantum Fisher information (QFI) $I$, the amount of information a quantum state carries about an unknown parameter. A larger QFI corresponds to a higher potential sensitivity, making its maximization a primary objective in the design of quantum sensors. So far, the field has seen remarkable experimental developments, such as phase measurements with entangled photons~\cite{Mitchell2004}, entanglement-enhanced microscopy~\cite{Ono2013}, and adaptive estimation~\cite{Okamoto2012}, showing the promise for quantum-enhanced metrology.

Much of the theoretical framework for quantum metrology has been developed within the context of interferometric setups~\cite{Pezze2007,Spagnolo2012}. In this paradigm, a prepared probe state undergoes a unitary evolution during which a parameter is encoded, followed by a measurement to extract information. The precision limit is tied to the available resources; in systems composed of $M$ discrete units, such as qubits, the system size $M$ itself is considered a primary resource. The standard quantum limit corresponds to a QFI that scales linearly with the resource $I\propto M$, a result achievable with uncorrelated probes. In contrast, the Heisenberg limit is customarily defined as a quadratic scaling $I\propto M^2$, which generally requires the use of highly entangled states like the Greenberger-Horne-Zeilinger state~\cite{Huang2024}. In bosonic systems, where the dimension of the local Hilbert space is infinite, the number of excitations $n$ can also serve as a key resource~\cite{Friis2015}, and squeezed vacuum or NOON states allow for Heisenberg-type scaling $I\propto n^2$~\cite{Chen2025}. Separately, the use of many-body states for quantum sensing has also attracted attention in recent years~\cite{Zanardi2008,Invernizzi2008,Wei2019,Mirkhalaf2021,Montenegro2021,Gietka2022,Sarkar2022,Sarkar2025}. Altogether, the aforementioned results have been relatively well-established in the interferometric settings, forming a mature subfield of quantum metrology.

In contrast, the vast potential of continuous quantum metrology, where a quantum system is continuously monitored over time, remains much less explored~\cite{Gambetta2001,Gammelmark2013,Catana2015}. In this alternative paradigm, the sensor is an open quantum system that continuously interacts with its environment~\cite{Kiilerich2014,Gammelmark2014,Macieszczak2016,Kiilerich2016,Cortez2017,Genoni2017,Gong2018,Albarelli2018,Clark2019,Ma2019,Liu2020,Rossi2020,Zhang2020,Wu2021,Ilias2022,Fallani2022,Clark2022,Turner2022,Khanahmadi2023,Yang2023,Boeyens2023,Cabot2024,Ilias2024,Mehboudi2025,Gorecki2025,Khan2025,Midha2025,Mattes2025,Radaelli2026,Yang2026,Cabot2025,Lee2025}, where information is progressively carried away by the emitted radiation, such as photons leaking from a cavity. The dynamics of the sensor is consequently nonunitary with an intertwining of coherent dynamics and measurement backaction. A potential advantage of this approach is its robustness and practicality; for instance, the need for preparing and maintaining fragile entangled states might be alleviated.

The fundamental precision limits in continuous sensing are characterized by two information quantities. The first is the global QFI $I_{\rm G}$, which represents the total information encoded in the joint quantum state of the sensor and its environment. This quantity sets the ultimate bound on precision, assuming one could perform the most general measurements on the combined system-environment state~\cite{Gammelmark2014,Ilias2022}. The second, and often more experimentally relevant, quantity is the environmental QFI $I_{\rm E}$, which quantifies the maximum information that can be extracted by monitoring only the emitted radiation~\cite{Yang2023}. Understanding the bounds and scaling behaviors of $I_{\rm G}$ and $I_{\rm E}$ is crucial for designing optimal sensing protocols, as the QFIs directly set the precision limits on the parameter $\theta$ to be estimated via
\begin{equation}
{\rm Var}[\theta_{\rm est}]\geq\frac{1}{I_{\rm E}}\geq\frac{1}{I_{\rm G}}.
\label{eq11}
\end{equation}

In practice, analytically evaluating the global and environmental QFIs is challenging due to the ever-growing Hilbert space of the environment. Nevertheless, theoretical progress has been made in developing efficient methods to compute these quantities. A notable advancement is the formulation of a generalized Lindblad equation that allows for the calculation of the global QFI by tracking an auxiliary operator that evolves solely within the system Hilbert space~\cite{Gammelmark2014}. This approach has been applied to analyze several continuous sensing scenarios, including those enhanced by dynamical phase transitions~\cite{Macieszczak2016} and criticality~\cite{Ilias2022}, and has been extended to formulate the environmental QFI as well~\cite{Yang2023,Midha2025,Lee2025}.

Despite these developments, many fundamental questions in continuous sensing remain open. A central issue is to understand how the global and environmental QFIs behave in quantum many-particle systems, where collective effects could lead to novel scaling behaviors. Furthermore, while much of the existing work has focused on systems with a finite-dimensional local Hilbert space, such as qubits, the realm of continuous sensing with bosonic systems has received considerably less attention. Addressing this gap is of significant importance, as bosonic systems are foundational to numerous quantum technologies, including quantum optics~\cite{Barbieri2022}, optomechanics~\cite{Aspelmeyer2014}, circuit quantum electrodynamics~\cite{Blais2021}, and levitated particles~\cite{Millen2020,Gonzalez2021}. Understanding the ultimate precision limits for continuously monitored bosonic many-particle systems is therefore a critical step towards harnessing their full metrological potential. This motivates us to pose the following two questions:

\begin{itemize}
\item[(A)]What are the precision limits of continuous sensing with multimode bosonic systems?
\item[(B)]How do the QFIs scale with resources in a physically relevant setup?
\end{itemize}

\begin{figure}[]
\includegraphics[width=8.5cm]{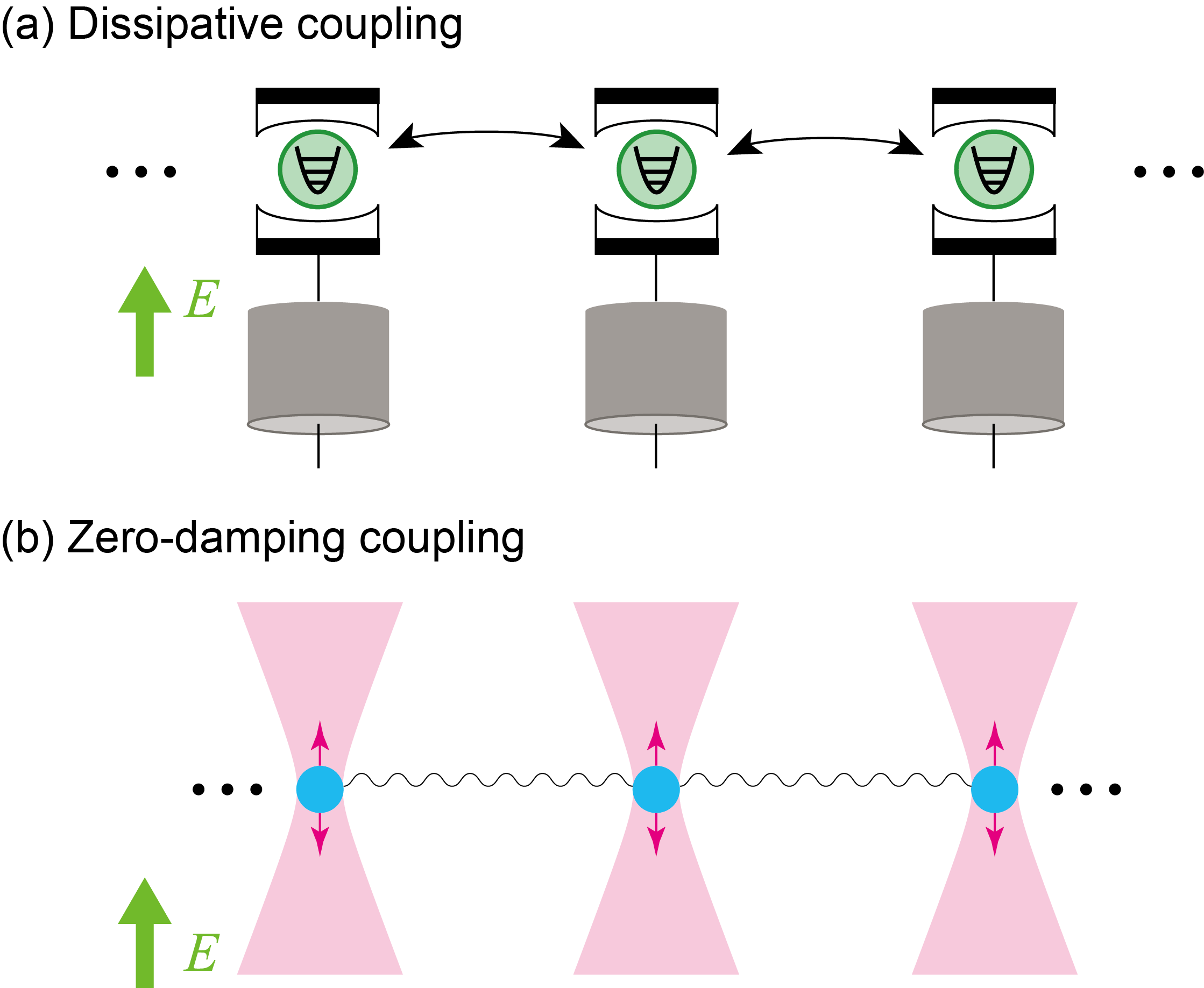}
\caption{\label{fig1}Schematic illustrations of continuous quantum sensing protocols with (a) dissipative or (b) zero-damping system-environment couplings. In both cases, the environment consists of the electromagnetic fields. A linear external field $E$ to be estimated is encoded in the array dynamics. (a) Coupled cavity array, where $M$ coupled cavity modes are connected to waveguides. Photons leaking from the cavities are collected by the waveguides. (b) Trapped particle array, where $M$ trapped particles are arranged periodically.}
\end{figure}

\begin{table*}[htbp]
\centering
\renewcommand{\arraystretch}{1.6} % Increased row height for more white space
\label{tab:combined}
\begin{tabular}{@{\hspace{4mm}}  l|l @{\hspace{13mm}} c@{\hspace{13mm}}c@{\hspace{13mm}}c@{\hspace{13mm}}c@{\hspace{7mm}}}
\toprule
\multicolumn{1}{l}{{\hspace{3mm}} {Coupling}} &{\hspace{5mm}} {Setup} & $I_{\rm G}$ & $I_{\rm E}$ & ${\bar{n}}$ & Asymptotes \\
\midrule
\multirow{2}{*}{Dissipative{\hspace{8mm}}}
&{\hspace{5mm}} local & $O(Mt)$ & $O(Mt)$ & $O(1)$ &\multirow{2}{*}{$\dot{I}_{\rm G,st}=\dot{I}_{\rm E,st}${\hspace{0mm}}} \\
&{\hspace{5mm}} local$\,+\,$global & $O(M^2t)$ & $O(M^2t)$ & $O(1)$ \\
\addlinespace[0.8em] % Adds extra vertical space between sections
\multirow{3}{*}{Zero-damping{\hspace{8mm}}}
&{\hspace{5mm}} local & $O(Mt)$ & $O(Mt)$ & $O(t)$ & \multirow{3}{*}{$\dot{I}_{\rm G,st}>\dot{I}_{\rm E,st}${\hspace{0mm}}} \\
& {\hspace{5mm}}global & $O(M^2t)$ & $O(M^2t)$ & $O(t)$ \\
& {\hspace{5mm}}nonreciprocal & $\exp(O(M))t$ & $O(Mt)$ & $\exp(O(M))t$ \\
\bottomrule
\end{tabular}
\caption{\label{tab1} Summary of the long-time asymptotic scalings of the global QFI $I_{\rm G}$, environmental QFI $I_{\rm E}$, and the number of bosonic excitations per mode ${\bar{n}}$. The table describes two classes of system-environment couplings: dissipative coupling where the system evolves towards a unique steady state, and zero-damping coupling leading to long-lasting oscillations and unbounded energy growth. In both cases, when couplings are local, the QFIs yield the linear scaling with the number of modes $M$ and time $t$, while the inclusion of a global coupling enhances the QFIs quadratically with respect to $M$. In the case of zero-damping couplings, a nonreciprocal setup can leverage the non-Hermitian skin effect to produce an exponentially enhanced global QFI at the cost of energy consumption. Notably, however, this enhancement is not reflected in the environmental QFI, revealing a qualitative distinction between the total information encoded in the joint state and the information radiated into the environment. The asymptotic rates of the global and environmental QFIs, denoted by $\dot{I}_{\rm G,st}$ and $\dot{I}_{\rm E,st}$, respectively, coincide in the case of dissipative couplings, while the former is generally larger than the latter in the case of zero-damping couplings.}
\end{table*}

To address these questions, we develop a comprehensive theoretical framework for the quantum metrology of multimode free bosons subjected to continuous Gaussian measurements. Throughout this work, we focus on the linear dynamics of a bosonic Gaussian state that allows for an analytical treatment. This situation arises when a system Hamiltonian is at most quadratic in quadrature operators and jump operators are given by linear combinations thereof. We consider a sensor driven by a linear field to be estimated and evaluate the maximum information about the parameter encoded in the joint system-environment state.

Within this framework, we analyze the asymptotic scaling behaviors of the global and environmental QFIs in the long-time limit. Indeed, we find that both QFIs grow linearly in time in the long-time limit, and our framework provides analytical expressions for the corresponding asymptotic growth rates. Using these expressions, we derive asymptotic resource-dependent bounds in terms of key physical resources in free-bosonic systems. Importantly, the asymptotic rate of the global QFI depends on the number of modes $M$ and the number of bosonic excitations per mode $\bar{n}$, whereas the asymptotic rate of the environmental QFI is controlled only by the mode number $M$.

More specifically, we consider continuous sensing in two broad classes of system-environment couplings. The first involves {\it dissipative} couplings, as exemplified by an array of coupled cavities [Fig.~\ref{fig1}(a)]. Since photons in each cavity leak into the attached waveguides, the system evolves towards a unique steady state. The emitted photons are continuously monitored via, for example, photon counting, homodyne detection, or heterodyne detection. We here restrict our attention to dissipations arising solely from measurement channels. The second class comprises {\it zero-damping} couplings, e.g., in an array of trapped particles measured by collecting scattered photons [Fig.~\ref{fig1}(b)]. These are situations in which the linear dynamics of mean values is purely oscillatory with unbounded energy growth, since no dissipative stabilization is present. Hence, the system state does not reach a steady state. While this is a rather ideal situation with no dissipative channels, this simplification allows for explicit analytical evaluations of the asymptotic QFI growth rates and their resource scalings. The simplest realization occurs when the coupling operators are linear and Hermitian. However, more exotic scenarios are also possible, where jump operators are non-Hermitian yet induce no net damping, analogue to a parity-time (PT) symmetry~\cite{Ashida2020} in the PT unbroken phase.

The theoretical model shown in Fig.~\ref{fig1}(b) is based on recent analyses of trapped particle array~\cite{Rudolph2022,Yokomizo2023,Rudolph2024v1,Rudolph2024v2,Yokomizo2025,Ho2025v2} and is directly motivated by experiments demonstrating coherent control and manipulation of levitated nanoparticles~\cite{Delic2019,Rieser2022,Yan2023,Vijayan2024}. This situation corresponds to direct probing of quadrature operators of the internal bosonic modes, without inducing any photon loss. Such a measurement scheme can be implemented, for instance, by using a backaction-evading measurement~\cite{Clerk2008,Hertzberg2010,Clerk2010}. We remark that there exists a third scenario in which the coupling to the environment induces exponential growth of mode amplitudes; this unstable regime is not considered in the present work.

We summarize the asymptotic scaling behaviors of the global and environmental QFIs, $\dot{I}_{\rm G,st}$ and $\dot{I}_{\rm E,st}$, for both dissipative and zero-damping couplings in Table~\ref{tab1}. Suppose that the strength of the system-environment coupling is a constant independent of the mode number. Interestingly, the comparison between $\dot{I}_{\rm G,st}$ and $\dot{I}_{\rm E,st}$ reveals that the information stored in the joint system-environment state and the information carried by the emitted radiation field can exhibit various asymptotic scalings depending on the nature of system-environment couplings. Specifically, in the case of dissipative couplings, we find that these asymptotic growth rates coincide:
\begin{equation}
\dot{I}_{\rm G,st}=\dot{I}_{\rm E,st}.
\label{eq12}
\end{equation}
A similar equivalence has also been found in discrete-variable systems with finite local Hilbert-space dimension~\cite{Midha2025}. We emphasize, however, that this equivalence concerns only the long-time asymptotic rates; the full time-dependent QFIs do not necessarily coincide. In contrast, in the case of zero-damping coupling, we find that the asymptotic growth rates of the global and environmental QFIs are generally separated,
\begin{equation}
\dot{I}_{\rm G,st}>\dot{I}_{\rm E,st},
\label{eq13}
\end{equation}
and can exhibit qualitatively different resource scalings. This separation reflects the absence of dissipative relaxations in the zero-damping case; the system can retain parameter-dependent coherent excitations, and the emitted radiation field does not exhaust the information contained in the joint system-environment state.

To further characterize the best-case performance of a given sensing setup~\cite{Lee2025}, we introduce the optimized QFIs, $\dot{I}_{\rm G,st}^\ast$ and $\dot{I}_{\rm E,st}^\ast$, defined by maximizing the corresponding asymptotic QFI rates over normalized linear signal profiles. We then derive asymptotic resource-dependent bounds on these optimized rates in terms of key physical resources for two distinct classes of system-environment couplings. In the derivation, we impose a set of physically motivated assumptions that specify the regime of validity of the results. For dissipative couplings, we assume strict dissipativity, ensured by the presence of decay channels. We then show that the optimized rates are bounded as
\begin{equation}
r_1\bar{n}_{\rm st}M\leq\dot{I}_{\rm G,st}^\ast=\dot{I}_{\rm E,st}^\ast\leq r_2\bar{n}_{\rm st}M^2,
\label{eq14}
\end{equation}
where $r_{1,2}>0$ are constants independent of the number of modes $M$, and $\bar{n}_{\rm st}$ is the steady-state value of the excitation number per mode. In contrast, for the zero-damping coupling, we assume a nonzero asymptotic excitation-growth rate, so that the generated bosonic excitations provide a well-defined energetic resource in the long-time regime. We then show that the optimized global QFI rate is bounded by
\begin{equation}
r_3\dot{\bar{n}}_{\rm st}M\leq\dot{I}_{\rm G,st}^\ast\leq r_4\dot{\bar{n}}_{\rm st}M^2,
\label{eq15}
\end{equation}
while the optimized environmental QFI is subject to a different constraint:
\begin{equation}
r_5M\leq\dot{I}_{\rm E,st}^\ast\leq\min(r_6,r_7\dot{\bar{n}}_{\rm st})\cdot M^2,
\label{eq16}
\end{equation}
with $r_{3\mathchar`-7}>0$ being constants. The bounds in Eqs.~(\ref{eq14}) and (\ref{eq15}) indicate that the optimized global QFI can exhibit a quadratic scaling with the mode number $M$, while it scales at most linearly with the energy resource, given by the average number of bosonic excitations $\bar{n}_{\rm st}$ or its rate $\dot{\bar{n}}_{\rm st}$. These bounds also highlight a qualitative difference between the optimized global and environmental QFIs in the zero-damping case. Indeed, the bounds in Eq.~(\ref{eq16}) for $\dot{I}_{\rm E,st}^\ast$, which is distinct from Eq.~(\ref{eq15}), clearly suggests that the information carried by the emitted field need not scale in the same way as the information contained in the joint system-environment state.

To demonstrate the asymptotic scaling, we analyze several concrete models relevant to a coupled cavity array or a trapped particle array. When the system-environment couplings are local, both the global and environmental QFIs scale linearly with the number of bosonic modes $M$. In contrast, the inclusion of global setups realizable, e.g., by placing the sensor array within a cavity~\cite{Ho2025,Peters2025}, enables the attainment of the quadratic scaling $I_{\rm G,E}\propto M^2$ in the long-time regime. We note that these scalings are obtained at a fixed coupling strength per mode, so that the total coupling strength can increase with $M$. Furthermore, for the zero-damping coupling, we investigate the setup where a nonreciprocal manipulation leads to a unidirectional energy flow inside the system. In other words, the engineered nonreciprocity induces a non-Hermitian skin effect, resulting in an exponential scaling of the asymptotic bosonic excitation rate $\dot{\bar{n}}_{\rm st}\propto\exp(O(M))$. The internally generated bosonic excitations can then be converted into metrological information, as indicated by an exponentially enhanced global QFI. We note that this behavior reflects the exponential growth of the excitation resource generated by the non-Hermitian skin effect. Interestingly, however, this enhancement does not appear in the environmental QFI, which retains a linear scaling with $M$. This discrepancy highlights a fundamental distinction between the global information and the portion of information radiated into the environment as described above. Thus, our analysis identifies collective couplings and nonreciprocal bosonic amplification as mechanisms that generate distinct QFIs scaling behaviors in continuously monitored bosonic systems.

From a broader perspective, our work is also related to previous studies on non-Hermitian sensing and many-body metrology. As described above, the exponentially enhanced global QFI we predict in the nonreciprocal setup is a direct consequence of the so-called non-Hermitian skin effect~\cite{Yao2018,Okuma2020,Zhang2020NH}. In this respect, our finding contributes to a growing body of research exploring the use of non-Hermitian skin effects for designing highly sensitive sensors~\cite{Mcdonald2020,Budich2020,Koch2022,Yuan2023,Sarkar2024,Deng2024,Xie2024,Parto2025,Clavero2025,Zeng2025,Blanchard2025}. A key distinction from these studies is that our analysis is based on the continuous sensing and demonstrates the nonreciprocal enhancement in the global QFI. Moreover, while the previous studies on non-Hermitian sensing have largely focused on classical or discrete quantum systems, our work extends the concepts to the realm of multimode continuous-variable quantum systems. Our study also offers a complementary perspective to ongoing studies of many-body sensing~\cite{Montenegro2025}, which have so far concentrated on the interferometric setups based on the unitary evolution rather than the nonunitary dynamics of continuously monitored open systems as considered here. Our results demonstrate that the interplay between coherent dynamics and measurement backaction in bosonic many-particle systems gives rise to a rich and largely unexplored landscape for continuous quantum metrology, such as unique scaling behaviors and resource trade-offs that fundamentally differ from their single-shot interferometric counterparts.

The remainder of this paper is organized as follows. In Sec.~\ref{sec2}, we review the quantum metrology based on continuous measurements and define the global and environment QFIs. In Sec.~\ref{sec3}, we present a framework for the Gaussian quantum metrology under continuous measurements. We then show that the quantum fluctuations determine the time evolution of the global and environmental QFIs. In Sec.~\ref{sec4}, for the dissipative coupling, we derive the bounds and scalings of the global and environmental QFIs in terms of resources and demonstrate those general results by studying the coupled cavity array. Similarly, in Sec.~\ref{sec5}, we derive the bounds and scalings of the QFIs for the zero-damping coupling, and demonstrate them by analyzing the trapped particle array. In Sec.~\ref{sec6}, we discuss the asymptotic growth rates of the QFIs in the convention of the fixed-total-coupling strength. Finally, in Sec.~\ref{sec7}, we summarize our results and outline future directions, including the experimental feasibility of our proposal.

%
% -------------------------------------------------------------------------------------------------------------------------------------------------------------------------------------
%

\section{\label{sec2}Quantum sensing under continuous measurements}
In this section, we review the theoretical framework for quantum metrology based on continuous measurements, primarily drawing upon the formalisms developed in Refs.~\cite{Gammelmark2014,Yang2023}. The ultimate precision in quantum continuous metrology is determined by the QFI of the global state, which encompasses both the sensor and the environmental degrees of freedom that carry the total information. This quantity, termed the {\it global} QFI represents the fundamental limit on precision achievable with the most general measurements on the combined system and environment. While a direct treatment of the global state is often intractable due to the growing complexity of the environment, an efficient method for calculating the global QFI exists by solving a generalized Lindblad equation that involves only the system degrees of freedom~\cite{Gammelmark2014}.

From a practical perspective, however, sensing protocols typically grant access only to the radiation emitted by the sensor. The maximal information that can be extracted from this output field is quantified by the {\it environmental} QFI, which is associated with the reduced quantum state of the environment and is necessarily upper-bounded by the global QFI~\cite{Yang2023}. One notable question is therefore to identify the conditions under which the environmental QFI saturates the global QFI, ensuring that measurements solely on the emitted radiation are sufficient to attain the precision limit as addressed below.

%
% -------------------------------------------------------------------------------------------------------------------------------------------------------------------------------------
%

\subsection{\label{sec2A}Setup}
We consider the dynamics of an open quantum system coupled to a Markovian environment, for which the Born-Markov approximation is valid. A central consequence of this approximation is that the environment can be treated as a sequence of independent, ancillary subsystems or modes, with each mode interacting with the system for a single time interval $[t_k,t_k+dt]$ with $t_k=kdt$. The total Hilbert space of the composite entity at time $t_k$ is thus given by the tensor product of ${\cal H}_S$ and $\otimes_{i=0}^{k-1}{\cal H}_{E,i}$, where ${\cal H}_S$ is the system Hilbert space and ${\cal H}_{E,i}$ is the Hilbert space of the environmental mode interacting with the system during the $i$th time step.

For each environmental mode, we associate an $(l+1)$-dimensional Hilbert space ${\cal H}_{E,i}$ spanned by an orthonormal basis $\{|m_i\rangle\}$, where $m_i=0,1,\dots,l$. We assume that, prior to the interaction, every environmental mode is prepared in a fiducial vacuum state $|0_i\rangle$. The system, meanwhile, starts in an arbitrary initial state $|\psi(0)\rangle\in{\cal H}_S$. The joint evolution of the system and the $i$th environmental mode during the interval $[t_i,t_i+dt]$ is described by a unitary operator $\hat{U}_i(\theta)$ that acts on ${\cal H}_S\otimes{\cal H}_{E,i}$. The global state of the system and all environmental modes after $k$ steps is then generated by the sequential application of these unitaries:
\begin{equation}
|\psi_\theta(t_k)\rangle=\hat{U}_{t_{k-1}}(\theta)\cdots\hat{U}_{t_0}(\theta)\left(|\psi(0)\rangle\bigotimes_{i=0}^{k-1}|0_i\rangle\right).
\label{eq2A1}
\end{equation}
Projective measurements performed on the environmental modes, characterized by a set of outcomes $\{m_0,m_1,\dots,m_{k-1}\}$, induce transitions from $|0_i\rangle$ to $|m_i\rangle$ in each environmental subspace. This process imparts a corresponding measurement backaction on the system. To make this explicit, we define the Kraus operators, $\hat{\Omega}_{m_i}(\theta)$, which act solely on ${\cal H}_S$ and are defined by the matrix elements of the unitary operator between environmental basis states, namely, $\hat{\Omega}_{m_i}(\theta)\equiv\langle m_i|\hat{U}_i(\theta)|0_i\rangle$. By inserting the identity $\hat{I}_{E,i}=\sum_{m_i}|m_i\rangle\langle m_i|$ for each environmental mode, the global state $|\Psi_\theta(t)\rangle$ can be expanded in the basis of measurement outcomes as
\begin{eqnarray}
&&|\Psi_\theta(t)\rangle=\sum_{m_0,\cdots,m_{k-1}}\hat{\Omega}_{m_{k-1}}(\theta)\cdots\hat{\Omega}_{m_0}(\theta)|\psi(0)\rangle \nonumber\\
&&\otimes|m_{k-1},\dots,m_0\rangle.
\label{eq2A2}
\end{eqnarray}
The unitarity of $\hat{U}_i(\theta)$ ensures that the Kraus operators for a single time step form a positive operator-valued measure, satisfying $\sum_{m_i}\hat{\Omega}_{m_i}^\dag(\theta)\hat{\Omega}_{m_i}(\theta)=\hat{I}_S$. A specific sequence of outcomes $\{m_0,m_1,\dots,m_{k-1}\}$ constitutes a single measurement record, and the corresponding unnormalized conditional state of the system, $\hat{\Omega}_{m_{k-1}}(\theta)\cdots\hat{\Omega}_{m_0}(\theta)|\psi(0)\rangle$, defines a stochastic quantum trajectory.

The ensemble of all such trajectories reproduces the unconditional dynamics of the reduced density matrix, which is described by the Lindblad master equation. To see this, we trace out the environmental degrees of freedom in Eq.~(\ref{eq2A2}), leading to the linear map given by
\begin{equation}
\hat{\rho}_{t_i}=\sum_n\hat{\Omega}_n(\theta)\hat{\rho}_{t_{i-1}}\hat{\Omega}_n^\dag(\theta).
\label{eq2A3}
\end{equation}
For an infinitesimal time step, the evolution becomes
\begin{equation}
\frac{d\hat{\rho}}{dt}=\frac{1}{dt}\left[\sum_n\hat{\Omega}_n(\theta)\hat{\rho}\hat{\Omega}_n^\dag(\theta)-\hat{\rho}\right].
\label{eq2A4}
\end{equation}
We remark that $n=0$ corresponds to the null measurement, while $n\neq0$ represents a quantum jump, i.e., the detection of the radiation emitted from the system. Correspondingly, the jump operators are defined as
\begin{eqnarray}
&&\hat{\Omega}_0(\theta)=\hat{1}-i\left(\hat{H}(\theta)-\frac{i}{2}\sum_n\hat{L}_n^\dag\hat{L}_n\right)dt, \label{eq2A5}\\
&&\hat{\Omega}_n=\hat{L}_n\sqrt{dt},\label{eq2A6}
\end{eqnarray}
where $\hat{H}(\theta)$ is a Hamiltonian of the system; throughout this work, we assume that the parameter $\theta$ to be estimated is encoded in the Hamiltonian and adopt units where $\hbar=1$. As a result, we obtain the Lindblad master equation as follows:
\begin{equation}
\frac{d\hat{\rho}}{dt}=-i[\hat{H}\left(\theta\right),\hat{\rho}]+\sum_n\left(\hat{L}_n\hat{\rho}\hat{L}_n^\dag-\frac{1}{2}\{\hat{L}_n^\dag\hat{L}_n,\hat{\rho}\}\right).
\label{eq2A7}
\end{equation}

%
% -------------------------------------------------------------------------------------------------------------------------------------------------------------------------------------
%

\subsection{\label{sec2B}Quantum Fisher information of the global state}
In the context of continuous quantum metrology, where an open quantum system continuously interacts with its environment, information about the parameter $\theta$ encoded in the system dynamics is progressively carried away by the emitted radiation. The ultimate precision limit for estimating $\theta$ is therefore determined by the total information contained within the combined state of the system and its environment. This limit is quantified by the global QFI, $I_{\rm G}(t)$, which accounts for any measurement strategy performed on the joint system-environment state. However, a direct computation of $I_{\rm G}(t)$ is in general challenging, as it would necessitate tracking the evolution of the global state in Eq.~(\ref{eq2A1}) within the continuously growing Hilbert space of the environment.

A tractable approach, however, allows for the calculation of the global QFI by considering only the system degrees of freedom~\cite{Gammelmark2014}. The methodology begins with the formal definition of the QFI, which, for a family of pure states, is intrinsically linked to the Bures fidelity. Specifically, the global QFI is given by the curvature of the quantum fidelity ${\cal F}_{\rm G}(\theta_1,\theta_2,t)=|\langle\Psi_{\theta_1}(t)|\Psi_{\theta_2}(t)\rangle|$ between two global states evolved under infinitesimally different parameter values, $\theta_1$ and $\theta_2$ (see Appendix~\ref{secA}):
\begin{equation}
I_{\rm G}(t)=\left.4\partial_{\theta_1}\partial_{\theta_2}\log{\cal F}_{\rm G}(\theta_1,\theta_2,t)\right|_{\theta_1=\theta_2=\theta}.
\label{eq2B1}
\end{equation}
The key insight is that the fidelity between global pure states can be simplified by tracing out the environmental degrees of freedom. This procedure maps the problem onto the system Hilbert space:
\begin{align}
{\cal F}_{\rm G}(\theta_1,\theta_2,t)&=|{\rm Tr}_{\rm Env,Sys}[|\Psi_{\theta_2}(t)\rangle\langle\Psi_{\theta_1}(t)|]| \nonumber\\
&=|{\rm Tr}_{\rm Sys}[\hat{\mu}_{\theta_1,\theta_2}(t)]|,
\label{eq2B2}
\end{align}
where we have introduced the pseudo-reduced density matrix,
\begin{equation}
\hat{\mu}_{\theta_1,\theta_2}(t)\equiv{\rm Tr}_{\rm Env}[|\Psi_{\theta_2}(t)\rangle\langle\Psi_{\theta_1}(t)|]
\label{eq2B3}
\end{equation}
This operator, acting solely on the system Hilbert space, can be considered as a generalization of the density matrix, which, for $\theta_1=\theta_2=\theta$, reduces to $\hat{\mu}_{\theta,\theta}(t)=\hat{\rho}$. When $\theta_1\neq\theta_2$, however, $\hat{\mu}_{\theta_1,\theta_2}(t)$ is not necessarily Hermitian or positive semidefinite. It serves as a mathematical object that quantifies the overlap between the two distinct evolutionary histories of the global state, projected back onto the system degrees of freedom. This formulation allows the global QFI to be calculated entirely from the system degrees of freedom:
\begin{equation}
I_{\rm G}(t)=\left.4\partial_{\theta_1}\partial_{\theta_2}\log|{\rm Tr}_{\rm Sys}[\hat{\mu}_{\theta_1,\theta_2}(t)]|\right|_{\theta_1=\theta_2=\theta}.
\label{eq2B4}
\end{equation}

The problem of determining the ultimate precision limit thus boils down to finding the time evolution of $\hat{\mu}_{\theta_1,\theta_2}(t)$. As described above, in an infinitesimal time step $dt$, the joint system-environment evolution is generated by the parameter-dependent Kraus operators $\{\hat{\Omega}_n(\theta)\}$. The pseudo-reduced density matrix at time $t+dt$ is then related to its value at time $t$ by the update rule,
\begin{equation}
\hat{\mu}_{\theta_1,\theta_2}(t+dt)=\sum_n\hat{\Omega}_n(\theta_1)\hat{\mu}_{\theta_1,\theta_2}(t)\hat{\Omega}_n^\dag(\theta_2).
%\frac{d\hat{\mu}_{\theta_1,\theta_2}}{dt}=\frac{1}{dt}\left[\sum_n\hat{\Omega}_m(\theta_1)\hat{\mu}_{\theta_1,\theta_2}\hat{\Omega}_n^\dag(\theta_2)-\hat{\mu}_{\theta_1,\theta_2}\right].
\label{eq2B5}
\end{equation}
By taking the continuous-time limit $dt\rightarrow0$ and identifying the jump operators analogously to Eqs.~(\ref{eq2A5}) and (\ref{eq2A6}), we obtain the generalized Lindblad equation as follows:
\begin{eqnarray}
&&\frac{d\hat{\mu}_{\theta_1,\theta_2}}{dt}=-i\hat{H}(\theta_1)\hat{\mu}_{\theta_1,\theta_2}+i\hat{\mu}_{\theta_1,\theta_2}\hat{H}(\theta_2) \nonumber\\
&&+\sum_n\left(\hat{L}_n\hat{\mu}_{\theta_1,\theta_2}\hat{L}_n^\dag-\frac{1}{2}\{\hat{L}_n^\dag\hat{L}_n,\hat{\mu}_{\theta_1,\theta_2}\}\right).
\label{eq2B6}
\end{eqnarray}
This equation provides a complete description of the dynamics required to calculate the global QFI. While the dissipative terms, which model the influence of the environment, retain the same form as in the standard Lindblad master equation for the reduced density matrix $\hat{\rho}$, the coherent evolution breaks the Hermiticity, with $\hat{H}(\theta_1)$ propagating the state forward in time from left and $\hat{H}(\theta_2)$ propagating it backward from the right. By solving this linear differential equation for $\hat{\mu}_{\theta_1,\theta_2}(t)$ in the neighborhood of $\theta_1=\theta_2=\theta$, the global QFI can be calculated without any explicit reference to the environmental state.

%
% -------------------------------------------------------------------------------------------------------------------------------------------------------------------------------------
%

\subsection{\label{sec2C}Quantum Fisher information of the emitted radiation}
In many sensing experiments, particularly within quantum optics and related fields, parameter estimation is performed by using the information obtained from monitoring the radiation emitted from the system. The quanta emitted by the system, such as photons leaking from a cavity or scattered by particles, carry away information about the parameter $\theta$. Continuous measurements of environment thus provide a nondemolition strategy for parameter estimation. The maximum amount of accessible information in such situation is quantified by the QFI of the emission field, which is represented by the reduced density matrix of the environment. To formalize this, we consider a pure joint system-environment state in Eq.~(\ref{eq2A2}) and construct the environment reduced density matrix by tracing out the system degrees of freedom,
\begin{equation}
\hat{\rho}_\theta^{\rm E}(t)={\rm Tr}_{\rm Sys}[|\Psi_\theta(t)\rangle\langle\Psi_\theta(t)|].
\label{eq2C1}
\end{equation}

The QFI associated with this environmental state, which we refer to as the environmental QFI and denote by $I_{\rm E}(t)$, sets the ultimate precision bound for any estimation protocol based solely on monitoring the emission field. Following the standard definition of the QFI, this quantity is derived from the Bures fidelity between two environmental states corresponding to infinitesimally different parameter values, $\theta_1$ and $\theta_2$,
\begin{equation}
I_{\rm E}(t)=\left.4\partial_{\theta_1}\partial_{\theta_2}\log{\cal F}_{\rm E}(\theta_1,\theta_2,t)\right|_{\theta_1=\theta_2=\theta},
\label{eq2C2}
\end{equation}
where ${\cal F}_{\rm E}(\theta_1,\theta_2,t)$ measures the distinguishability between the two environmental state $\hat{\rho}_{\theta_1}^{\rm E}(t)$ and $\hat{\rho}_{\theta_2}^{\rm E}(t)$, and is defined as
\begin{equation}
{\cal F}_{\rm E}(\theta_1,\theta_2,t)={\rm Tr}_{\rm Env}\left[\sqrt{\sqrt{\hat{\rho}_{\theta_1}^{\rm E}(t)}\hat{\rho}_{\theta_2}^{\rm E}(t)\sqrt{\hat{\rho}_{\theta_1}^{\rm E}(t)}}\right].
\label{eq2C3}
\end{equation}
While these definitions are formally exact, a direct calculation of $I_{\rm E}(t)$ from them is in general intractable because the dimension of the environment Hilbert space grows with time.

A useful method to circumvent this difficulty is to reformulate the environmental quantum fidelity in terms of the pseudo-reduced density matrix in Eq.~(\ref{eq2B3}), which acts solely on the system Hilbert space. Specifically, the environmental quantum fidelity can be expressed as the trace norm of this operator~\cite{Yang2023} (see Appendix~\ref{secB}),
\begin{equation}
{\cal F}_{\rm E}(\theta_1,\theta_2,t)={\rm Tr}_{\rm Sys}\left[\sqrt{\hat{\mu}_{\theta_1,\theta_2}(t)\hat{\mu}_{\theta_1,\theta_2}^\dag(t)}\right],
\label{eq2C4}
\end{equation}
where the dynamics of $\hat{\mu}_{\theta_1,\theta_2}(t)$ is governed by Eq.~(\ref{eq2B6}). This approach provides a practical and efficient pathway to compute the environmental QFI and analyze the performance of sensing protocols based on continuous measurements.

Finally, the environmental QFI can be related to the global QFI by the inequality $I_{\rm E}(t)\leq I_{\rm G}(t)$, which follows from monotonicity of the QFI under partial trace. This relation has a clear physical interpretation: the information accessible from the environment is at most equal to the total information encoded in the joint system-environment state. The difference $\delta I=I_{\rm G}(t)-I_{\rm E}(t)$ arises from the information that remains localized within the system or is stored in system-environment correlations.

%
% -------------------------------------------------------------------------------------------------------------------------------------------------------------------------------------
%

\subsection{\label{sec2D}Remark on classical Fisher information}
The operational meaning of the environmental QFI should be distinguished from that of the classical Fisher information associated with a specific measurement record. The environmental QFI is interpreted as an information-theoretic upper bound on the amount of parameter information contained in the emitted radiation field. By contrast, the actually attainable sensitivity under a specified detection scheme is characterized by the classical Fisher information of the corresponding measurement record, such as a homodyne, heterodyne, or photon-counting record. In general, saturating the environmental QFI may require an optimal measurement of the full temporal multimode output state, which can be nonlocal in time.

%
% -------------------------------------------------------------------------------------------------------------------------------------------------------------------------------------
%

\subsection{\label{sec2E}Redundancy condition}
We here derive the condition under which any information that remains in the sensor system is redundant with the information already transferred to the environment. From an experimental standpoint, environmental monitoring is often the only feasible measurement strategy, and identifying such condition should be of practical importance. To this end, we recall that the global QFI is obtained through the modulus of the trace of the pseudo-density matrix $\hat{\mu}_{\theta_1,\theta_2}$, while the environmental QFI is characterized its trace norm. Let us denote the eigenvalues and singular values of $\hat{\mu}_{\theta_1,\theta_2}$ as $\{\lambda_d\}$ and $\{\sigma_d\}$, respectively, for a system with a $D$-dimensional Hilbert space. The two fidelities can then be expressed as
\begin{eqnarray}
&&{\cal F}_{\rm G}=|{\rm Tr}_{\rm Sys}[\hat{\mu}_{\theta_1,\theta_2}]|=\left|\sum_{d=1}^D\lambda_d\right|, \label{eq2D1}\\
&&{\cal F}_{\rm E}={\rm Tr}_{\rm Sys}\left[\sqrt{\hat{\mu}_{\theta_1,\theta_2}\hat{\mu}_{\theta_1,\theta_2}^\dagger}\right]=\sum_{d=1}^D\sigma_d. \label{eq2D2}
\end{eqnarray}
Since the QFIs are determined by the second-order expansion of these fidelities for $\theta_1$ and $ \theta_2$, the condition $I_G=I_E$ is met if and only if ${\cal F}_{\rm G}={\cal F}_{\rm E}$ is satisfied around $\theta_1=\theta_2$.

For any matrix, the modulus of the sum of its eigenvalues is bounded by the sum of its singular values through the general inequalities $|\sum_{d}\lambda_d|\leq\sum_{d}|\lambda_d|\leq\sum_{d}\sigma_d$. The equality ${\cal F}_{\rm G}={\cal F}_{\rm E}$ holds if and only if both inequalities  are saturated. The first inequality, $|\sum_{d} \lambda_d|=\sum_{d}|\lambda_d|$, is saturated if all nonzero eigenvalues $\{\lambda_d\}$ share a common phase. The second inequality follows from Weyl's majorant theorem and is saturated when the matrix $\hat{\mu}_{\theta_1,\theta_2}$ is normal, i.e., $[\hat{\mu}_{\theta_1,\theta_2},\hat{\mu}_{\theta_1,\theta_2}^\dag]=0$. These two requirements are jointly satisfied iff $\hat{\mu}_{\theta_1,\theta_2}$ is proportional to a positive semidefinite operator. Specifically, we can state the necessary and sufficient condition for ${\cal F}_{\rm G}={\cal F}_{\rm E}$ and consequently $I_G=I_E$ as
\begin{equation}
\hat{\mu}_{\theta_1,\theta_2}(t)=e^{-\eta(\theta_1,\theta_2,t)}\hat{\rho}(t),
\label{eq2D3}
\end{equation}
where $\eta(\theta_1,\theta_2,t)\in{\mathbb C}$ is a complex scalar and $\hat{\rho}$ is a valid density matrix, i.e., it is a Hermitian and positive semidefinite matrix with unit trace.

It is noteworthy that an auxiliary-qubit formulation of the generalized Lindblad equation (\ref{eq2B6}) provides a clear physical interpretation of the redundancy condition (\ref{eq2D3}). The key observation is that the pseudo-density matrix $\hat{\mu}_{\theta_1,\theta_2}$ can be interpreted as an off-diagonal block of a density matrix describing a sensor system coupled to an auxiliary qubit via a dispersive interaction~\cite{Molmer2015,Lee2025}. In this picture, the real part of $\eta(\theta_1,\theta_2,t)$ in Eq.~(\ref{eq2D3}) characterizes the dephasing of the auxiliary qubit. This implies that there does not exist any operation acting solely on the sensor system that can restore the coherence of the auxiliary qubit. Consequently, the information acquired by the sensor system is lost to the environment, leading to the equivalence between the global and environmental QFIs.

%
% -------------------------------------------------------------------------------------------------------------------------------------------------------------------------------------
%

\section{\label{sec3}General framework}
We provide a theoretical framework to analyze the precision bounds for a continuously monitored free-bosonic system. We begin by introducing our model, which consists of a system of multimode free bosons governed by a quadratic Hamiltonian and subjected to a continuous Gaussian measurement. In this setup, a quantum state is maintained as a bosonic Gaussian state, allowing its dynamics to be fully characterized by the evolution of its first and second statistical moments. Building upon this formalism, we delineate the methodology for calculating the global QFI for a parameter encoded in the system. Our analysis reveals that the behavior of the global QFI is predominantly governed by the intrinsic quantum fluctuations of the bosonic state. We also derive an analytical formula of the difference between the global and environmental QFIs, which allows for efficiently evaluating the environmental QFI. For the sake of notational simplicity, in the following, ${\rm Tr}[\cdots]$ denotes the trace over the system degrees of freedom.

%
% -------------------------------------------------------------------------------------------------------------------------------------------------------------------------------------
%

\subsection{\label{sec3A}Free bosons under continuous Gaussian measurements}
We consider a system composed of $M$ bosonic modes described within a continuous-variable formalism, where the canonical observables are the position and momentum quadrature operators, $\hat{x}_j$ and $\hat{p}_j$, respectively, for each mode $j=1,2,\dots,M$. For the sake of convenience, we arrange these operators into a $2M$-component vector,
\begin{equation}
\hat{\bm\phi}=(\hat{x}_1,\dots,\hat{x}_M,\hat{p}_1,\dots,\hat{p}_M)^{\rm T}.
\label{eq3A1}
\end{equation}
These operators satisfy the canonical commutation relations, which can be expressed in a matrix form as
\begin{eqnarray}
[\hat{\phi}_j,\hat{\phi}_k]=i\sigma_{jk},~\sigma=\left( \begin{array}{cc}
O    & 1_M \vspace{3pt}\\
-1_M & O
\end{array}\right),
\label{eq3A2}
\end{eqnarray}
where $1_M$ and $O$ denote the $M\times M$ identity matrix and zero matrices, respectively. The dynamics of a free-bosonic system is governed by a Hamiltonian that is, at most, quadratic in the quadrature operators. We consider a general form given by
\begin{equation}
\hat{H}=\frac{1}{2}\hat{\bm\phi}^{\rm T}{\mathbb H}\hat{\bm\phi}+\theta{\bm a}^{\rm T}\cdot\hat{\bm\phi},
\label{eq3A3}
\end{equation}
where ${\mathbb H}$ is a $2M\times2M$ real, symmetric matrix, and ${\bm a}$ is a $2M$-element real vector. The second term represents a linear field driving the system, whose strength is proportional to a parameter $\theta$ to be estimated. For instance, in a setup involving an array of trapped particles, this term can model the displacement of the particles due to an external field, which is a scenario we will discuss in Sec.~\ref{sec5}.

We next introduce the class of bosonic Gaussian states. A Gaussian state, denoted by a density matrix $\hat{\rho}_{\rm G}$, is fully characterized by the first and second moments of the quadrature operators. A key tool in the Gaussian formalism is the displacement operator, which acts on the state in phase space as
\begin{equation}
\hat{D}_{\bm v}\equiv\exp(i\hat{\bm\phi}^{\rm T}\sigma{\bm v}),
\label{eq3A4}
\end{equation}
where ${\bm v}\in{\mathbb R}^{2M}$ is a vector of real phase-space coordinates. Using this, the characteristic function for a general operator $\hat{A}$ is defined as its expectation value,
\begin{equation}
\chi_A({\bm v})={\rm Tr}[\hat{A}\hat{D}_{\bm v}].
\label{eq3A5}
\end{equation}
A bosonic state is then defined as Gaussian if its characteristic function has a Gaussian form~\cite{Weedbrook2012}:
\begin{equation}
\chi_{\rho_{\rm G}}({\bm v})=\exp\left(-\frac{1}{2}{\bm v}^{\rm T}\sigma^{\rm T}\Gamma_\phi\sigma{\bm v}+i{\bm\phi}^{\rm T}\sigma{\bm v}\right).
\label{eq3A6}
\end{equation}
This distribution is completely specified by the vector of mean values,
\begin{equation}
{\bm\phi}=\langle\hat{\bm\phi}\rangle_{\rm G},
\label{eq3A7}
\end{equation}
and the $2M\times2M$ real, symmetric covariance matrix,
\begin{equation}
\left(\Gamma_\phi\right)_{jk}=\frac{1}{2}\langle\{\delta\hat{\phi}_j,\delta\hat{\phi}_k\}\rangle_{\rm G},~(j,k=1,2,\dots,2M),
\label{eq3A8}
\end{equation}
where $\langle\cdots\rangle_{\rm G}$ denotes the expectation value with respect to $\hat{\rho}_{\rm G}$, the operator $\delta\hat{\phi}_j=\hat{\phi}_j-\langle\hat{\phi}_j\rangle_{\rm G}$ represents the quantum fluctuation around the mean, and $\{\cdot,\cdot\}$ is the anticommutator. The diagonal elements of $\Gamma_\phi$ correspond to the variances of the individual quadratures, while the off-diagonal elements describe the correlations between them. For any physical quantum state, the covariance matrix must satisfy the generalized uncertainty relation $\Gamma_\phi+i\sigma/2\geq0$; for pure Gaussian states, this inequality is saturated, indicating that they are minimum uncertainty states.

In addition to the unitary evolution generated by $\hat{H}$, we consider a continuous Gaussian measurement, which is described by a set of jump operators that are linear in the quadrature operators:
\begin{equation}
\hat{L}_n=({\mathbb L}\hat{\bm\phi})_n,
\label{eq3A9}
\end{equation}
where ${\mathbb L}$ is a complex-valued matrix. Such measurements are physically relevant and can be realized, for example, via homodyne detection of light scattered from the system. A key feature of this setup is that the combination of a quadratic Hamiltonian and linear jump operators preserves the Gaussian nature of the state. That is, if the system is initialized in a Gaussian state, its state remains Gaussian at all subsequent times, and the evolution can be described simply by an affine transformation of ${\bm\phi}$ and $\Gamma_\phi$. This closure property allows us to describe the entire quantum dynamics solely in terms of the evolution of the mean vector and the covariance matrix. As shown below, the description of a Gaussian state through its mean and covariance matrix renders a theoretical treatment of the QFIs tractable.

For later discussions, we also give the equations of motion for a bosonic Gaussian state governed by the Lindblad master equation. By substituting the general quadratic Hamiltonian (\ref{eq3A3}) and the linear jump operators (\ref{eq3A9}) into the master equation (\ref{eq2A7}), we obtain
\begin{eqnarray}
&&\frac{d\hat{\rho}}{dt}=-\frac{i}{2}\sum_{j,k}{\mathbb H}_{jk}[\hat{\phi}_j\hat{\phi}_k,\hat{\rho}]-i\sum_ja_j[\hat{\phi}_j,\hat{\rho}] \nonumber\\
&&+\sum_{j,k}{\mathbb M}_{jk}\left(\hat{\phi}_k\hat{\rho}\hat{\phi}_j-\frac{1}{2}\{\hat{\phi}_j\hat{\phi}_k,\hat{\rho}\}\right),
\label{eq3A10}
\end{eqnarray}
where the matrix ${\mathbb M}$ is defined as
\begin{equation}
{\mathbb M}={\mathbb L}^\dag{\mathbb L}.
\label{eq3A11}
\end{equation}
Being the product of a matrix and its conjugate transpose, ${\mathbb M}$ is necessarily Hermitian and positive semidefinite. The corresponding equations of motion for the mean vector is found by computing $d\langle\hat{\bm\phi}\rangle/dt={\rm Tr}[\hat{\bm\phi}\,d\hat{\rho}/dt]$, which yields
\begin{equation}
\frac{d{\bm\phi}}{dt}=\sigma{\mathbb H}_{\rm eff}{\bm\phi}+\theta\sigma{\bm a}.
\label{eq3A12}
\end{equation}
This evolution is driven by an effective Hamiltonian matrix ${\mathbb H}_{\rm eff}$, defined as
\begin{equation}
{\mathbb H}_{\rm eff}={\mathbb H}+{\mathbb M}_I,
\label{eq3A13}
\end{equation}
where ${\mathbb M}_I=-i({\mathbb M}-{\mathbb M}^\ast)/2$ is the imaginary part of the matrix ${\mathbb M}$.

Similarly, the equation of motion for the covariance matrix is derived by Wick's theorem for Gaussian states, which allows for expressing higher-order moments of the quadrature operators by the first and second moments. This procedure closes the hierarchy of equations and leads to a deterministic evolution equation for the covariance matrix, given by
\begin{equation}
\frac{d\Gamma_\phi}{dt}=\sigma{\mathbb H}_{\rm eff}\Gamma_\phi+\Gamma_\phi(\sigma{\mathbb H}_{\rm eff})^{\rm T}-\sigma{\mathbb M}_R\sigma,
\label{eq3A14}
\end{equation}
where ${\mathbb M}_R=({\mathbb M}+{\mathbb M}^\ast)/2$ is the real part of ${\mathbb M}$. The first two terms on the right-hand side describe the evolution of the covariance matrix under the effective Hamiltonian ${\mathbb H}_{\rm eff}$, while the final term $-\sigma{\mathbb M}_R\sigma$ represents diffusion or heating introduced by the coupling to the environment.% We remark that an initial bosonic Gaussian state remains to be Gaussian during the time evolution because Eqs.~(\ref{eq3A12}) and (\ref{eq3A14}) are the closed evolution equations.

As outlined in the introduction, we will focus on two representative classes of the system-environment couplings; in Sec.~\ref{sec4}, we consider a class of dynamically stable systems where all eigenvalues of the matrix $\sigma{\mathbb H}_{\rm eff}$ have negative real parts. This condition leads to dissipative contractive dynamics, in which any initial state evolves towards a unique steady state. In contrast, in Sec.~\ref{sec5}, we consider a class of oscillatory dynamics where all eigenvalues of $\sigma{\mathbb H}_{\rm eff}$ are purely imaginary. Physically, this implies a long-lasting oscillation with unbounded energy growth as in, e.g., a backaction-evading measurement of quadrature operators in the system.

%
% -------------------------------------------------------------------------------------------------------------------------------------------------------------------------------------
%

\subsection{\label{sec3B}Global quantum Fisher information}
We now analyze the global QFI for the free-bosonic system under the continuous Gaussian measurement by employing the characteristic function; a similar approach has been used to study backaction on a qubit coupled dispersively to quadratic bosonic systems~\cite{Clerk2007,Wang2024}. A key observation is that the pseudo-reduced density matrix $\hat{\mu}_{\theta_1,\theta_2}$ in Eq.~(\ref{eq2B3}) admits a factorized form, which is a consequence of the dynamics preserving the Gaussian character of the state. Specifically, we can show that $\hat{\mu}_{\theta_1,\theta_2}$ is written as
\begin{equation}
\hat{\mu}_{\theta_1,\theta_2}(t)=e^{-\eta(\theta_1,\theta_2,t)}\hat{\nu}_{\theta_1,\theta_2}(t),
\label{eq3B1}
\end{equation}
where $\eta(\theta_1,\theta_2,t)\in{\mathbb C}$ is a complex-valued scalar that determines the trace of the pseudo-reduced density matrix, i.e., $e^{-\eta(\theta_1,\theta_2,t)}={\rm Tr}[\hat{\mu}_{\theta_1,\theta_2}(t)]$. The Gaussian operator $\hat{\nu}_{\theta_1,\theta_2}(t)$ is in general non-Hermitian but is normalized to have unit trace ${\rm Tr}[\hat{\nu}_{\theta_1,\theta_2}(t)]=1$. Its properties are fully captured by its first and second moments, i.e., the characteristic function retains a Gaussian form throughout the time evolution:
\begin{equation}
\chi_\nu({\bm v})=\exp\left[-\frac{1}{2}{\bm v}^{\rm T}\sigma^{\rm T}\Gamma_\phi\sigma{\bm v}+i({\bm\phi}^{\rm T}+i{\bm\kappa}^{\rm T})\sigma{\bm v}\right],
\label{eq3B2}
\end{equation}
where ${\bm v}$ is a real vector in the $2M$-dimensional phase space. In contrast to a valid Gaussian density operator, we emphasize that the operator $\hat{\nu}_{\theta_1,\theta_2}$ has the complex-valued first moment,
\begin{equation}
{\rm Tr}[\hat{\bm\phi}\,\hat{\nu}_{\theta_1,\theta_2}]={\bm\phi}+i{\bm\kappa},
\label{eq3B3}
\end{equation}
where ${\bm\phi}$ and ${\bm\kappa}$ are real-valued vectors representing the real and imaginary parts of the moment, respectively. The second moment of $\hat{\nu}_{\theta_1,\theta_2}$ is characterized by the covariance matrix,
\begin{equation}
(\Gamma_\phi)_{jk}=\frac{1}{2}{\rm Tr}[\{\delta\hat{\phi}_j,\delta\hat{\phi}_k\}\,\hat{\nu}_{\theta_1,\theta_2}],~(j,k=1,2,\dots,2M),
\label{eq3B4}
\end{equation}
where $\delta\hat{\phi}_j=\hat{\phi}_j-(\phi_j+i\kappa_j)$. As shown below, $\Gamma_\phi$ obeys the same equation as Eq.~(\ref{eq3A14}). This is because the parameter to be estimated appears only in the Hamiltonian (\ref{eq3A3}) as a linear driving field. Consequently, for an initial density matrix of a Gaussian state, $\Gamma_\phi$ remains real and symmetric throughout the evolution. The non-Hermitian nature of $\hat{\nu}_{\theta_1,\theta_2}$ is instead encoded in a nonzero imaginary part of the mean vector, ${\bm\kappa}\neq{\bm0}$, which results in an asymmetric characteristic function, i.e., $\chi_\nu({\bm v})\neq\chi_\nu^\ast(-{\bm v})$. Therefore, from the saturation condition discussed in Sec.~\ref{sec2E}, we see that there exists nonzero information difference $\delta I=I_{\rm G}-I_{\rm E}>0$ iff ${\bm\kappa}\neq{\bm0}$.

To validate the Gaussian form (\ref{eq3B2}), we substitute Eq.~(\ref{eq3B1}) into the generalized Lindblad equation (\ref{eq2B6}). By equating terms of the same order in the quadrature operators $\hat{\bm\phi}$, we derive a closed set of deterministic, first-order differential equations that govern the time evolution of the parameters $\eta,{\bm\phi},{\bm\kappa}$, and $\Gamma_\phi$:
\begin{eqnarray}
&&\frac{d}{dt}\eta=i(\theta_1-\theta_2){\bm a}^{\rm T}\cdot({\bm\phi}+i{\bm\kappa}), \label{eq3B5}\\
&&\frac{d}{dt}{\bm\phi}=\sigma{\mathbb H}_{\rm eff}{\bm\phi}+\frac{1}{2}(\theta_1+\theta_2)\sigma{\bm a}, \label{eq3B6} \\
&&\frac{d}{dt}{\bm\kappa}=\sigma{\mathbb H}_{\rm eff}{\bm\kappa}-(\theta_1-\theta_2)\Gamma_\phi{\bm a}, \label{eq3B7} \\
&&\frac{d}{dt}\Gamma_\phi=\sigma{\mathbb H}_{\rm eff}\Gamma_\phi+\Gamma_\phi(\sigma{\mathbb H}_{\rm eff})^{\rm T}-\sigma{\mathbb M}_R\sigma. \label{eq3B8}
\end{eqnarray}
The structure of these equations confirms the consistency of our ansatz; the dynamics of the first and second moments are described by affine shifts, ensuring the Gaussian character of $\hat{\nu}_{\theta_1,\theta_2}$ over time. As mentioned above, the evolution of the covariance matrix $\Gamma_\phi$ is governed by a Lyapunov equation that is entirely independent of the parameters $\theta_{1,2}$. The dynamics of the mean-vector components ${\bm\phi}$ and ${\bm\kappa}$ are driven by both the effective Hamiltonian and the $\theta$-dependent terms proportional to ${\bm a}$. The normalization factor $\eta(\theta_1,\theta_2,t)$ is determined by integrating Eq.~(\ref{eq3B5}) with respect to time, which shows its direct dependence on the trajectory of the complex-valued mean vector ${\bm\phi}+i{\bm\kappa}$.

Within this framework, the global QFI $I_{\rm G}(t)$ can be computed directly from the normalization factor $\eta(\theta_1,\theta_2,t)$. The relationship between the global QFI and the trace of the pseudo-reduced density matrix (cf. Eq.~(\ref{eq2B4})) leads to the expression:
\begin{equation}
I_{\rm G}(t)=\left.-4\partial_{\theta_1}\partial_{\theta_2}{\rm Re}[\eta(\theta_1,\theta_2,t)]\right|_{\theta_1=\theta_2=\theta}.
\label{eq3B9}
\end{equation}
By taking the time derivative of Eq.~(\ref{eq3B9}) and using Eq.~(\ref{eq3B5}), we obtain
\begin{equation}
\frac{d}{dt}I_{\rm G}(t)=\left.-4{\bm a}^{\rm T}\cdot({\partial_{\theta_1}{\bm\kappa}}-{\partial_{\theta_2}{\bm\kappa}})\right|_{\theta_1=\theta_2=\theta}.
\label{eq3B10}
\end{equation}
To proceed, we need the derivatives of ${\bm\kappa}$ with respect to $\theta_l$ for $l=1,2$. These can be found by formally solving Eq.~(\ref{eq3B7}) and then differentiating with respect to the parameters:
\begin{equation}
\partial_{\theta_l}{\bm\kappa}=(-1)^l\int_0^td\tau e^{\sigma{\mathbb H}_{\rm eff}(t-\tau)}\Gamma_\phi(\tau){\bm a}.
\label{eq3B11}
\end{equation}
Substituting this result into the QFI growth rate yields the final form:
\begin{equation}
\frac{d}{dt}I_{\rm G}(t)=8{\bm a}^{\rm T}\int_0^td\tau\,e^{\sigma{\mathbb H}_{\rm eff}(t-\tau)}\Gamma_\phi(\tau){\bm a}.
\label{eq3B12}
\end{equation}
This result shows that the growth rate of the global QFI is determined by a time-integrated correlation function. The integrand consists of the covariance matrix $\Gamma_\phi(\tau)$ at a past time $\tau$, which is then propagated forward in time by the effective non-Hermitian generator $\sigma{\mathbb H}_{\rm eff}$, and finally projected onto the measurement direction ${\bm a}$. In all the scenarios discussed below, the system reaches a long-time regime with the linear growth in time $I_{\rm G}(t)\propto t$, indicating a continuous improvement in the estimation precision.

%
% -------------------------------------------------------------------------------------------------------------------------------------------------------------------------------------
%

\subsection{\label{sec3C}Environmental quantum Fisher information}
A general solution (\ref{eq3B1}) provides a direct pathway to derive an analytical expression for the environmental QFI defined in Eq.~(\ref{eq2C2}). The environmental quantum fidelity in Eq.~(\ref{eq2C4}) can be written as
\begin{equation}
{\cal F}_{\rm E}(\theta_1,\theta_2,t)=e^{-{\rm Re}(\eta(\theta_1,\theta_2,t))}{\rm Tr}\left[\sqrt{\hat{\nu}_{\theta_1,\theta_2}(t)\hat{\nu}_{\theta_1,\theta_2}^\dag(t)}\right].
\label{eq3C1}
\end{equation}
This expression highlights a distinction from the global QFI; the trace norm of $\hat{\nu}_{\theta_1,\theta_2}$ accounts for the difference
\begin{align}
\delta I(t)&=I_{\rm G}(t)-I_{\rm E}(t) \label{eq3C2}\\
&=\left.-4\partial_{\theta_1}\partial_{\theta_2}\log{\rm Tr}\left[\sqrt{\hat{\nu}_{\theta_1,\theta_2}(t)\hat{\nu}_{\theta_1,\theta_2}^\dag(t)}\right]\right|_{\theta_1=\theta_2=\theta}, \label{eq3C3}
\end{align}
which quantifies the metrological information that is inaccessible through environmental monitoring. To analytically compute this quantity, we note that the operator product $\hat{\nu}_{\theta_1,\theta_2}\hat{\nu}_{\theta_1,\theta_2}^\dag$ corresponds to an unnormalized Gaussian state. To see this, we point out that its characteristic function is given by the convolution:
\begin{equation}
\chi_{\nu\nu^\dag}({\bm v})=\int\frac{d^{2M}u}{(2\pi)^M}\chi_\nu({\bm u})\chi_{\nu^\dag}({\bm v}-{\bm u})\exp\left(\frac{i}{2}{\bm u}^{\rm T}\sigma{\bm v}\right),
\label{eq3C4}
\end{equation}
where the characteristic function of $\hat{\nu}_{\theta_1,\theta_2}^\dag$ is obtained by replacing ${\bm\kappa}\rightarrow-{\bm\kappa}$ in $\chi_\nu$. Performing this Gaussian integration yields
\begin{eqnarray}
&&\chi_{\nu\nu^\dag}({\bm v})=\frac{\exp({\bm\kappa}^{\rm T}\Gamma_\phi^{-1}{\bm\kappa})}{2^M\sqrt{\det\Gamma_\phi}} \nonumber\\
&&\times\exp\left[-\frac{1}{4}{\bm v}^{\rm T}\sigma^{\rm T}\left(\Gamma_\phi+\frac{1}{4}\sigma\Gamma_\phi^{-1}\sigma^{\rm T}\right)\sigma{\bm v}\right. \nonumber\\
&&\left.+i\left({\bm\phi}^{\rm T}+\frac{1}{2}{\bm\kappa}^{\rm T}\Gamma_\phi^{-1}\sigma^{\rm T}\right)\sigma{\bm v}\right],
\label{eq3C5}
\end{eqnarray}
which confirms that $\hat{\nu}_{\theta_1,\theta_2}\hat{\nu}_{\theta_1,\theta_2}^\dag$ is indeed an unnormalized Gaussian state. In general, the square root of a Gaussian operator is also a Gaussian operator~\cite{Banchi2015}. Thus, $\chi_{\sqrt{\nu\nu^\dag}}({\bm v})$ is still the Gaussian function, and its value at the origin gives the desired quantity in Eq.~(\ref{eq3C3}),
\begin{equation}
{\rm Tr}\left[\sqrt{\hat{\nu}_{\theta_1,\theta_2}\hat{\nu}_{\theta_1,\theta_2}^\dag}\right]=\chi_{\sqrt{\nu\nu^\dag}}({\bm0}).
\label{eq3C6}
\end{equation}
Since $\chi_{\sqrt{\nu\nu^\dag}}({\bm0})$ depends on the parameters $\theta_{1,2}$ only through ${\bm\kappa}$, the logarithm of the trace takes the form $\log\chi_{\sqrt{\nu\nu^\dag}}({\bm0})=\frac{1}{2}{\bm\kappa}^{\rm T}(\Gamma_\phi)^{-1}{\bm\kappa}+C$, where $C$ is a constant independent of $\theta_{1,2}$. This leads to a concise final expression for the information difference:
\begin{equation}
\delta I=\left.-2\partial_{\theta_1}\partial_{\theta_2}[{\bm\kappa}^{\rm T}\Gamma_\phi^{-1}{\bm\kappa}]\right|_{\theta_1=\theta_2=\theta}.
\label{eq3C7}
\end{equation}
We note that ${\bm\kappa}$ depends on $\theta_{1,2}$, whereas $\Gamma_\phi$ is independent of them. This result allows for analytically quantifying the portion of quantum metrological information inaccessible to the environment monitoring. Since the covariance matrix $\Gamma_\phi$ is positive definite, so is its inverse, ensuring that $\delta I\geq0$ for all times. The relation (\ref{eq3C7}) is also consistent with the saturation condition in Sec.~\ref{sec2E}; a nonzero ${\bm\kappa}$ causes the deviation of $\hat{\nu}_{\theta_1,\theta_2}$ from the normalized, valid Gaussian density operator, which inevitably leads to a nonvanishing information difference $\delta I>0$. As discussed below, in the long-time regime, the information difference $\delta I$ can either converge to a constant value independent of time or grow linearly in time, depending on whether the dynamics is dissipative or oscillatory.

%
% -------------------------------------------------------------------------------------------------------------------------------------------------------------------------------------
%

\subsection{\label{sec3D}Quantum fluctuations}
As discussed in Secs.~\ref{sec3B} and \ref{sec3C}, both the global and environmental QFIs are determined by the covariance matrix. Hence, analyzing its time evolution is essential to elucidate the scaling behaviors of the QFIs. We recall that the equation of motion for the covariance matrix as
\begin{equation}
\frac{d\Gamma_\phi}{dt}=X\Gamma_\phi+\Gamma_\phi X^{\rm T}+Y,
\label{eq3D1}
\end{equation}
where the matrices $X$ and $Y$ are defined as
\begin{equation}
X=\sigma{\mathbb H}_{\rm eff},~Y=-\sigma{\mathbb M}_R\sigma.
\label{eq3D2}
\end{equation}
The structure of Eq.~(\ref{eq3D1}) reveals the two processes: the matrix $X$ corresponds to the drift terms governed by the effective Hamiltonian ${\mathbb H}_{\rm eff}$, which consists of the real symmetric matrix ${\mathbb H}$ and the real antisymmetric matrix ${\mathbb M}_I$ (cf. Eq.~(\ref{eq3A13})). Meanwhile, the real symmetric matrix $Y$ describes diffusion or heating due to the noise arising from the coupling to the environment.

Once again, we emphasize that the dynamics (\ref{eq3D1}) is primarily characterized by a structure of the matrix $X$ (see also Ref.~\cite{Behr2019}); in the following sections, we focus on two broad classes of system-environment couplings. In Sec.~\ref{sec4}, we firstly consider dissipative couplings leading to contractive dynamics, in which Eq.~(\ref{eq3D1}) admits a unique steady solution. In this case, the covariance matrix always converges to the same stationary configuration in the long-time limit. Secondly, in Sec.~\ref{sec5}, we discuss zero-damping couplings leading to oscillatory dynamics, where a general solution of Eq.~(\ref{eq3D1}) exhibits long-lasting oscillations and unbounded growth over time. We show that these qualitatively different couplings lead to the distinct asymptotic behaviors of the QFIs.

%
% -------------------------------------------------------------------------------------------------------------------------------------------------------------------------------------
%

\subsection{\label{sec3E}Characteristic quantities}
Below we introduce quantities that are useful for discussions in the following sections. These quantities characterize the asymptotic long-time metrological performances of continuously monitored free-bosonic systems.

%
% -------------------------------------------------------------------------------------------------------------------------------------------------------------------------------------
%

\subsubsection{\label{sec3E1}Optimized quantum Fisher information}
As mentioned in Secs.~\ref{sec3B} and \ref{sec3C}, both the global and environmental QFIs exhibit a linear growth $I_{\rm G,E}(t)\propto t$ in the long-time regime, i.e., the growth rates of the QFIs converge to steady values. We can thus define the asymptotic rates of the QFIs as
\begin{align}
\dot{I}_{\rm G,st}&\equiv\lim_{t\rightarrow\infty}\frac{I_{\rm G}(t)}{t}, \label{eq3E11}\\
\dot{I}_{\rm E,st}&\equiv\lim_{t\rightarrow\infty}\frac{I_{\rm E}(t)}{t}. \label{eq3E12}
\end{align}
We point out that these asymptotic rates are not intrinsic properties of the setup alone; Eqs.~(\ref{eq3E11}) and (\ref{eq3E12}) depend on the vector ${\bm a}$, which characterizes how the unknown parameter $\theta$ couples to the individual bosonic modes of the sensor. To isolate the performance of the setup itself, it is useful to define a figure of merit that is independent of the particular choice of ${\bm a}$~\cite{Lee2025}. Accordingly, we introduce the optimized QFIs, which are defined as the maximum achievable rates over possible vectors:
\begin{align}
\dot{I}_{\rm G,st}^\ast&\equiv\max_{|{\bm a}|^2=M}[\dot{I}_{\rm G,st}({\bm a})], \label{eq3E13}\\
\dot{I}_{\rm E,st}^\ast&\equiv\max_{|{\bm a}|^2=M}[\dot{I}_{\rm E,st}({\bm a})]. \label{eq3E14}
\end{align}
We note that, for any fixed admissible profile ${\bm a}_0$, the corresponding rates satisfy $\dot{I}_{\rm G,st}({\bm a}_0)\leq\dot{I}_{\rm G,st}^\ast$ and $\dot{I}_{\rm E,st}({\bm a}_0)\leq\dot{I}_{\rm E,st}^\ast$. These bounds are attained by the respective optimizing profiles and thus quantify the best-case performance that a given sensor can possibly achieve under the stated normalization. Therefore, although the optimized rates do not give the actual sensitivity for every particular sensing target, they provide exact ceilings against which the performance for any fixed signal profile can be assessed.

%
% -------------------------------------------------------------------------------------------------------------------------------------------------------------------------------------
%

\subsubsection{\label{sec3E2}Bosonic excitation}
In the context of bosonic quantum metrology, the number of bosonic excitations inside the sensor system is often customarily considered as a physical resource~\cite{Friis2015}, since practical implementations are typically constrained by the maximum tolerable excitation level, above which nonlinear effects or additional dissipation mechanisms degrade the sensor performance~\cite{Lau2018}. To elucidate the connection between the optimized QFIs and this energetic limitation, we define the average number of bosonic excitations per mode as
\begin{equation}
\bar{n}\equiv\frac{1}{M}\sum_{j=1}^M\langle\hat{b}_j^\dag\hat{b}_j\rangle,
\label{eq3E21}
\end{equation}
where $\hat{b}_j=(\hat{x}_j+i\hat{p}_j)/\sqrt{2}$ is the annihilation operator for the $j$th bosonic mode. As detailed below, we will show that, in the long-time limit, the number of bosonic excitations exhibits either a convergence to a steady value or an unbounded growth that is asymptotically linear in time, depending on whether the dynamics is contractive or oscillatory.

%
% -------------------------------------------------------------------------------------------------------------------------------------------------------------------------------------
%

\section{\label{sec4}Dissipative coupling}
Building on the formalism developed in the previous sections, we first consider dissipative couplings where the nonunitary dynamics is contractive, i.e., the dissipative time-evolution equation (\ref{eq3D1}) admits a unique steady-state solution, leading to simple analytical expressions for both the global and environmental QFIs. Importantly, we demonstrate that the long-time asymptotic rates of these QFIs coincide. These findings allow us to derive bounds on the scaling behaviors of the QFIs in terms of metrological resources.

%
% -------------------------------------------------------------------------------------------------------------------------------------------------------------------------------------
%

\subsection{\label{sec4A}General solution of Eq.~(\ref{eq3D1})}
In many physical scenarios, particularly when jump operators are linear in the annihilation operators $\hat{b}_j$, the corresponding Lindblad master equation (\ref{eq3A10}) typically has a unique steady-state solution. In the context of Gaussian systems, this means that the time evolution of the covariance matrix is stable, namely, all the eigenvalues of the matrix $X$ in Eq.~(\ref{eq3D2}) have negative real parts, ensuring the contractive dynamics where any perturbation from the steady state will exponentially decay over time. Since $X$ is a real matrix, its eigenvalues must appear in complex-conjugate pairs. Accordingly, we can express its spectral decomposition as
\begin{equation}
X=\sum_{\alpha=1}^M(\lambda_\alpha P_\alpha+\lambda_\alpha^\ast P_\alpha^\ast),
\label{eq4A1}
\end{equation}
where $\lambda_\alpha$ are the eigenvalues satisfying ${\rm Re}(\lambda_\alpha)<0$ for all $\alpha$. The operators $P_\alpha$ are the projectors onto the corresponding eigenspaces.

To ensure the physical relevance and mathematical tractability of our derivation, we introduce a set of assumptions regarding the matrix $X$. First, we assume that the spectrum of $X$ is bounded as $0<c_1\leq|\lambda_\alpha|\leq c_2$, where $c_{1,2}$ are constants independent of $M$. Second, we assume that the condition number of $X$ is bounded as $1\leq\kappa(X)\leq c_3$, where $c_3$ is a constant independent of $M$. Third, we assume that $X$ satisfies
\begin{equation}
X+X^{\rm T}\leq-c_41_{2M},
\label{eq4A2}
\end{equation}
where $c_4$ is a positive constant independent of $M$. Physically, Eq.~(\ref{eq4A2}) means that $X$ is strictly dissipative, and this condition is satisfied in the scenario of system-environment couplings involving, e.g., one-body decaying channels as discussed below.

The differential equation (\ref{eq3D1}) can be systematically solved by projecting it onto the eigenbasis of $X$. This procedure decouples the dynamics into a set of simpler differential equations for the projected components of the covariance matrix, such as $P_\alpha\Gamma_\phi P_{\alpha^\prime}^\dag$. By solving these equations and forming their linear combinations, we obtain a general solution for the time-dependent covariance matrix $\Gamma_\phi(t)$. The full solution, describing the evolution from an initial state $\Gamma_\phi(0)$, is given by
\begin{eqnarray}
&&\Gamma_\phi(t)=2\sum_{\alpha,\alpha^\prime}{\rm Re}\Bigg\{e^{(\lambda_\alpha+\lambda_{\alpha^\prime}^\ast)t}P_\alpha\Gamma_\phi(0)P_{\alpha^\prime}^\dag \nonumber\\
&&+e^{(\lambda_\alpha+\lambda_{\alpha^\prime})t}P_\alpha\Gamma_\phi(0)P_{\alpha^\prime}^{\rm T}+\frac{P_\alpha YP_{\alpha^\prime}^\dag}{(\lambda_\alpha+\lambda_{\alpha^\prime}^\ast)}\left[e^{(\lambda_\alpha+\lambda_{\alpha^\prime}^\ast)t}-1\right] \nonumber\\
&&+\frac{P_\alpha YP_{\alpha^\prime}^{\rm T}}{(\lambda_\alpha+\lambda_{\alpha^\prime})}\left[e^{(\lambda_\alpha+\lambda_{\alpha^\prime})t}-1\right]\Bigg\},
\label{eq4A3}
\end{eqnarray}
where the terms proportional to $\Gamma_\phi(0)$ decay exponentially due to the stability condition ${\rm Re}(\lambda_\alpha)<0$, representing the damping of initial fluctuations. Conversely, the terms involving the noise matrix $Y$ describe the continuous generation of correlations and fluctuations driven by the system-environment coupling. The exponential terms in Eq.~(\ref{eq4A3}) vanish in the long-time limit $t\rightarrow\infty$, and the asymptotic form is given by the steady-state covariance matrix $\Gamma_{\rm st}$:
\begin{align}
\Gamma_{\rm st}&\equiv\lim_{t\rightarrow\infty}\Gamma_\phi(t) \nonumber\\
&=-2\sum_{\alpha,\alpha^\prime}{\rm Re}\left[\frac{P_\alpha YP_{\alpha^\prime}^\dag}{(\lambda_\alpha+\lambda_{\alpha^\prime}^\ast)}+\frac{P_\alpha YP_{\alpha^\prime}^{\rm T}}{(\lambda_\alpha+\lambda_{\alpha^\prime})}\right]. \label{eq4A4}
\end{align}
This expression is the solution to the continuous-time Lyapunov equation $X\Gamma_{\rm st}+\Gamma_{\rm st}X^{\rm T}=-Y$, written in the eigenbasis of $X$ \footnote{The existence of this solution is a direct consequence of the uniqueness of the steady-state density matrix of the underlying Lindblad master equation (\ref{eq2A7}); this connection holds for systems whose quantum dynamical semigroup is Davies irreducible, which precludes the existence of multiple or nonfaithful steady states~\cite{Barthel2022}.}.

%
% -------------------------------------------------------------------------------------------------------------------------------------------------------------------------------------
%

\subsection{\label{sec4B}Quantum Fisher information}
The steady-state solution derived above allows us to investigate the asymptotic behaviors of both the global and environmental QFIs in the long-time regime. First, we consider the global QFI $I_{\rm G}(t)$, which encapsulates the total information about the parameter $\theta$ encoded in the joint state. By substituting the steady-state correlator (\ref{eq4A3}) into the general expression (\ref{eq3B12}) for the QFI rate, we find that the global QFI exhibits a linear growth $I_{\rm G}(t)\simeq\dot{I}_{\rm G,st}t$ at long times, where the asymptotic rate of information is
\begin{align}
\dot{I}_{\rm G,st}&=8{\bm a}^{\rm T}(-X^{-1}\Gamma_{\rm st}){\bm a} \label{eq4B1}\\
&=16{\bm a}^{\rm T}\sum_{\alpha,\alpha^\prime}{\rm Re}\left[\frac{1}{\lambda_\alpha}\left(\frac{P_\alpha YP_{\alpha^\prime}^\dag}{\lambda_\alpha+\lambda_{\alpha^\prime}^\ast}+\frac{P_\alpha YP_{\alpha^\prime}^{\rm T}}{\lambda_\alpha+\lambda_{\alpha^\prime}}\right)\right]{\bm a}. \label{eq4B2}
\end{align}
The optimized global QFI is subsequently found by maximizing this expression over the control vector ${\bm a}$, as prescribed by Eq.~(\ref{eq3E13}).

Similarly, we can analyze the long-time behavior of the environmental QFI $I_{\rm E}(t)$. It follows from Eqs.~(\ref{eq3C7}) and (\ref{eq4A4}) that the discrepancy $\delta I(t)=I_{\rm G}(t)-I_{\rm E}(t)$ converges to a time-independent constant value $\delta I_{\rm st}$ in the long-time limit. Thus, the time derivative of $\delta I(t)$ vanishes at long times, and the environmental QFI must grow linearly with an asymptotic rate identical to that of the global QFI:
\begin{equation}
\dot{I}_{\rm E,st}=\lim_{t\rightarrow\infty}\frac{I_{\rm E}(t)}{t}=\lim_{t\rightarrow\infty}\frac{I_{\rm G}(t)-\delta I(t)}{t}=\dot{I}_{\rm G,st}.
\label{eq4B3}
\end{equation}
This saturation indicates that, as the system approaches the steady state, its capacity to store additional information about the parameter $\theta$ becomes exhausted. Consequently, any new information gained from the interaction with the parameter-dependent Hamiltonian must be continuously transferred to the environment. The optimization for both quantities yields the same maximal rate, as determined by Eqs.~(\ref{eq3E13}) and (\ref{eq3E14}), resulting in
\begin{equation}
\dot{I}_{\rm G,st}^\ast=\dot{I}_{\rm E,st}^\ast.
\label{eq4B4}
\end{equation}

An illustrative example of the present scenario is a continuous leak of photons from cavities into waveguides (cf. Fig.~\ref{fig1}(a)). In this case, the jump operator effectively projects the system state upon each detection event, preventing the coherent accumulation of information within the system itself. As a result, nearly all information is carried away by the leaking photons, while the retained information $\delta I_{\rm st}$ will be negligible.

%
% -------------------------------------------------------------------------------------------------------------------------------------------------------------------------------------
%

\subsection{\label{sec4C}Bounds on the optimized quantum Fisher information}
As discussed in Sec.~\ref{sec3E}, the total number of bosonic excitations $\bar{n}$ in Eq.~(\ref{eq3E21}) constitutes a resource for the quantum metrology, as it quantifies the capacity to respond to the external field being estimated. In our setup, the mean of $\hat{b}_j$ for each mode vanishes when no external fields are applied to the sensor system. As a result, under the evolution described by the Hamiltonian (\ref{eq3A3}), we obtain $\langle\hat{b}_j\rangle\propto\theta$, which leads to a contribution to $\bar{n}$ that scales as $\theta^2$. At a sufficiently small $\theta$, we can thus ignore the contributions from the first-order moments ${\bm\phi}$ to bosonic excitations. Consequently, we characterize the asymptotic value of the number of excitations by the steady-state covariance matrix as
\begin{equation}
\bar{n}_{\rm st}\equiv\frac{1}{2M}{\rm Tr}[\Gamma_{\rm st}].
\label{eq4C1}
\end{equation}
This value represents the average number of excitations sustained due to the fluctuations.

Building on this, we derive bounds for the optimized steady-state QFI rates. As shown in Appendix~\ref{secC}, and using the equivalence $\dot{I}_{\rm G,st}^\ast=\dot{I}_{\rm E,st}^\ast$ from Eq.~(\ref{eq4B4}), these bounds are determined by the two resources $\bar{n}_{\rm st}$ and $M$ as
\begin{equation}
r_1\bar{n}_{\rm st}M\leq\dot{I}_{\rm G,st}^\ast=\dot{I}_{\rm E,st}^\ast\leq r_2\bar{n}_{\rm st}M^2,
\label{eq4C2}
\end{equation}
where $r_{1,2}>0$ are constants that do not depend on the mode number $M$. This relation not only quantifies the precision limit but also indicates the asymptotic scaling laws in terms of metrological resources. For instance, Eq.~(\ref{eq4C2}) characterizes how the precision limit scales with the number of bosonic modes $M$ for a fixed excitation number $\bar{n}_{\rm st}$; the lower bound corresponds to the standard scaling limit, where each of the $M$ modes contributes independently to the information gain, while the upper bound signifies the potential for a collective enhancement. This result implies that an properly designed setup can harness the mode number $M$ as a resource to increase the QFIs that scale quadratically with $M$ (see also Table~\ref{tab1}).

%
% -------------------------------------------------------------------------------------------------------------------------------------------------------------------------------------
%

\subsection{\label{sec4D}Numerical results}

%
% -------------------------------------------------------------------------------------------------------------------------------------------------------------------------------------
%

\subsubsection{\label{sec4D1}Coupled cavity array}
To provide a concrete demonstration, we consider a one-dimensional array of $M$ coupled bosonic modes, a platform realizable in various architectures, such as coupled optical or microwave cavities and phononic modes of a trapped-ion chain. We model the coherent dynamics by a quadratic Hamiltonian
\begin{eqnarray}
&&\hat{H}=\omega_0\sum_{j=1}^M\hat{b}_j^\dag\hat{b}_j-\frac{\Delta}{2}\sum_{j=1}^{M-1}(\hat{b}_{j+1}^\dag\hat{b}_j+\hat{b}^\dag_j\hat{b}_{j+1}) \nonumber\\
&&-\frac{\Delta}{4}\sum_{j=1}^{M-1}(\hat{b}_j\hat{b}_{j+1}+\hat{b}_{j+1}\hat{b}_j+\hat{b}^\dag_{j+1}\hat{b}^\dag_j+\hat{b}^\dag_j\hat{b}^\dag_{j+1}) \nonumber\\
&&-\frac{E}{\sqrt{2}}\sum_{j=1}^M(\hat{b}_j+\hat{b}_j^\dag),
\label{eq4D11}
\end{eqnarray}
where the first term represents the on-site energy $\omega_0$ for each mode, the second term describes a hopping process between adjacent modes, the third term corresponds to a two-mode squeezing, or parametric amplification, which creates or annihilates boson pairs in neighboring sites. The second and third terms can be interpreted as a harmonic potential between adjacent modes, which ensures that the system does not have dynamical instability. The final term models the linear interaction of each mode with an external classical field $E$, which we treat as the unknown parameter to be estimated. Without loss of generality, we set $\omega_0=1$ to define the characteristic energy scale; all other parameters will be expressed in units of $\omega_0$. For the sake of convenience, the initial condition is chosen to be the vacuum state $\Gamma_\phi(0)=(1/2)1_M$, although the asymptotic state in the long-time limit is independent of this choice.

%
% -------------------------------------------------------------------------------------------------------------------------------------------------------------------------------------
%

\subsubsection{\label{sec4D2}Local setup}
\begin{figure}[b]
\includegraphics[width=8.5cm]{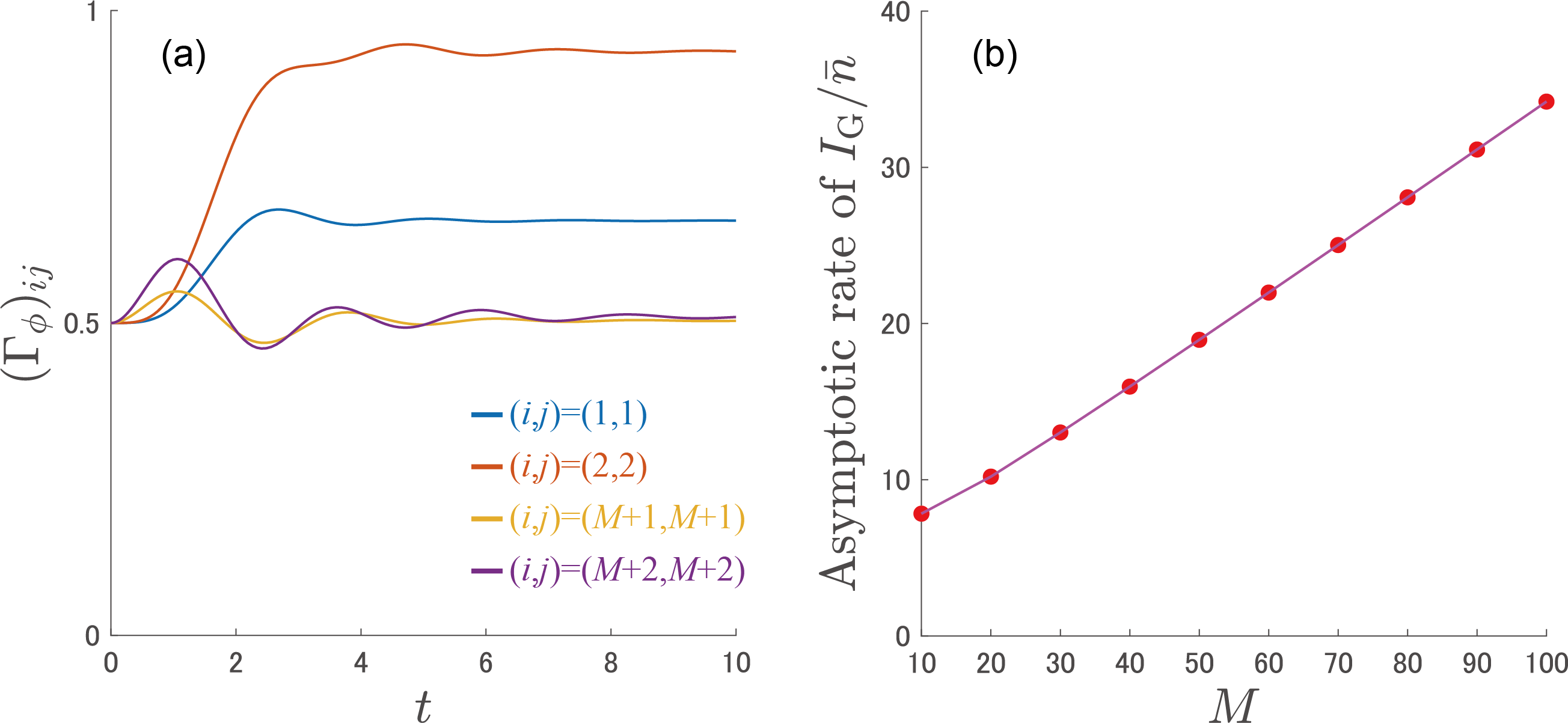}
\caption{\label{fig2}Local setup. (a) Time evolution of diagonal elements of the covariance matrix. The variances of the quadrature operators converge to steady-state values, indicating that the system reaches a unique steady state. (b) Mode-number scaling of the asymptotic rate of the global QFI, normalized by the bosonic excitation number. The dots are obtained from numerical simulations, while the solid curve represents the analytical result obtained from the steady-state expression $\dot{I}_{\rm G,st}$ (cf. Eq.~(\ref{eq4B2})). The parameters are set to be $M=60$ for (a), with $\Delta=0.5,\zeta=0.3$, and $E=0.1$ for (b).}
\end{figure}
We first analyze the case of local dissipative couplings, where photons from each cavity continuously leak into waveguides [Fig.~\ref{fig1}(a)]. This process is described by a set of $M$ local jump operators, each corresponding to the annihilation of a bosonic excitation:
\begin{equation}
\hat{L}_n=\sqrt{\zeta}\hat{b}_n,~(n=1,2,\dots,M).
\label{eq4D21}
\end{equation}
The dissipative nature of this monitoring ensures that the system evolves towards a unique steady state. This is numerically demonstrated in, e.g., Fig.~\ref{fig2}(a), which shows the time evolution of the diagonal elements of the covariance matrix $\Gamma_\phi$. In the long-time limit, the matrix $\Gamma_\phi$ converges to its steady-state solution, and the continuous accumulation leads to a linear growth of both the global and environmental QFIs over time, which have the same asymptotic growth rates as shown in Eq.~(\ref{eq4B3}).

We next analyze how the asymptotic rate of information gain scales with the mode number $M$. Figure~\ref{fig2}(b) presents the mode-number scaling of the normalized asymptotic rate of the global QFI. The numerical results, shown as dots, reveal a linear dependence on $M$. Given that the asymptotic value of the total excitation number $\bar{n}$ is independent of $M$, these scalings can be summarized as follows:
\begin{equation}
I_{\rm G,E}(t)\propto\bar{n}_{\rm st}Mt,~\bar{n}_{\rm st}\propto M^0,
\label{eq4D22}
\end{equation}
where the leaking signals from individual modes contribute independently to the total information content. Such behavior corresponds to the lower bound of the performance inequality (\ref{eq4C2}), which is analogous to the standard quantum limit for parameter estimation using $M$ independent probes. The numerical results agree with the analytical prediction (solid line) obtained by the steady-state expression in Eq.~(\ref{eq4B2}).

%
% -------------------------------------------------------------------------------------------------------------------------------------------------------------------------------------
%

\subsubsection{\label{sec4D3}Hybrid local-and-global setup}
\begin{figure}[b]
\includegraphics[width=8.5cm]{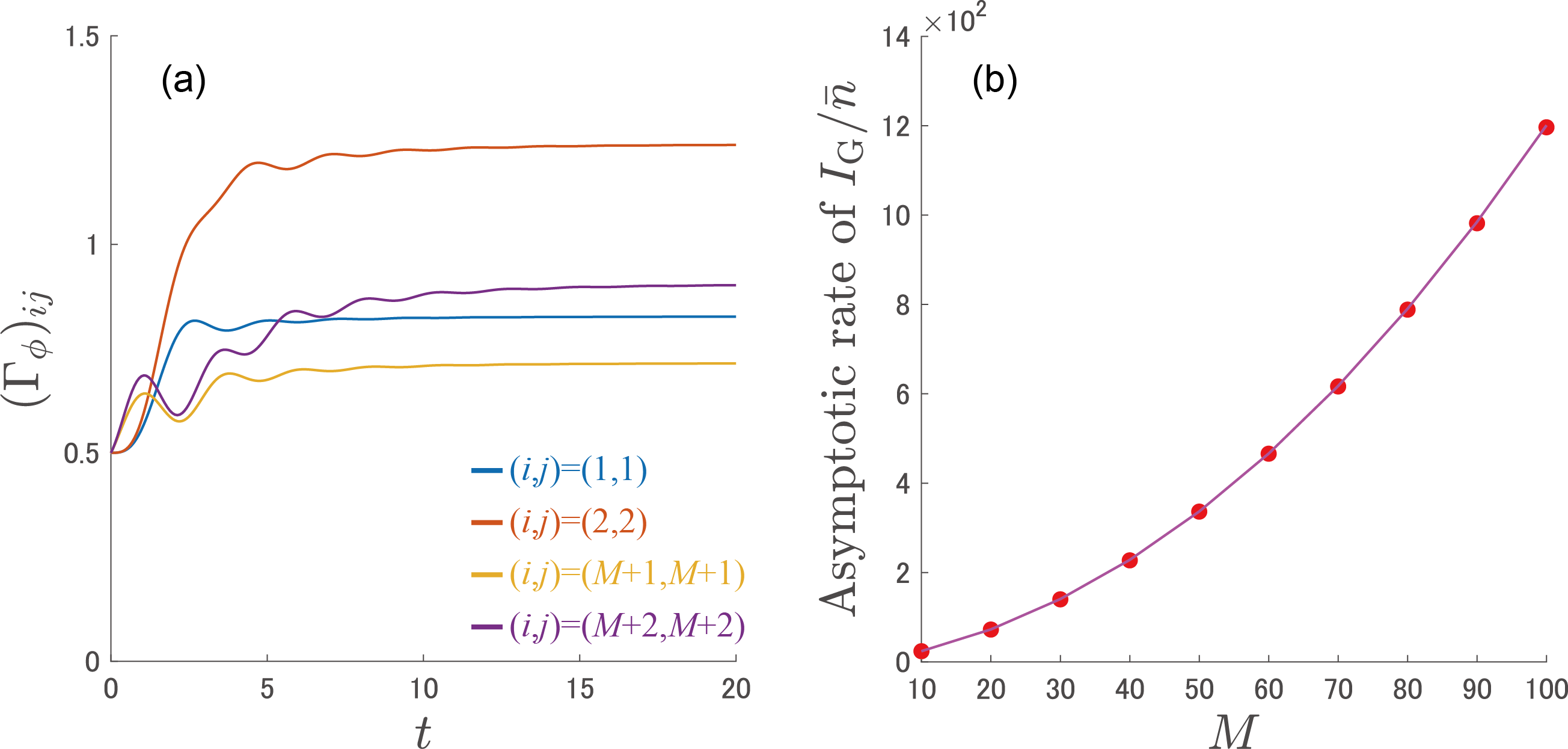}
\caption{\label{fig3}Hybrid local-and-global setup. (a) Time evolution of diagonal elements of the covariance matrix. As in the local-only setup, the system converges to a unique steady state. (b) Mode-number scaling of the asymptotic rate of the global QFI, normalized by the bosonic excitation number. The dots are obtained from numerical simulations, while the solid curve represents the analytical result from Eq.~(\ref{eq4B2}). The parameters are set to be $M=60$ for (a), with $\Delta=0.5,\zeta=0.1,\gamma=0.3$, and $E=0.1$ for (b).}
\end{figure}
We next study a hybrid process that incorporates a global coupling across the entire array in addition to the individual local couplings. This process is described by the following set of $M+1$ jump operators:
\begin{eqnarray}
&&\hat{L}_n=\sqrt{\zeta}\hat{b}_n,~(n=1,2,\dots,M), \label{eq4D31}\\
&&\hat{L}_{M+1}=\sqrt{\frac{\gamma}{2}}\sum_{j=1}^M(\hat{b}_j+\hat{b}_j^\dag). \label{eq4D32}
\end{eqnarray}
The local operators in Eq.~(\ref{eq4D31}) are identical to those in the previous case, representing one-body loss channels. The additional operator in Eq.~(\ref{eq4D32}) corresponds to a global system-environment coupling. The persistence of the local dissipative channels ensures that the covariance matrix $\Gamma_\phi$ still converges to a steady-state solution in the long-time limit [Fig.~\ref{fig3}(a)]. Consequently, both the global and environmental QFIs continue to exhibit the same asymptotic linear growth in time.

The inclusion of the collective coupling dramatically enhances the metrological capabilities. Specifically, we find the following scalings in the long-time regime [Fig.~\ref{fig3}(b)]:
\begin{equation}
I_{\rm G,E}(t)\propto\bar{n}_{\rm st}M^2t,~\bar{n}_{\rm st}\propto M^0.
\label{eq4D33}
\end{equation}
This scaling behavior demonstrates that the global coupling is suitable in that it saturates the upper bound of the inequality presented in Eq.~(\ref{eq4C2}). The agreement between the numerical data (dots) and the analytical result (solid curve) for the steady value in Fig.~\ref{fig3}(b) also validates our theoretical framework.

Physically, the quadratic scaling of the QFIs originates from collective bosonic excitations at low momenta. To clarify this point, we define the Fourier mode of the annihilation operator:
\begin{equation}
\hat{B}_k=\frac{1}{\sqrt{M}}\sum_{j=1}^M\hat{b}_je^{-ikj}.
\label{eq4D34}
\end{equation}
The global jump operator (\ref{eq4D32}) can then be expressed solely in terms of the $k=0$ mode as
\begin{equation}
\hat{L}_{M+1}=\sqrt{\frac{\gamma M}{2}}(\hat{B}_0+\hat{B}_0^\dag),
\label{eq4D35}
\end{equation}
indicating that this coupling induces heating in the $k=0$ mode. Meanwhile, in the long-time regime, the scaling of the global QFI is given by~\cite{Lee2025}
\begin{equation}
I_{\rm G}(t)\propto2S(\omega=0)t,
\label{eq4D36}
\end{equation}
where $S(\omega=0)$ is the steady-state noise spectrum of $\hat{L}_{M+1}$ at zero frequency. We can then estimate the noise spectrum as
\begin{equation}
S(\omega=0)\propto\frac{\langle\hat{L}_{M+1}^2\rangle_{\rm st}}{\tau}\propto M\langle(\hat{B}_0+\hat{B}_0^\dag)^2\rangle_{\rm st},
\label{eq5C7}
\end{equation}
where $\langle\dots\rangle_{\rm st}$ denotes the expectation value with respect to the steady state of the sensor system, and $\tau$ is a relaxation time of $O(1)$. Since the total boson number in the $k=0$ mode should scale with $M\bar{n}_{\rm st}$, we obtain $S(\omega=0)\propto M^2\bar{n}_{\rm st}$, which explains the quadratic scaling of the global QFI in Eq.~(\ref{eq4D33}).

%
% -------------------------------------------------------------------------------------------------------------------------------------------------------------------------------------
%

\section{\label{sec5}Zero-damping coupling}
In Sec.~\ref{sec4}, we have analyzed a class of metrological protocols where the system evolves towards a unique steady state. We now turn our attention to a distinct class of protocols characterized by zero-damping couplings, wherein the covariance matrix described in Eq.~(\ref{eq3D1}) does not converge but instead exhibits an unbounded linear growth with time. We begin by providing a description of the physically relevant setup under consideration. Subsequently, we derive analytical expressions for the asymptotic rates of the global and environmental QFIs. These results serve as the foundation for an analysis of the distinct behaviors across several illustrative examples discussed later. Furthermore, the derived formula allows us to establish useful inequalities that bound the optimized QFIs.

%
% -------------------------------------------------------------------------------------------------------------------------------------------------------------------------------------
%

\subsection{\label{sec5A}General solution of Eq.~(\ref{eq3D1})}
We consider the case of zero-damping couplings, where the system does not reach a steady state but rather exhibits persistent oscillations accompanied by unbounded energy growth. For example, this situation arises in continuous measurements of the quadrature operators of the sensor system, in which jump operators are Hermitian, and the backaction leads to heating. More generally, such oscillatory dynamics is realized when the eigenvalues of ${\mathbb H}_{\rm eff}$ are entirely real, or equivalently, those of $X=\sigma{\mathbb H}_{\rm eff}$ are purely imaginary \footnote{While the matrix ${\mathbb H}_{\rm eff}$ is generally non-Hermitian, the real-valuedness of the spectrum of ${\mathbb H}_{\rm eff}$ can be robust against the perturbation introduced by the non-Hermitian term ${\mathbb M}_I$ provided that the Hermitian component ${\mathbb H}$ has a sufficiently large energy gap~\cite{Ashida2020}.}. Under this assumption, the matrix $X$ admits the spectral decomposition
\begin{equation}
X=\sum_{\alpha=1}^M i\lambda_\alpha(P_\alpha-P_\alpha^\ast),
\label{eq5A1}
\end{equation}
where $\lambda_\alpha$ are positive real numbers determining the oscillation frequencies of the eigenmodes, and $P_\alpha$ is the corresponding projection operator onto the eigenspace.

As derived in Appendix~\ref{secD}, a general solution of Eq.~(\ref{eq3D1}) under the condition (\ref{eq5A1}) can be expressed as
\begin{widetext}
\begin{eqnarray}
&&\Gamma_\phi(t)=2{\rm Re}\Bigg\{\sum_{\alpha,\alpha^\prime}e^{i(\lambda_\alpha-\lambda_{\alpha^\prime})t}P_\alpha\Gamma_\phi(0)P_{\alpha^\prime}^\dag+\sum_{\alpha,\alpha^\prime}e^{i(\lambda_\alpha+\lambda_{\alpha^\prime})t}P_\alpha\Gamma_\phi(0)P_{\alpha^\prime}^{\rm T} \nonumber\\
&&+\sum_{\alpha\neq\alpha^\prime}\frac{P_\alpha YP_{\alpha^\prime}^\dag}{i(\lambda_\alpha-\lambda_{\alpha^\prime})}\left[e^{i(\lambda_\alpha-\lambda_{\alpha^\prime})t}-1\right]+\sum_{\alpha,\alpha^\prime}\frac{P_\alpha YP_{\alpha^\prime}^{\rm T}}{i(\lambda_\alpha+\lambda_{\alpha^\prime})}\left[e^{i(\lambda_\alpha+\lambda_{\alpha^\prime})t}-1\right]+t\sum_\alpha P_\alpha YP_\alpha^\dag\Bigg\},
\label{eq5A2}
\end{eqnarray}
\end{widetext}
where $\Gamma_\phi(0)$ represents the initial condition. The terms on the right-hand side have distinct physical interpretations. The first two terms describe the transient evolution of the initial fluctuations, which oscillate at frequencies determined by the differences or the sums of $\lambda_\alpha$. The subsequent two terms represent the build-up of coherences between different eigenmodes due to the noise $Y$. The last term, which grows linearly in time, describes the incoherent accumulation of noise within each eigenmode.

In the long-time regime, oscillatory contributions from the first four terms undergo dephasing. Unless eigenvalues are degenerate, these terms interfere destructively and their net contribution averages to zero over a certain characteristic timescale, which is reminiscent of relaxation dynamics in quantum many-body systems~\cite{De2018}. Consequently, the long-time dynamics of the covariance matrix is dominated by the last, linearly growing term. This asymptotic behavior allows for a simple characterization of the quantum fluctuations in the long-time regime. Specifically, we can define a steady rate of increase for the covariance matrix as
\begin{equation}
\dot{\Gamma}_{\rm st}\equiv\lim_{t\rightarrow\infty}\frac{\Gamma_\phi(t)}{t}=2{\rm Re}\left(\sum_{\alpha=1}^MP_\alpha YP_\alpha^\dag\right).
\label{eq5A3}
\end{equation}
This quantity represents the asymptotic diffusion rate of the quantum state in phase space driven by the continuous quantum measurement.

%
% -------------------------------------------------------------------------------------------------------------------------------------------------------------------------------------
%

\subsection{\label{sec5B}Continuous position measurement}
We consider a physically relevant scenario where the parameter $\theta$ to be estimated is coupled exclusively to the position degrees of freedom of the $M$ bosonic modes. This constraint leads to the following block form of the vector ${\bm a}$ in Eq.~(\ref{eq3A3}):
\begin{eqnarray}
{\bm a}=\left( \begin{array}{c}
{\bm b} \vspace{5pt}\\
{\bm0}
\end{array}\right),
\label{eq5B1}
\end{eqnarray}
where ${\bm b}$ is an $M$-component column vector that characterizes the connection of the parameter to each position quadrature, while the lower block corresponding to the momentum quadrature is a zero vector. We consider a class of dynamics where the matrices ${\mathbb H}$ and ${\mathbb M}$, which respectively govern the Hamiltonian evolution and the measurement backaction, adopt a block-diagonal structure:
\begin{eqnarray}
{\mathbb H}=\left( \begin{array}{cc}
h & O \vspace{5pt}\\
O & \Omega1_M
\end{array}\right),~{\mathbb M}=\left( \begin{array}{cc}
m_R+im_I & O \vspace{5pt}\\
O   & O
\end{array}\right).
\label{eq5B2}
\end{eqnarray}
Here, $h\in{\mathbb R}^{M\times M}$ and $m_R\in{\mathbb R}^{M\times M}$ are symmetric matrices that characterize the internal potential energy and the dissipative backaction, respectively, and $m_I\in{\mathbb R}^{M\times M}$ is an antisymmetric matrix that describes the conservative backaction on the position quadratures. We note that $m_I$ vanishes if all the jump operators are Hermitian. The parameter $\Omega$ is a scalar that represents a uniform kinetic energy, and the off-diagonal blocks are zero matrices. This class of systems covers all the physical setups discussed in later sections.

Under the condition (\ref{eq5A1}), the effective Hamiltonian matrix for the position quadratures admits a spectral decomposition as follows (cf. Eq.~(\ref{eq3A13})):
\begin{equation}
h_{\rm eff}\equiv h+m_I=\sum_{\alpha=1}^M\tilde{\lambda}_\alpha\tilde{P}_\alpha,
\label{eq5B3}
\end{equation}
where $\tilde{\lambda}_\alpha$ are real, positive, and nondegenerate eigenvalues characterizing the oscillation frequencies of the normal modes under the measurement, and $\tilde{P}_\alpha\in\mathbb{R}^{M\times M}$ are the corresponding projectors. The condition for the eigenvalues is consistent with that for dynamical stability, as instabilities typically arise at exceptional points, where eigenvalues and eigenvectors coalesce and the system exhibits the spectral transition.

To ensure the physical relevance and mathematical tractability of our derivation, we introduce a set of assumptions regarding the properties of the matrices. First, we assume that the spectrum of $h_{\rm eff}$ is bounded, $0<c_1\leq\tilde{\lambda}_\alpha\leq c_2$ and $\|h_{\rm eff}\|\leq c_2^\prime$, where $c_{1,2}$ and $c_2^\prime$ are constants independent of $M$. Second, we assume $c_3 1_M\leq h$ with $c_3$ being a positive constant independent of $M$, which guarantees the rigidity of the conservative couplings. Third, we assume that all the elements of the matrix $m_R$ are constants independent of $M$, and thus the operator norm of $m_R$ is bounded as $0<c_4\leq\|m_R\|\leq c_5M$ with $c_{4,5}$ being constants independent of $M$. Physically, the upper bound $\|m_R\|\propto M$ can be satisfied when all the bosonic modes are monitored simultaneously as realized in, e.g., the global setup as discussed below. Fourth, we assume
\begin{equation}
0<c_6\leq\frac{1}{M}{\rm Tr}\left[\sum_{\alpha=1}^M\tilde{P}_\alpha m_R\tilde{P}_\alpha^{\rm T}\right]
\label{eq5B4}
\end{equation}
with $c_6$ being a positive constant independent of $M$; the physical motivation for this assumption will be clarified below.

We remark that the matrix $h_{\rm eff}$ is real-valued but not necessarily Hermitian. The non-Hermitian character can arise from, e.g., the nonreciprocal nature of the continuous measurement process, which can induce effective interactions that do not conserve an energy. One can, however, still diagonalize it by introducing generally distinct right and left eigenvectors, denoted by ${\bm r}_\alpha$ and ${\bm l}_\alpha$, respectively. In this basis, the projector $\tilde{P}_\alpha$ in Eq.~(\ref{eq5B3}) is constructed as an outer product:
\begin{equation}
\tilde{P}_\alpha={\bm r}_\alpha{\bm l}_\alpha^{\rm T}.
\label{eq5B5}
\end{equation}
The right and left eigenvectors satisfy the biorthogonality condition ${\bm r}_\alpha^{\rm T}{\bm l}_{\alpha^\prime}=\delta_{\alpha\alpha^\prime}$. Without loss of generality, we can adopt the normalization convention ${\bm l}_\alpha^{\rm T}{\bm l}_\alpha=1$ for the left eigenvectors~\cite{Ashida2020}.

%
% -------------------------------------------------------------------------------------------------------------------------------------------------------------------------------------
%

\subsection{\label{sec5C}Bosonic excitation}
The eigenvalues $\lambda_\alpha$ of the matrix $X$ in Eq.~(\ref{eq5A1}) are given by $\lambda_\alpha=\sqrt{\Omega\tilde{\lambda}_\alpha}$, while the corresponding projectors $P_\alpha$ are
\begin{eqnarray}
P_\alpha=\frac{1}{2}\left( \begin{array}{cc}
1                                     & -i\sqrt{\Omega/\tilde{\lambda}_\alpha} \vspace{5pt}\\
i\sqrt{\tilde{\lambda}_\alpha/\Omega} & 1
\end{array}\right)\otimes\tilde{P}_\alpha.
\label{eq5C1}
\end{eqnarray}
We start from analyzing the long-time asymptotic behavior of the bosonic excitation number $\bar{n}$ in Eq.~(\ref{eq3E21}). In the long-time regime, $\bar{n}$ is predominantly determined by the covariance matrix, simplifying to $\bar{n}\simeq{\rm Tr}[\Gamma_\phi]/(2M)$. The asymptotic growth rate of the covariance matrix, which follows from Eq.~(\ref{eq5A3}), is given by
\begin{eqnarray}
\dot{\Gamma}_{\rm st}=\frac{1}{2}\sum_{\alpha=1}^M\left( \begin{array}{cc}
\Omega/\tilde{\lambda}_\alpha & 0 \vspace{5pt}\\
0                             & 1
\end{array}\right)\otimes\tilde{P}_\alpha m_R\tilde{P}_\alpha^{\rm T}.
\label{eq5C2}
\end{eqnarray}
Consequently, the asymptotic rate of bosonic excitations is
\begin{equation}
\dot{\bar{n}}_{\rm st}\equiv\lim_{t\rightarrow\infty}\frac{\bar{n}(t)}{t}=\frac{1}{4M}\sum_{\alpha=1}^M{\rm Tr}\left[\left(\frac{\Omega}{\tilde{\lambda}_\alpha}+1\right)\tilde{P}_\alpha m_R\tilde{P}_\alpha^{\rm T}\right],
\label{eq5C3}
\end{equation}
which represents the energetic cost of the sensing protocol per unit of time. We note that the assumption (\ref{eq5B4}) leads to the inequality
\begin{equation}
\dot{\bar{n}}_{\rm st}\geq\frac{c_6}{4}\left(1+\frac{1}{c_2}\right).
\label{eq5C4}
\end{equation}
Physically, this assumption guarantees that the nonvanishing energy per mode is consumed in the long-time regime.

%
% -------------------------------------------------------------------------------------------------------------------------------------------------------------------------------------
%

\subsection{\label{sec5D}Bounds on the optimized global quantum Fisher information}
Equation~(\ref{eq5C1}) enables us to analyze the long-time asymptotic behavior of the global QFI $I_{\rm G}(t)$. Specifically, we find that $I_{\rm G}(t)$ exhibits a linear growth in time with the following asymptotic rate (see Appendix~\ref{secE}) \footnote{We note that one cannot merely replace $\Gamma_\phi(\tau)$ by $\dot{\Gamma}_{\rm st}(\tau)$ in Eq.~(\ref{eq3B12}) to obtain the correct asymptotic information rate. The reason is that it is the subleading oscillating term in $\Gamma_\phi(\tau)$ that leads to Eq.~(\ref{eq5D1}) via interference with the exponential factor $e^{-\sigma{\mathbb H}_{\rm eff}\tau}$, while the contribution from the leading term $\dot{\Gamma}_{\rm st}\tau$ in fact vanishes as shown in Appendix~\ref{secE}.}:
\begin{equation}
\dot{I}_{\rm G,st}=4{\bm b}^{\rm T}\left(\sum_{\alpha,\alpha^\prime}\frac{\tilde{P}_\alpha m_R\tilde{P}_{\alpha^\prime}^{\rm T}}{\tilde{\lambda}_\alpha\tilde{\lambda}_{\alpha^\prime}}+\sum_{\alpha}\frac{\tilde{P}_\alpha m_R\tilde{P}_\alpha^{\rm T}}{2\tilde{\lambda}_\alpha^2}\right){\bm b}.
\label{eq5D1}
\end{equation}
We note that the second term in Eq.~(\ref{eq5D1}) is the mode-diagonal contribution that remains after the intermode dephasing. The comparison between the rate of excitation generation in Eq.~(\ref{eq5C3}) and the optimized information rate of Eq.~(\ref{eq5D1}) leads to a relation between the precision limit and the number of bosonic excitations as follows (see Appendix~\ref{secF}):
\begin{equation}
r_3\dot{\bar{n}}_{\rm st}M\leq\dot{I}_{\rm G,st}^\ast\leq r_4\dot{\bar{n}}_{\rm st}M^2,
\label{eq5D2}
\end{equation}
where $r_{3,4}>0$ are constants independent of the number of bosonic modes $M$.

The precision bounds (\ref{eq5D2}) exhibit the same $M$ scalings as those in the inequality (\ref{eq4C2}) for the dissipative dynamics. For instance, the upper bound $\dot{I}_{\rm G,st}^\ast\propto M^2$ points to the possibility of a quadratic enhancement in the similar manner as discussed in Sec.~\ref{sec4C}; we will later show that a collective setup can indeed achieve this scaling. Meanwhile, the relation (\ref{eq5D2}) also demonstrates that bosonic excitations are a key resource for enhancing the global QFI. This naturally suggests a strategy for enhancing the global QFI by increasing the rate of bosonic excitation creation $\dot{\bar{n}}_{\rm st}$, for instance, by employing the nonreciprocity; as we will demonstrate, the use of nonreciprocal couplings can lead to an exponential scaling of the excitation rate $\dot{\bar{n}}_{\rm st}\propto\exp(O(M))$, which can be converted into the global QFI, albeit at the cost of increased energy consumption.

%
% -------------------------------------------------------------------------------------------------------------------------------------------------------------------------------------
%

\subsection{\label{sec5E}Bounds on the optimized environmental quantum Fisher information}
We next analyze the environmental QFI $I_{\rm E}$ through the information difference $\delta I=I_{\rm G}-I_{\rm E}$. As shown in Eq.~(\ref{eq3C7}), the long-time evolution of this difference is determined by the asymptotic growth rate of the covariance matrix in Eq.~(\ref{eq5A3}). Thus, following a line of reasoning analogous to that used for the global QFI, we can deduce the asymptotic growth rate for this difference term. By subtracting this contribution from Eq.~(\ref{eq5D1}), one obtains the asymptotic acquisition rate of the environmental QFI in Eq.~(\ref{eq3E12}). The result, derived in Appendix~\ref{secG}, is
\begin{equation}
\dot{I}_{\rm E,st}=4{\bm b}^{\rm T}\sum_{\alpha,\alpha^\prime}\frac{\tilde{P}_\alpha m_R\tilde{P}_{\alpha^\prime}^{\rm T}}{\tilde{\lambda}_\alpha\tilde{\lambda}_{\alpha^\prime}}{\bm b}.
\label{eq5E1}
\end{equation}
A key distinction from the global QFI becomes clear when considering the bounds on its optimized rate $\dot{I}_{\rm E,st}^\ast$ defined in Eq.~(\ref{eq3E14}); namely, $\dot{I}_{\rm E,st}^\ast$ is subject to the bounds different from Eq.~(\ref{eq5D2}). This is because, as inferred from Eq.~(\ref{eq5E1}), $\dot{I}_{\rm E,st}^\ast$ depends on the measurement-backaction matrix $m_R$ without the dephasing between distinct eigenmodes that is present in the global one $\dot{I}_{\rm G,st}^\ast$ [see the second term in Eq.~(\ref{eq5D1})]. In other words, the parameter information initially stored in the system is transferred to the environment as the relative phases between distinct eigenmodes are averaged out. Thus, the environmental QFI lacks the constraints that lead to the bounds for the global one, and instead, the bounds are mainly governed by the operator norm $\|m_R\|$, leading to (see Appendix~\ref{secG})
\begin{equation}
r_5M\leq\dot{I}_{\rm E,st}^\ast\leq\min(r_6,r_7\dot{\bar{n}}_{\rm st})M^2,
\label{eq5E2}
\end{equation}
where $r_{5,6,7}>0$ are constants independent of $M$. The bounds (\ref{eq5E2}) reveal that, while the energy must be consumed for information acquisition, the scaling of the maximumly extracted environmental information is mainly characterized by the number of bosonic modes rather than the excitation number.

According to Eq.~(\ref{eq5D2}), increasing the rate of boson excitations enhances the total information acquired by the joint system and environment. This implies that the bosonic excitations at all energy scales can contribute to the global QFI. In contrast, the bosonic excitation does not directly contribute to the environmental QFI as in Eq.~(\ref{eq5E2}), which can be understood from the fact that only low-energy excitations inside the sensor can be converted into emitted quanta in the environment. As a concrete demonstration, we will show that nonreciprocal backaction leads to the qualitatively distinct scalings of the global and environmental QFIs in the following sections.

%
% -------------------------------------------------------------------------------------------------------------------------------------------------------------------------------------
%

\subsection{\label{sec5F}Scaling behaviors}
Before delving into a detailed analysis of case studies, we first outline the overarching scaling behaviors that emerge from our framework (see also Table~\ref{tab1}). A central finding, applicable across all the scenarios considered later, is that the qualitative scaling of the asymptotic growth rates for both the global QFI and the excitation number can be captured by a `dephased' measurement-backaction matrix given by (cf. Eqs.~(\ref{eq5C3}) and (\ref{eq5D1}))
\begin{equation}
\tilde{m}_R=\sum_{\alpha=1}^M\tilde{P}_\alpha m_R\tilde{P}_\alpha^{\rm T},
\label{eq5F1}
\end{equation}
where $m_R$ is projected onto the individual eigenspaces of the effective non-Hermitian Hamiltonian in Eq.~(\ref{eq5B3}). The matrix $\tilde{m}_R$ encapsulates the net effect of the measurement backaction as distributed among the effective eigenmodes, thereby governing the long-time response of internal system dynamics. In contrast, the asymptotic growth rate of the environmental QFI lacks such dephasing structure (see Eq.~(\ref{eq5E1})) and is governed by the original matrix $m_R$ rather than $\tilde{m}_R$. The physical interpretation is that the information carried away by the environment is encoded in the emitted quanta, a process directly governed by the system-environment coupling described by $m_R$.

%
% -------------------------------------------------------------------------------------------------------------------------------------------------------------------------------------
%

\subsubsection{\label{sec5F1}Local setup}
We first analyze the scenario of a local, reciprocal setup, where each bosonic mode is monitored independently and symmetrically. In this configuration, the measurement-backaction matrix $m_R$ is diagonal and proportional to the identity matrix, i.e., $(m_R)_{jk}\propto\delta_{jk}$. Consequently, its operator norm $\|m_R\|$ is bounded and remains independent of $M$. The projection onto the eigenmodes results in a dephased matrix $\tilde{m}_R$ whose matrix elements are also of order one. In this regime, the sensitivity is constrained by the lower bound on the global QFI presented in Eq.~(\ref{eq5D2}). This leads to a linear scaling of the global QFI with the number of bosonic modes $M$ in the long-time limit:
\begin{equation}
I_{\rm G}(t)\propto\dot{\bar{n}}_{\rm st}Mt,~\dot{\bar{n}}_{\rm st}\propto M^0.
\label{eq5F11}
\end{equation}
This scaling, along with a steady excitation rate $\dot{\bar{n}}_{\rm st}$ that is independent of $M$, is characteristic of the standard quantum limit. It signifies a scenario where each bosonic mode functions as an independent sensor, and the total information is simply the sum of the incoherent contributions from each mode, without any collective enhancement. Concurrently, the environmental QFI is governed by the properties of $m_R$ as explained above. Since $\|m_R\|$ is independent of $M$, the lower bound in Eq.~(\ref{eq5E2}) yields a similar linear scaling:
\begin{equation}
I_{\rm E}(t)\propto Mt.
\label{eq5F12}
\end{equation}
We conclude that, for a local reciprocal setup, both the global and environmental QFIs exhibit linear scalings with $M$. This indicates that the acquired metrological information is distributed between the system and the environment in a manner that does not fundamentally change with the mode number.

%
% -------------------------------------------------------------------------------------------------------------------------------------------------------------------------------------
%

\subsubsection{\label{sec5F2}Global setup}
Next, we examine the case of a global setup, where a collective observable involving all bosonic modes is monitored. In this scenario, all the elements in $m_R$ are comparable and nonvanishing, reflecting the global nature of the measurement. A key consequence of this structure is that, after projection onto the eigenmodes, the dephased matrix $\tilde{m}_R$ possesses a single, dominant eigenvalue that scales linearly with the mode number $\propto M$. The dominant eigenvalue corresponds to the collective mode being measured. The existence of such a large eigenvalue allows the system to approach the upper bound on the global QFI in Eq.~(\ref{eq5D2}). This results in a quadratic scaling of the global QFI in the long-time regime:
\begin{equation}
I_{\rm G}(t)\propto\dot{\bar{n}}_{\rm st}M^2t,~\dot{\bar{n}}_{\rm st}\propto M^0.
\label{eq5F21}
\end{equation}
The quadratic enhancement with $M$ arises because the collective measurement scheme effectively correlates the modes, allowing them to contribute coherently to the information gain. Despite this collective enhancement in sensitivity, the rate of excitation generation per mode still remains independent of $M$, similar to the local setup.

In the environmental QFI, the scaling is determined by $\|m_R\|$, which scales as $O(M)$ for a dense matrix. As indicated in the upper bound in Eq.~(\ref{eq5E2}), we find a quadratic scaling for the environmental QFI:
\begin{equation}
I_{\rm E}(t)\propto M^2t.
\label{eq5F22}
\end{equation}
Thus, as in the local setup, both the global and environmental QFIs exhibit the similar scaling with $M$.

%
% -------------------------------------------------------------------------------------------------------------------------------------------------------------------------------------
%

\subsubsection{\label{sec5F3}Nonreciprocal local setup}
Non-Hermitian jump operators induce qualitative changes in the system behavior via the imaginary-part matrix $m_I$ in the matrix $h_{\rm eff}$ (see Eqs.~(\ref{eq5B2}) and (\ref{eq5B3})). To be concrete, we consider a system whose coupling to the environment is described by the following jump operator as detailed later:
\begin{equation}
\hat{L}^-_j=\sqrt{\gamma}(e^{i\Delta\phi/2}\hat{x}_j-e^{-i\Delta\phi/2}\hat{x}_{j+1}),
\label{eq5F31}
\end{equation}
where $\gamma$ and $\Delta\phi$ are real-valued parameters that characterize the coupling strength and the nonreciprocity, respectively. The key feature is that the corresponding matrix $m_I$ is nonvanishing when $\Delta\phi\neq 0$ and takes an antisymmetric form,
\begin{eqnarray}
m_I=\left( \begin{array}{ccc}
0                     & \gamma\sin\Delta\phi &        \vspace{5pt}\\
-\gamma\sin\Delta\phi & 0                    & \ddots \vspace{5pt}\\
                      & \ddots               & \ddots
\end{array}\right).
\label{eq5F32}
\end{eqnarray}
Physically, this leads to a spatially asymmetric energy exchange between the neighboring modes, which induces a unidirectional net energy flow. While the locality of the jump operator implies the asymptotic scaling $I_{\rm G}(t)\propto M\dot{\bar{n}}_{\rm st}t$, dictated by the lower bound in Eq.~(\ref{eq5D2}), the nonreciprocity fundamentally alters the structure of the projectors $\tilde{P}_\alpha$ through the non-Hermitian skin effect, as reviewed in Appendix~\ref{secH1}. This phenomenon, arising from the non-Hermitian nature of $h_{\rm eff}$ in Eq.~(\ref{eq5B3}), leads to a qualitative modification of the mode-number scaling of the excitation number $\bar{n}_{\rm st}$.

More specifically, due to the skin effect, a macroscopic fraction of the eigenmodes of $h_{\rm eff}$ becomes localized at one of the boundaries, say at site $j=1$. The corresponding right eigenvectors exhibit an exponential profile:
\begin{equation}
({\bm r}_\alpha)_j\propto\exp\left(\frac{M-j}{\xi}\right),
\label{eq5F33}
\end{equation}
where $\xi$ is a localization length; the left eigenvectors localize at the opposite boundary. As inferred from Eq.~(\ref{eq5B5}), this spatial separation of left and right eigenvectors causes the projection operators $\tilde{P}_\alpha$ to have norms that grow exponentially with $M$. This exponential scaling is inherited by the dephased matrix $\tilde{m}_R$ and, crucially, by the excitation rate. Consequently, in the long-time regime, we find
\begin{equation}
I_{\rm G}(t)\propto\dot{\bar{n}}_{\rm st}Mt,~\dot{\bar{n}}_{\rm st}\propto e^{2M/\xi}.
\label{eq5F34}
\end{equation}
Physically, the nonreciprocal dynamics acts as a directional amplifier for quantum fluctuations, which are then continuously funneled towards one boundary due to the skin effect. This accumulation leads to an exponential growth in the response to perturbations, thereby providing the exponential enhancement in the global QFI at the cost of energy consumption.

Meanwhile, however, the environmental QFI exhibits a fundamentally different scaling. Since the jump operator is local, the operator norm $\|m_R\|$ remains bounded and independent of $M$. Thus, as dictated by the lower bound in Eq.~(\ref{eq5E2}), the environmental QFI retains the linear scaling
\begin{equation}
I_{\rm E}(t)\propto Mt.
\label{eq5F35}
\end{equation}
Namely, while the internal dynamics is exponentially amplified, the information leaked to the environment is not, and the exponential enhancement is entirely contained in the correlations within the system. This unique feature distinguishes nonreciprocal continuous sensing from both local and global reciprocal setups.

%
% -------------------------------------------------------------------------------------------------------------------------------------------------------------------------------------
%

\subsection{\label{sec5G}Numerical results}
In this section, we explicitly demonstrate and verify the analytical results derived above. To this end, we study an array of trapped particles subjected to continuous Gaussian measurements. This model provides a platform to investigate the dynamics and mode-number scaling of both the global and environmental QFIs. We will show that, for different setups, the QFIs exhibit the distinct scaling behaviors predicted by our analytical arguments above, showcasing the rich phenomenology of many-particle continuous metrology. In the following, we chose the vacuum state as the initial condition, similarly to Sec.~\ref{sec4D}.

%
% -------------------------------------------------------------------------------------------------------------------------------------------------------------------------------------
%

\subsubsection{\label{sec5G1}Trapped particle array}
We consider a one-dimensional array of $M$ particles trapped in harmonic potentials. The collective dynamics is described by the canonical position and momentum operators, which we group into a $2M$-component vector $\hat{\bm\phi}=(\hat{x}_1,\dots,\hat{x}_M,\hat{p}_1,\dots,\hat{p}_M)^{\rm T}$. The Hamiltonian governing the array is given by
\begin{equation}
\hat{H}=\frac{\Omega}{2}\sum_{j=1}^M(\hat{p}_j^2+\hat{x}_j^2)+\frac{K}{2}\sum_{j=1}^{M-1}(\hat{x}_j-\hat{x}_{j+1})^2-E\sum_{j=1}^M\hat{x}_j,
\label{eq5G11}
\end{equation}
where the first term represents the sum of the harmonic trapping potentials with frequency $\Omega$, the second term describes a nearest-neighbor harmonic coupling with strength $K$, and the third term models the interaction with a uniform external field $E$ that couples to the position of each particle. The central objective of our analysis is to determine the ultimate precision limit for estimating this field, which is treated as the unknown parameter. For simplicity and to focus on the scaling properties, below we set the characteristic frequencies of the system to be equal, $\Omega=K=1$.

%
% -------------------------------------------------------------------------------------------------------------------------------------------------------------------------------------
%

\subsubsection{\label{sec5G2}Local setup}
\begin{figure}[]
\includegraphics[width=8.5cm]{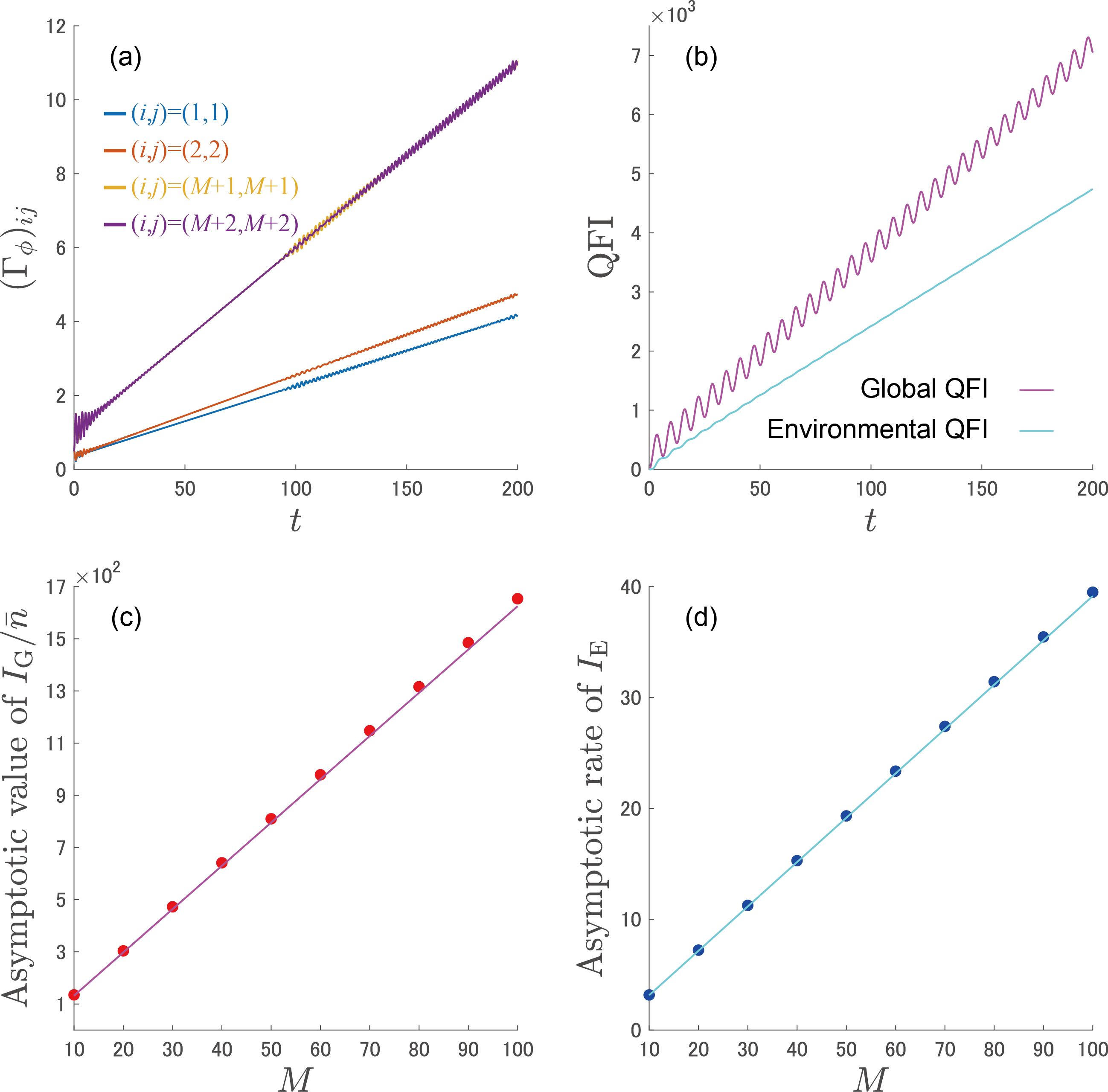}
\caption{\label{fig4}Local setup. (a) Time evolution of diagonal elements of the covariance matrix. (b) Time evolution of the global and environmental QFIs. (c) Mode-number scaling of the asymptotic rate of the global QFI, normalized by the bosonic excitation number. (d) Mode-number scaling of the asymptotic rate of the environmental QFI. In panels (c) and (d), the dots are obtained from numerical simulations, while the solid lines represent the analytical results for the steady values $\dot{I}_{\rm G,st},\dot{\bar{n}}_{\rm st}$, and $\dot{I}_{\rm E,st}$ (cf. Eqs.~(\ref{eq5C3}), (\ref{eq5D1}), and (\ref{eq5E1})). The parameters are set to be $M=60$ for (a) and (b), with $\gamma=0.1$ and $E=0.1$ for all the panels.}
\end{figure}
We first consider the simplest setup, where the position of each trapped particle is monitored independently. This corresponds to a set of $M$ local jump operators given by
\begin{equation}
\hat{L}_n=\sqrt{\gamma}\hat{x}_n,~(n=1,2,\dots,M).
\label{eq5G21}
\end{equation}
The numerical results for this scenario are presented in Fig.~\ref{fig4}. As shown in Fig.~\ref{fig4}(a), the diagonal elements of the covariance matrix $\Gamma_\phi$ exhibit a sustained linear growth in time. This behavior arises because dynamics is dominated by the linearly growing term in the general solution for $\Gamma_\phi(t)$ (cf. Eq.~(\ref{eq5A2})). Similarly, both the global QFI $I_{\rm G}$ and the environmental QFI $I_{\rm E}$ also grow linearly as shown in Fig.~\ref{fig4}(b).

We next analyze the mode-number scaling of these quantities. As discussed above, for a local setup, the asymptotic rate of bosonic excitations per mode $\dot{\bar{n}}_{\rm st}$ is independent of the mode number $M$, while the growth rate of both the global and environmental QFIs scale linearly with $M$ [Figs.~\ref{fig4}(c) and (d)]. These scaling behaviors demonstrate that each particle acts as an independent sensor, and the total information is the incoherent sum of contributions from each site, which corresponds to the lower bounds in the inequalities (\ref{eq5D2}) and (\ref{eq5E2}). Moreover, the numerically estimated values of the growth rates are in excellent agreement with the analytical results obtained for the long-time regime in Eqs.~(\ref{eq5C3}), (\ref{eq5D1}), and (\ref{eq5E1}) [solid lines in Figs.~\ref{fig4}(c) and (d)]. These numerical results validate the analytical predictions in Sec.~\ref{sec5F1}.

%
% -------------------------------------------------------------------------------------------------------------------------------------------------------------------------------------
%

\subsubsection{\label{sec5G3}Global setup}
\begin{figure}[]
\includegraphics[width=8.5cm]{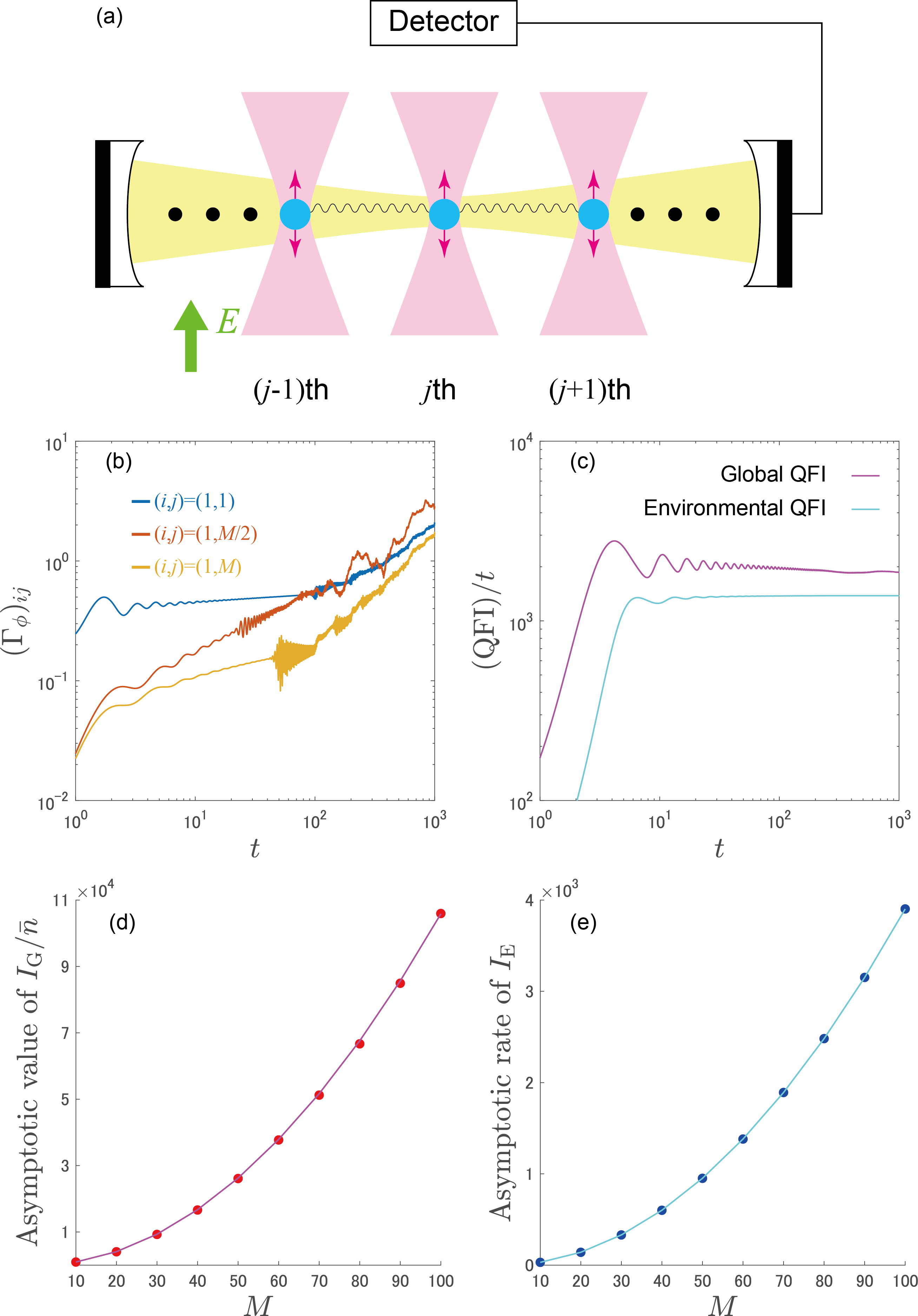}
\caption{\label{fig5}Global setup. (a) Array of $M$ trapped particles that is placed inside an optical cavity. A collective observable of all particles is monitored by detecting the light leaking from the cavity. (b) Time evolution of the covariance matrix in a log-log plot. (c) Time evolution of the global and environmental QFIs normalized by evolved time $t$ in a log-log plot. (d) Mode-number scaling of the asymptotic rate of the global QFI, normalized by the bosonic excitation number. (e) Mode-number scaling of the asymptotic rate of the environmental QFI. In panels (d) and (e), the dots are extracted from numerical results, while the solid curves represent the analytical results for the stationary regime. The parameters are $M=60$ for (b) and (c), with $\gamma=0.1$ and $E=0.1$ for all the panels.}
\end{figure}
Next, we investigate a global setup where a collective observable of the array is monitored. Such a process can be experimentally realized by placing the trapped particle array inside a cavity and monitoring the light leaking from it, as illustrated in Fig.~\ref{fig5}(a). This corresponds to a single jump operator that acts on all particles simultaneously:
\begin{equation}
\hat{L}=\sqrt{\gamma}\sum_{j=1}^M\hat{x}_j.
\label{eq5G31}
\end{equation}
The time evolution of the relevant quantities is shown in Figs.~\ref{fig5}(b) and (c). After an initial transient period, the covariance matrix $\Gamma_\phi$ settles into a regime of linear growth [Fig.~\ref{fig5}(b)]. This transition occurs after a dephasing time, beyond which the oscillatory components of $\Gamma_\phi(t)$ average out. The steady increase in quantum fluctuations directly translates into a linear growth of both $I_{\rm G}$ and $I_{\rm E}$ in the long-time limit, as shown in Fig.~\ref{fig5}(c).

As discussed in Sec.~\ref{sec5F2}, the asymptotic rate of bosonic excitations per mode $\dot{\bar{n}}_{\rm st}$ is found to be independent of the mode number $M$, similar to the local setup. However, the scalings of the QFIs are markedly different. As shown in Figs.~\ref{fig5}(d) and (e), the growth rates of both the global and environmental QFIs exhibit the quadratic scalings with the mode number $M$. These scaling behaviors are suitable in that they saturate the upper bounds in the inequalities (\ref{eq5D2}) and (\ref{eq5E2}). Also, the numerically estimated values of the growth rates again agree with the analytical results for the steady values indicated by the solid curves. These findings confirm the theoretical discussion in Sec.~\ref{sec5F2}, demonstrating that collective coupling can provide a quantum enhancement in continuous metrology.

%
% -------------------------------------------------------------------------------------------------------------------------------------------------------------------------------------
%

\begin{figure*}[]
\includegraphics[width=17cm]{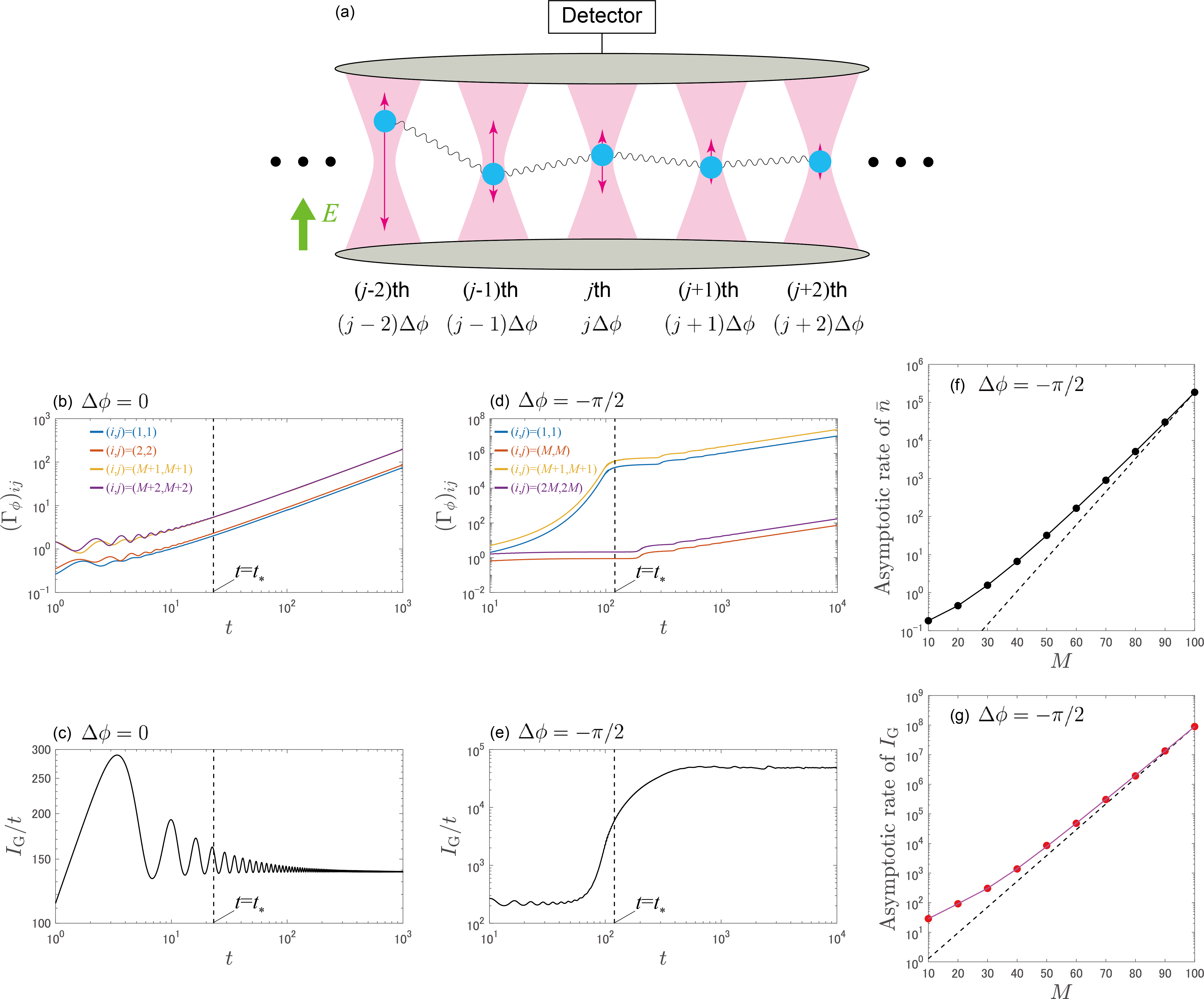}
\caption{\label{fig6}Nonreciprocal setup. The reciprocal ($\Delta\phi=0$, panels (b) and (c)) and nonreciprocal ($\Delta\phi\neq0$, panels (d)-(g)) scenarios are compared. (a) Array of $M$ trapped particles in which adjacent particles are trapped by lasers with a controlled phase difference $\Delta\phi$. This induces nonreciprocal backaction, leading to the exponentially enhanced quantum fluctuations of the trapped particles near the edge. (b) and (d) Time evolutions of diagonal elements of the covariance matrix $\Gamma_\phi$ in log-log plots. (c) and (e) Time evolutions of the global QFI in log-log plots. The vertical dashed lines indicate the dephasing time $t_\ast$ estimated by Eq.~(\ref{eq5G49}). (f) and (g) Asymptotic rates of the bosonic excitations in (f) and the global QFI in (g) plotted against the mode number $M$. In both panels, the dots correspond to the results estimated from numerical simulations, while the solid curves represent the analytical predictions. The black dashed lines show the exponential scaling $\exp(2M/\xi)$, with $\xi$ defined in Eq.~(\ref{eq5G45}). The parameters are $M=60$ in (b)-(e), with $\gamma=0.1$ and $E=0.1$ in all the panels.}
\end{figure*}

%
% -------------------------------------------------------------------------------------------------------------------------------------------------------------------------------------
%

\subsubsection{\label{sec5G4}Nonreciprocal setup}
We now turn to a more intricate scenario involving a nonreciprocal local setup. As depicted in Fig.~\ref{fig6}(a), such a scheme can be realized by introducing a controlled optical phase difference $\Delta\phi\in[-\pi/2,\pi/2]$ between the lasers trapping adjacent particles. This phase difference is imprinted onto the light scattered by the particles, and its subsequent detection leads to a spatially nonreciprocal measurement backaction. The corresponding jump operators are given by symmetric and antisymmetric combinations of position operators of adjacent particles:
\begin{eqnarray}
&&\hat{L}_j^-=\sqrt{\gamma}(e^{i\Delta\phi/2}\hat{x}_j-e^{-i\Delta\phi/2}\hat{x}_{j+1}), \label{eq5G41}\\
&&\hat{L}_j^+=\sqrt{\gamma}(\hat{x}_j+\hat{x}_{j+1}). \label{eq5G42}
\end{eqnarray}
Here, the index $n$ in the general formalism of Eq.~(\ref{eq2B6}) encompasses the site index $j$ and the type of combination $\pm$. For the system not to exhibit dynamical instability, we assume $K>\gamma|\sin\Delta\phi|$.

A key consequence of the phase difference $\Delta\phi$ is the emergence of effectively non-Hermitian dynamics, which manifests as the non-Hermitian skin effect. This can be seen by examining the classical equations of motion corresponding to the mean-field dynamics (cf. Eq.~(\ref{eq3A12})), which take the form:
\begin{eqnarray}
&&\frac{1}{\Omega}\frac{d^2x_j}{dt^2}=-(\Omega+2K)x_j+\frac{E}{\Omega} \nonumber\\
&&+(K+\gamma\sin\Delta\phi)x_{j-1}+(K-\gamma\sin\Delta\phi)x_{j+1}. \nonumber\\
\label{eq5G43}
\end{eqnarray}
The measurement backaction induces nonreciprocal exchanges $K\pm\gamma\sin\Delta\phi$ between adjacent particles, where $\Delta\phi$ controls the degree of nonreciprocity. For instance, with $\Delta\phi<0$, oscillations are preferentially transported to the left in Fig.~\ref{fig6}(a), leading to an exponential amplification of motion at the left boundary. The amplitude of this boundary oscillation increases with $|\Delta\phi|$.

In the following subsections, we shall analyze this system by applying non-Hermitian spectral theory to characterize the properties of the matrix $h_{\rm eff}$ defined in Eq.~(\ref{eq5B3}). We then demonstrate that this nonreciprocity leads to an exponential scaling of the global QFI with the mode number. We also show that the environmental QFI scales differently from the global QFI, highlighting the unique nature of information flow in the nonreciprocal system. We finally analyze the dephasing time, which exhibits an unconventional scaling with $M$.

%
% -------------------------------------------------------------------------------------------------------------------------------------------------------------------------------------
%

\subsubsection*{Non-Hermitian spectral theory}
A spectral theory for one-dimensional non-Hermitian tight-binding models (as detailed in Appendix~\ref{secH2}) allows for the exact determination of the eigenvalues of the matrix $h_{\rm eff}$, which governs the coherent part of the dynamics. The eigenvalues are found to be
\begin{equation}
\tilde{\lambda}_\alpha=\Omega+2K+2\sqrt{K^2-\gamma^2\sin^2\Delta\phi}\cos\theta_\alpha,
\label{eq5G44}
\end{equation}
where $\theta_\alpha=\pi\alpha/(M+1)$ for $\alpha=1,2,\dots,M$. In the parameter regime $K>\gamma|\sin\Delta\phi|$, the system preserves a PT symmetry associated with the non-Hermitian skin effect~\cite{Longhi2019,Xiao2021,Hu2024}, which ensures that all eigenvalues $\tilde{\lambda}_\alpha$ are real. This guarantees that the system exhibits purely oscillatory dynamics without exponential growth or decay.

As discussed in Sec.~\ref{sec5F3}, the spatial nonreciprocity leads to the non-Hermitian skin effect, which causes a macroscopic number of the bulk eigenmodes of $h_{\rm eff}$ to become exponentially localized at one of the boundaries. The non-Hermitian spectral theory provides not only the eigenvalues but also a precise characterization of this localization. For the trapped particle array, all eigenstates localize at the same boundary with a common localization length $\xi$, given by
\begin{equation}
\frac{1}{\xi}=\left|\log\sqrt{\frac{K+\gamma\sin\Delta\phi}{K-\gamma\sin\Delta\phi}}\right|.
\label{eq5G45}
\end{equation}
The localization length $\xi$ decreases as the nonreciprocity strength $|\Delta\phi|$ increases, signifying that the skin modes become more tightly confined to the edge of the array.

%
% -------------------------------------------------------------------------------------------------------------------------------------------------------------------------------------
%

\subsubsection*{Global quantum Fisher information}
We begin by examining the reciprocal case ($\Delta\phi=0$) as a baseline. In this case, all the jump operators are Hermitian, and the backaction from measurements of internal quadratures leads to heating. Figure~\ref{fig6}(b) shows the time evolution of several diagonal elements of the covariance matrix $\Gamma_\phi$. In the long-time regime $t>t_\ast$, where $t_\ast$ is a characteristic time scale for dephasing discussed below, these elements grow linearly with time, and their growth rates are uniform across the array. This behavior directly leads to an asymptotic linear growth of the global QFI $I_{\rm G}$ as shown in Fig.~\ref{fig6}(c). The mode-number scaling of the growth rates of both $\Gamma_\phi$ and $I_{\rm G}$ in this reciprocal limit is identical to that observed in the simple local-measurement case discussed before.

The dynamics qualitatively changes in the nonreciprocal case ($\Delta\phi\neq0$). As shown in Fig.~\ref{fig6}(d), the diagonal elements of $\Gamma_\phi$ undergo a drastic change in their growth around $t=t_\ast$, reflecting the influence of the non-Hermitian skin effect on the quantum fluctuations. This behavior originates from the overlap between the right and left eigenvectors of the matrix $h_{\rm eff}$, which exhibits the spatially exponential decay. Specifically, in Eq.~(\ref{eq5A2}), the linearly growing term is suppressed relative to the oscillatory terms in the short-time regime $t<t_\ast$. Moreover, this dynamical crossover is also manifested as a sharp increase of $I_{\rm G}$ around the dephasing time $t=t_\ast$ [Fig.~\ref{fig6}(e)], which is absent in the reciprocal case [Fig.~\ref{fig6}(c)]. We note that this behavior indicates a crossover in the mode-number dependence of $I_{\rm G}$. In the following section, we analyze the mode-number scaling of the characteristic time scale $t_\ast$ in terms of the dephasing of the oscillatory terms in Eq.~(\ref{eq5A2}).

In the long-time regime $t>t_\ast$, the quantum fluctuations of the particles near one edge (here, the leftmost particle, as $\Delta\phi<0$) are exponentially enhanced due to the non-Hermitian skin effect, while those at the opposite edge are suppressed [Fig.~\ref{fig6}(d)]. Consequently, the rate of bosonic excitations $\dot{\bar{n}}$ scales as $\exp(2M/\xi)$, where $\xi$ is the localization length in Eq.~(\ref{eq5G45}) [Fig.~\ref{fig6}(f)]. Furthermore, in $t>t_\ast$, the growth rate of $I_{\rm G}$ also exhibits the exponential scaling $\exp(2M/\xi)$ [Fig.~\ref{fig6}(g)]. These results demonstrate that the exponentially large number of bosonic excitations generated by the nonreciprocal dynamics serves as a resource that can be converted into metrological information, confirming the scaling relation in Eq.~(\ref{eq5F34}).

We note that, for a small $M$ in Figs.~\ref{fig6}(f) and (g), the numerical data (dots) deviate from the pure exponential scaling $e^{2M/\xi}$ (dashed black line) due to finite-size effects; this is because the localization length $\xi$ is not sufficiently smaller than the system size $M$. Nevertheless, the results still precisely agree with the steady values predicted by the analytical formulas in Eqs.~(\ref{eq5C3}) and (\ref{eq5D1}) (solid curves).

%
% -------------------------------------------------------------------------------------------------------------------------------------------------------------------------------------
%

\subsubsection*{Environmental quantum Fisher information}
\begin{figure}[]
\includegraphics[width=8.5cm]{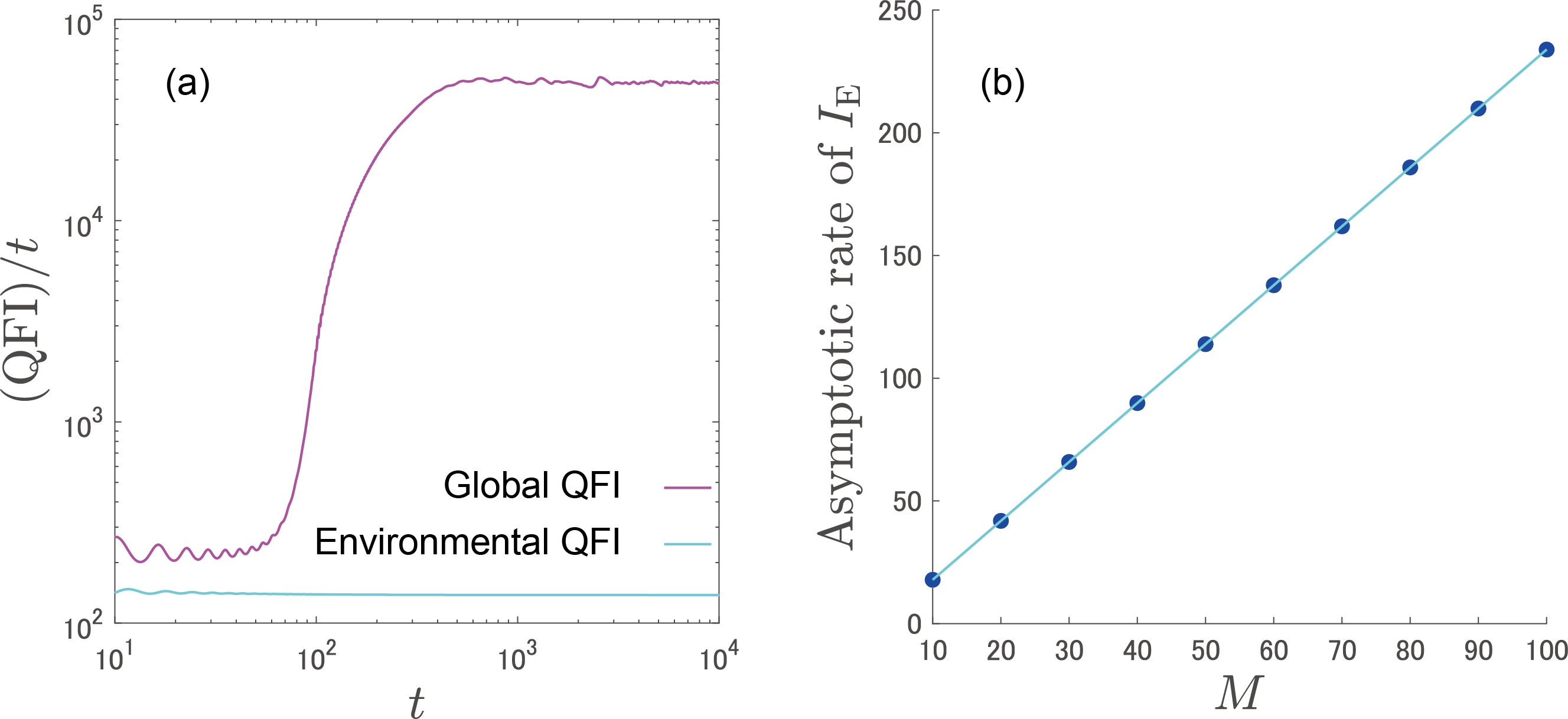}
\caption{\label{fig7}Global and environmental QFIs in the nonreciprocal local setup. (a) Time evolution of the global and environmental QFIs normalized by evolved time $t$. (b) Asymptotic rate of the environmental QFI plotted against the mode number $M$. The dots are extracted from numerical results, while the solid line represents the analytical prediction. The parameters are set to be $M=60$ in (a), with $\gamma=0.1,E=0.1$, and $\Delta\phi=-\pi/2$ in both panels.}
\end{figure}

We next analyze the behavior of the environmental QFI $I_{\rm E}$ in the nonreciprocal setup. As shown in Fig.~\ref{fig7}(a), $I_{\rm E}$ grows linearly in time throughout the evolution, in sharp contrast to the behavior of $I_{\rm G}$, which exhibits a dramatic increase around the dephasing time $t_\ast$. The result indicates that the asymptotic rate of $I_{\rm E}$ does not exhibit the exponential scaling.

This point is confirmed in a more explicit way by the mode-number scaling analysis in Fig.~\ref{fig7}(b), which shows that the asymptotic rate of $I_{\rm E}$ in the long-time regime is proportional to $M$. Such linear scaling, $I_{\rm E}\propto Mt$, is consistent with the theoretical discussion in Sec.~\ref{sec5F3} and corresponds to the lower bound in Eq.~(\ref{eq5E2}). Physically, this indicates that, while the internal system dynamics leads to an exponential accumulation of quantum fluctuations at the boundary, this enhanced resource is not transferred to the environment; the information accessible through environmental monitoring remains proportional to the number of local particles. This highlights qualitatively distinct roles between the information contained within the system and the information radiated into the environment in the presence of nonreciprocal dynamics.

%
% -------------------------------------------------------------------------------------------------------------------------------------------------------------------------------------
%

\subsubsection*{Dephasing time}
We have observed that the covariance matrix enters a linear growth regime after a characteristic dephasing time $t_\ast$ [Figs.~\ref{fig6}(b) and (d)]. This behavior stems from the dephasing of the oscillatory terms of $\Gamma_\phi(t)$ in Eq.~(\ref{eq5A2}). Here, we provide an estimate for this dephasing time by analyzing the spectral properties of the matrix $X$ (cf. Eq.~(\ref{eq5A1})).

The dephasing time can be inferred from the dispersion of the eigenvalue gaps of $X$, in a manner similar to relaxation in quantum many-body dynamics~\cite{De2018}. We first define an expansion factor $V_{\cal S}$ that quantifies the contribution of each oscillatory term, indexed by a pair of eigenmodes ${\cal S}=(\alpha,\alpha^\prime)$,
\begin{equation}
V_{\cal S}=V_{(\alpha,\alpha^\prime)}=\left\Vert\frac{P_\alpha YP_{\alpha^\prime}^\dag}{i(\lambda_\alpha-\lambda_{\alpha^\prime})}\right\Vert_{\rm tr},
\label{eq5G46}
\end{equation}
where ${\cal S}\in\{(\alpha,\alpha^\prime):\alpha,\alpha^\prime\in\{1,2,\dots,M\},\alpha\neq\alpha^\prime\}$, and $\|A\|_{\rm tr}\equiv\sqrt{{\rm Tr}[A^\dag A]}$ is the Hilbert-Schmidt norm. Let $\tilde{\lambda}_{\cal S}=\lambda_\alpha-\lambda_{\alpha^\prime}$ be the corresponding eigenvalue gap. The eigenvalue-gap dispersion is then defined as the weighted variance of these gaps:
\begin{equation}
\sigma_{\rm gap}^2\equiv\sum_{\cal S} q_{\cal S}(\tilde{\lambda}_{\cal S}-m_{\rm gap})^2,
\label{eq5G47}
\end{equation}
where
\begin{equation}
q_{\cal S}=\frac{V_{\cal S}^2}{\displaystyle\sum_{{\cal S}^\prime}V_{{\cal S}^\prime}^2}
\label{eq5G48}
\end{equation}
is the normalized weight, and $m_{\rm gap}=\sum_{\cal S}q_{\cal S}\tilde{\lambda}_{\cal S}$ is the average gap. In the present case, due to the symmetry $\tilde{\lambda}_{(\alpha,\alpha^\prime)}=-\tilde{\lambda}_{(\alpha^\prime,\alpha)}$, the average gap $m_{\rm gap}$ must vanish. The dephasing time can then be estimated as the inverse of the standard deviation of the gap distribution:
\begin{equation}
t_\ast\simeq\frac{\pi}{\sigma_{\rm gap}}.
\label{eq5G49}
\end{equation}
\begin{figure}[]
\includegraphics[width=5cm]{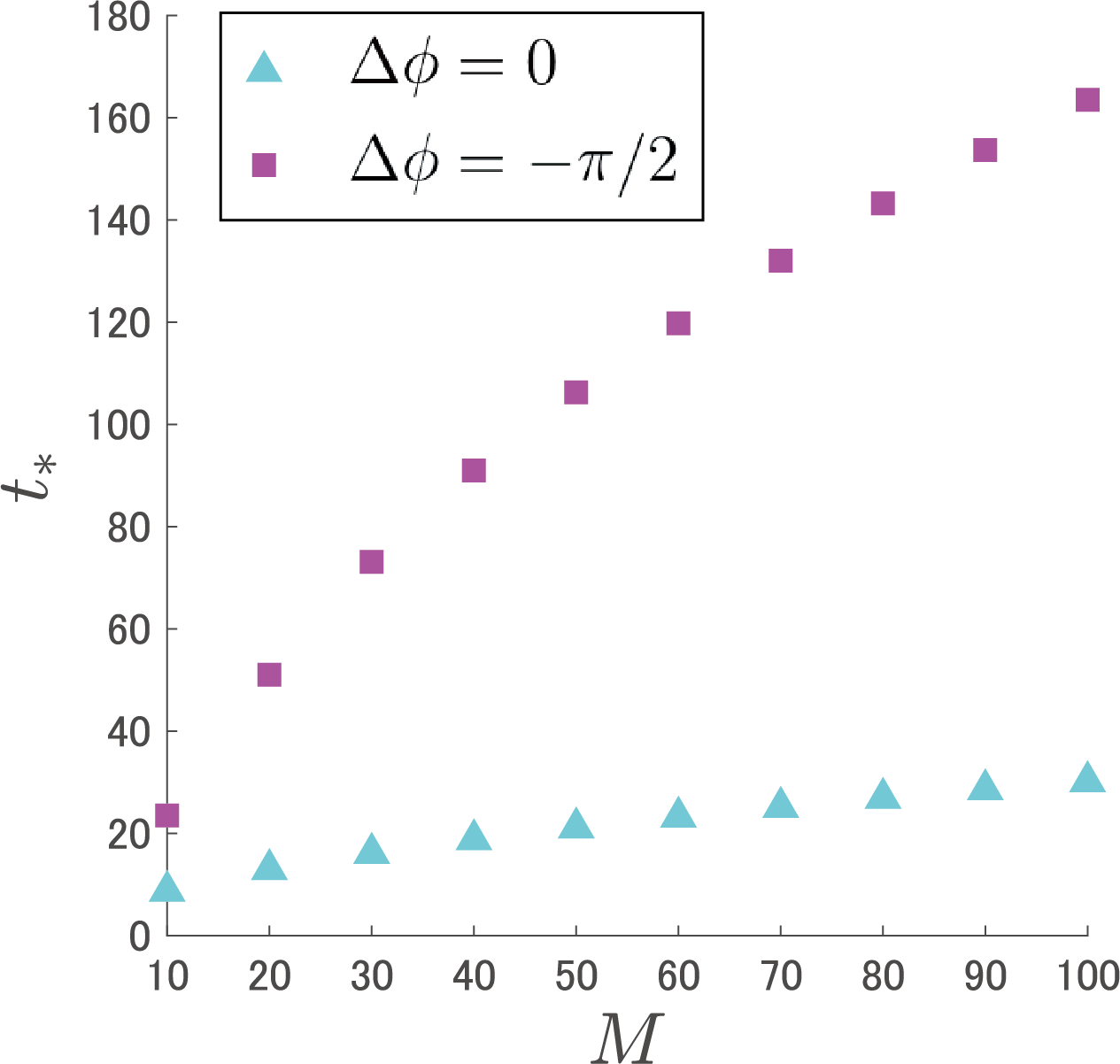}
\caption{\label{fig8}Mode-number scaling of the dephasing time $t_\ast$, calculated from Eq.~(\ref{eq5G49}), for reciprocal ($\Delta\phi=0$) and nonreciprocal ($\Delta\phi\neq0$) cases. The parameters are set to be $\gamma=0.1$ and $E=0.1$.}
\end{figure}

Figure~\ref{fig8} shows the mode-number dependence of the dephasing time $t_\ast$ obtained from Eq.~(\ref{eq5G49}). While $t_\ast$ is nearly independent of the mode number $M$ in the reciprocal case ($\Delta\phi=0$), we find that it grows linearly $\propto M$ in the nonreciprocal case ($\Delta\phi\neq0$). This theoretical estimate aligns with the timescale at which the quantum fluctuation of the boundary particle undergoes a sharp change in its growth rate (see Fig.~\ref{fig6}(e)). Such unconventional scaling of $t_\ast$ originates from the non-Hermitian skin effect; the exponential localization of eigenmodes reduces their spatial overlap, which in turn suppresses the magnitude of the expansion factors $V_{\cal S}$ for many pairs of modes. This effectively narrows the distribution of eigenvalue gaps, decreasing the dispersion $\sigma_{\rm gap}$ and thus prolonging the dephasing time.

We remark that the exponential scaling of the global QFI rate sets in only after a crossover time $t^\ast$. Thus, the asymptotic scaling applies in the regime $t\gg t^\ast$. Equivalently, the mode-number scaling should be understood in an iterated-limit sense; the long-time limit is first taken for each fixed $M$, and the resulting asymptotic rate is then analyzed as a function of $M$. In realistic sensing scenarios, the accessible interrogation time may be restricted by finite coherence windows, unmonitored loss, and other technical imperfections. Thus, the exponential scaling of the asymptotic QFI rate should be interpreted together with this order of limits and crossover condition.

%
% -------------------------------------------------------------------------------------------------------------------------------------------------------------------------------------
%

\section{\label{sec6}Fixed coupling strength}
In the preceding sections, the global setup has been introduced with a fixed system-environment coupling strength per mode. With this convention, the effective coupling to the collective bright mode increases with the number of modes $M$. To distinguish genuine mode-number scaling from the trivial increase of the global setup, it is also useful to discuss a scaling at a fixed total coupling strength, i.e., the total backaction strength of the collective channel is kept fixed as $M$ is varied. This is implemented by rescaling the collective jump operator by $1/\sqrt{M}$. Specifically, in the dissipative case, we consider the jump operators
\begin{eqnarray}
&&\hat{L}_n=\sqrt{\zeta}\hat{b}_n,~(n=1,2,\dots,M), \label{eq61}\\
&&\hat{L}_n=\sqrt{\frac{\gamma}{2M}}\sum_{j=1}^M(\hat{b}_j+\hat{b}_j^\dag). \label{eq62}
\end{eqnarray}
Similarly, in the zero-damping coupling, we use the normalized collective jump operator
\begin{equation}
\hat{L}=\sqrt{\frac{\gamma}{M}}\sum_{j=1}^M\hat{x}_j. \label{eq63}
\end{equation}
This normalization keeps the total strength of the collective monitoring channel fixed as $M$ is valid.

\begin{figure}[]
\includegraphics[width=8.5cm]{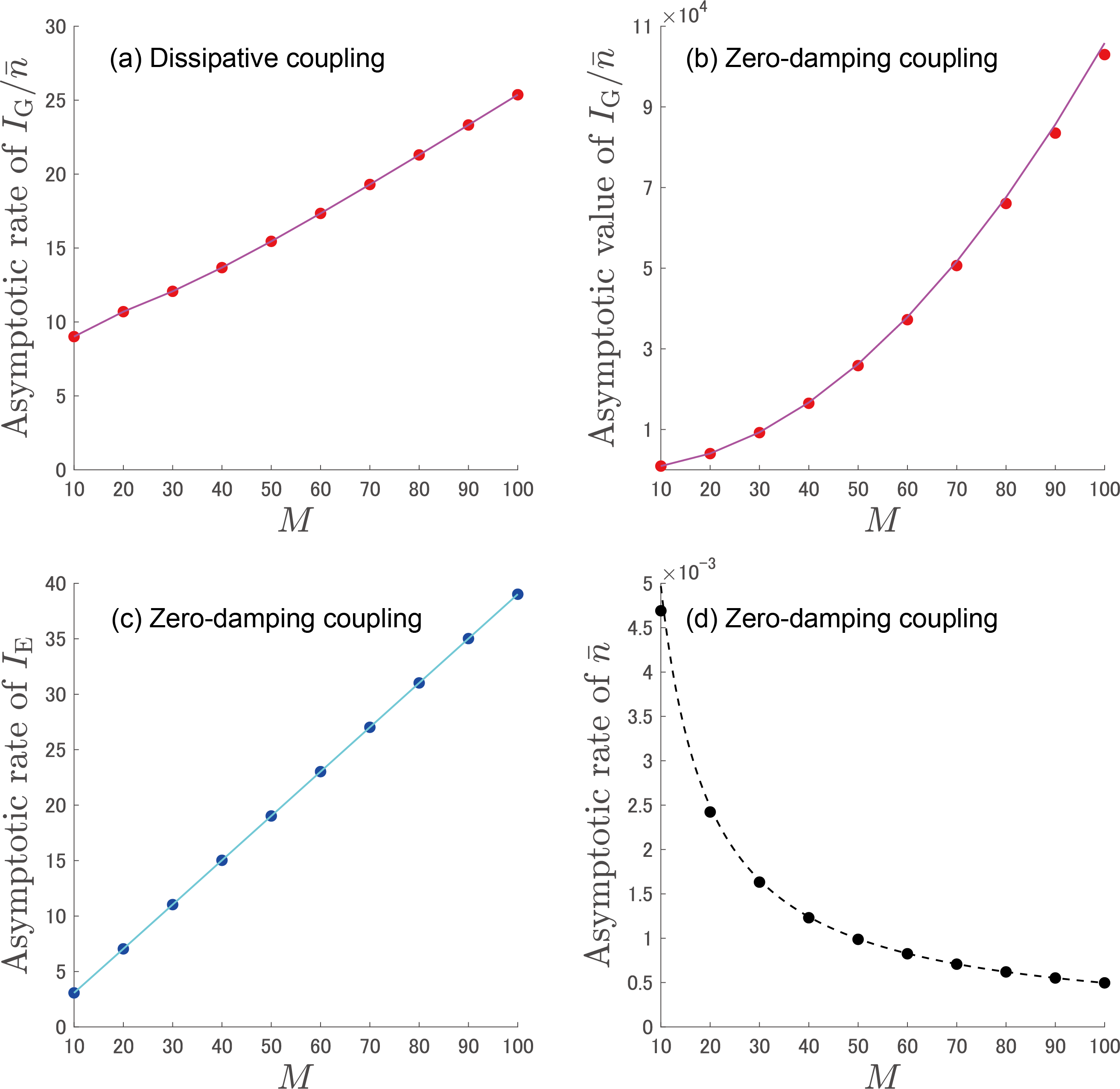}
\caption{\label{fig9}Asymptotic scalings of the global QFI $I_{\rm G}$, the environmental QFI $I_{\rm E}$, and the number of bosonic excitations per mode $\bar{n}$ under the normalization convention with the fixed coupling strength. (a) and (b) Asymptotic rates of the global QFI, normalized by the bosonic excitation number (c) Asymptotic rate of the environmental QFI. (d) Asymptotic rate of the bosonic excitation number. In (a), the parameters are $\Delta=0.5,\zeta=0.3$, and $E=0.1$. In (b), (c), and (d), the parameters are $\gamma=0.1$ and $E=0.1$.}
\end{figure}

With this convention, in the dissipative coupling, we find that the scaling of the asymptotic rate of $I_{\rm G}/\bar{n}$ behaves as $O(M)$; see Fig.~\ref{fig9}(a). This occurs because, in the large-$M$ limit, the local dissipative channels in Eq.~(\ref{eq61}) become dominant in determining the system dynamics. By contrast, in the zero-damping coupling, the asymptotic value of $I_{\rm G}/\bar{n}$ retains the same quadratic scaling with $M$, even after the collective jump operator is normalized by $1/\sqrt{M}$; see Fig.~\ref{fig9}(b). Meanwhile, the asymptotic rate of $I_E$ is proportional to $M$, as shown in Fig.~\ref{fig9}(c). The key point is that the asymptotic rate of bosonic excitations is also reduced by the same normalization, as shown in Fig.~\ref{fig9}(d), scaling as $1/M$. Consequently, the reductions of $I_{\rm G}$ and $\bar{n}$ compensate each other in the resource-normalized quantity $I_{\rm G}/\bar{n}$. As a result, the global resource-normalized sensitivity in this setup retains the same quadratic scaling with $M$.

%
% -------------------------------------------------------------------------------------------------------------------------------------------------------------------------------------
%

\section{\label{sec7}Summary and Outlook}
In this paper, we have developed a theoretical framework for the metrological analysis of a multimode free-bosonic system subjected to continuous Gaussian measurement. This formalism enables the precise calculation of both the global and environmental QFIs by leveraging the properties of bosonic Gaussian states. In particular, we derive analytical expressions for the asymptotic growth rates of the global and environmental QFIs (Eqs.~(\ref{eq4B2}), (\ref{eq5D1}), and (\ref{eq5E1})), which facilitate a detailed analysis of their scaling behaviors.

We investigated two broad classes of system-environment couplings. The first class comprises dissipative couplings leading to contractive dynamics, where the system evolves towards a unique steady state. A key finding is the asymptotic equivalence of the global and environmental QFI rates, $\dot{I}_{\rm G,st}=\dot{I}_{\rm E,st}$, showing that all acquired information can be faithfully transferred to the environment in the long-time limit. We also derived the bound (\ref{eq4C2}), which demonstrates that, while a quadratic scaling with the mode number $M$ is attainable, the asymptotic growth rates of the QFIs scales at most linearly with the energy resource. The second class consists of zero-damping couplings, which generate oscillatory dynamics without dissipative relaxation and can lead to unbounded growth of the bosonic excitation number. In this case, the global and environmental QFI rates are generally separated, $\dot{I}_{\mathrm{G,st}}>\dot{I}_{\mathrm{E,st}}$. This separation arises because the global QFI includes parameter information coherently retained in the system, whereas the environmental QFI captures only the information radiated into the emitted field. Accordingly, the global QFI rate obeys the resource-dependent bound in Eq.~(\ref{eq5D2}), which has the same form as in the dissipative case, while the environmental QFI rate obeys the distinct constraint in Eq.~(\ref{eq5E2}), whose lower bound is independent of the excitation rate. Thus, unlike in the dissipative case, the growth of internal bosonic excitations is not fully reflected in the information accessible from the environment.

As a concrete demonstration, we analyzed several concrete physical setups, which provide a platform for realizing the aforementioned scaling behaviors. For both dissipative and zero-damping couplings, we showed that protocols restricted to local setups on each mode yield the linear scaling of the QFIs $I_{\rm G,E}\propto Mt$. These scaling behaviors correspond to the lower bounds of our derived inequalities (cf. Eqs.~(\ref{eq4C2}), (\ref{eq5E2}), and (\ref{eq5D2})) and indicate a limited extraction of information serving as a baseline. In contrast, the inclusion of a global setup, which collectively probes all modes, can lead to the quadratic scaling $I_{\rm G,E}\propto M^2t$. This signifies that collective couplings are a key ingredient for enabling quantum-enhanced performance in continuous bosonic sensing. Finally, in the nonreciprocal local setup, the global and environmental QFIs adhere to the scalings $I_{\rm G}\propto e^{2M/\xi}t$ and $I_{\rm E}\propto Mt$, respectively. A notable feature is the exponential enhancement of the global QFI at the cost of energy consumption, which is rooted in the non-Hermitian skin effect and the unbounded nature of the local Hilbert space in bosonic systems. Thus, the entire population of bosonic excitations can in principle be harnessed for sensing purposes, provided that measurements act on not only environment but also the probe itself. All these scaling behaviors are summarized in Table~\ref{tab1}.

We remark that, in the global setups, the quadratic scalings are obtained under the convention that the system-environment coupling strength per mode is kept fixed. When normalizing the collective jump operators with a fixed-total-coupling strength, the quadratic scaling changes to the linear one in the dissipative case. This reduction occurs because the local dissipative channels become dominant in the limit of a large mode number. In contrast, in the zero-damping case, the quadratic scaling of the global QFI rate persists after the renormalization, whereas the environmental QFI rate changes from the quadratic to linear scaling. This behavior can be understood from the fact that the renormalization also reduces the bosonic excitation number generated inside the system.

In any practical implementation, a quantum sensor will inevitably be subject to noise, leading to, e.g., the loss of bosonic excitations. A full calculation of the QFI associated with the fixed bosonic excitations in such a noisy many-particle sensor remains a formidable challenge. Indeed, recent theoretical efforts in this direction have been mainly confined to the detection of single-mode photons emitted from quantum sensors with a bounded Hilbert space~\cite{Khan2025,Yang2026}. In the nonreciprocal setup, however, we expect that the mode number can still serve as a resource for enhancing the metrological sensitivity even in the presence of noise. The reason is that the bosonic excitation number should maintain its exponential scaling owing to the topological robustness of the non-Hermitian skin effect against dissipation~\cite{Okuma2020,Zhang2020NH}. Consequently, the QFI should be still exponentially enhanced by increasing the mode number at the cost of energy consumption. In contrast, under other setups including the local or global setup, substantial loss of bosonic excitations and/or coherence should be detrimental to metrological performance.

Beyond this qualitative discussion, a quantitative analysis of realistic sensing performance requires taking into account various experimental imperfections. A systematic treatment of unmonitored loss, finite detection efficiency, imperfect backaction evasion, and technical noise would require extending the present framework to noisy detection channels and finite-time sensing protocols. Furthermore, it will be useful to evaluate not only the global and environmental QFIs, but also the classical Fisher information associated with concrete measurement protocols. We leave these practically important problems for future work.

Finally, we mention a possible experimental realization of the zero-damping-coupling setup. We envision that the trapped particle array can be implemented by using tweezer arrays of ultracold atoms or levitated nanoparticles. In this case, the interaction between the polarizations of nanoparticles and an applied external electric field produces a force that shifts the positions of nanoparticles. The electric field strength can be then measured by monitoring this displacement. For instance, the model depicted in Fig.~\ref{fig5}(a) is motivated by recent experiments involving the manipulation of ultracold atoms~\cite{Ho2025,Peters2025} or levitated nanoparticles within an optical cavity~\cite{Delic2019,Vijayan2024}. Similarly, the model shown in Fig.~\ref{fig6}(a) is based on the experimental demonstration of nonreciprocal exchanges between two levitated nanoparticles mediated by an optical phase difference~\cite{Rieser2022}; a similar nonreciprocal exchange can be induced by cavity-mediated interactions~\cite{Ho2025v2}. Furthermore, continuous measurements should be achievable through direct monitoring of the quadrature operators by using, e.g., a backaction-evading measurement. These experimental advancements suggest that tweezer arrays of ultracold atoms or levitated nanoparticles are a promising platform for the development of practical sensors capable of continuously probing external fields with high sensitivity.

%
% -------------------------------------------------------------------------------------------------------------------------------------------------------------------------------------
%

\begin{acknowledgments}
We are grateful to Zongping Gong, Theodoros Ilias, Kensuke Kobayashi, Masaya Nakagawa, Ryota Yutani, and Siro Yanai for valuable discussions. K.Y. is supported by JSPS KAKENHI through Grant No.~JP23K13027. A.C. acknowledges support from the U.S. Department of Energy Office of Science National Quantum Information Science Research Centers as part of the Q-NEXT center. Y.A. acknowledges support from the Japan Society for the Promotion of Science through Grant No.~JP19K23424 and from JST FOREST Program (Grant No.~JPMJFR222U, Japan) and JST CREST (Grant No.~JPMJCR23I2, Japan). This research was supported in part by grant NSF PHY-2309135 to the Kavli Institute for Theoretical Physics (KITP).
\end{acknowledgments}

%
% -------------------------------------------------------------------------------------------------------------------------------------------------------------------------------------
%

\appendix

%
% -------------------------------------------------------------------------------------------------------------------------------------------------------------------------------------
%

\section{\label{secA}Quantum Fisher information}
In this section, we show that the QFI is represented in terms of the second partial derivative of the quantum fidelity in general. The QFI is given by~\cite{Zhou2019}
\begin{equation}
I(\theta)=8\lim_{\delta\theta\rightarrow0}\frac{1-{\cal F}(\theta+\delta\theta/2,\theta-\delta\theta/2)}{(\delta\theta)^2},
\label{eqappA1}
\end{equation}
where $\theta$ is a parameter to be estimated, and ${\cal F}(\theta_1,\theta_2)$ is the quantum fidelity characterizing the overlap between two quantum states. Specifically, it is defined as
\begin{equation}
{\cal F}(\theta_1,\theta_2)={\rm Tr}\left[\sqrt{\sqrt{\hat{\rho}_{\theta_1}}\hat{\rho}_{\theta_2}\sqrt{\hat{\rho}_{\theta_1}}}\right],
\label{eqappA2}
\end{equation}
where $\hat{\rho}_{\theta_1}$ and $\hat{\rho}_{\theta_2}$ are density matrices. We note that the quantum fidelity satisfies the bounded condition
\begin{equation}
0\leq{\cal F}\left(\theta_1,\theta_2\right)\leq{\cal F}\left(\theta,\theta\right)=1
\label{eqappA3}
\end{equation}
and the symmetry
\begin{equation}
{\cal F}\left(\theta_1,\theta_2\right)={\cal F}\left(\theta_2,\theta_1\right).
\label{eqappA4}
\end{equation}
By using these properties, we can obtain
\begin{eqnarray}
&&2\left[1-{\cal F}(\theta+\delta\theta/2,\theta-\delta\theta/2)\right] \nonumber\\
&&={\cal F}(\theta+\delta\theta/2,\theta+\delta\theta/2)-{\cal F}(\theta+\delta\theta/2,\theta-\delta\theta/2) \nonumber\\
&&-{\cal F}(\theta-\delta\theta/2,\theta+\delta\theta/2)+{\cal F}(\theta-\delta\theta/2,\theta-\delta\theta/2). \nonumber\\
\label{eqappA5}
\end{eqnarray}
Equation~(\ref{eqappA5}) thus allows us to reformulate Eq.~(\ref{eqappA1}) as a second partial derivative, which is given by
\begin{equation}
I(\theta)=\left.4\partial_{\theta_1}\partial_{\theta_2}{\cal F}(\theta_1,\theta_2)\right|_{\theta_1=\theta_2=\theta}.
\label{eqappA6}
\end{equation}

We here remark that, if an analytical function of two real variables $f(x_1,x_2)$ satisfies $0\leq f(x_1,x_2)\leq f(x,x)=1$ and $f(x_1,x_2)=f(x_2,x_1)$ for $\forall~x_1,x_2$, then
\begin{eqnarray}
&&\left.\partial_{x_1}\partial_{x_2}f(x_1,x_2)\right|_{x_1=x_2=x} \nonumber\\
&&=\left.\partial_{x_1}\partial_{x_2}\log f(x_1,x_2)\right|_{x_1=x_2=x}.
\label{eqappA7}
\end{eqnarray}
By using Eqs.~(\ref{eqappA3}) and (\ref{eqappA4}), we can rewrite Eq.~(\ref{eqappA6}) as
\begin{equation}
I(\theta)=\left.4\partial_{\theta_1}\partial_{\theta_2}\log{\cal F}(\theta_1,\theta_2)\right|_{\theta_1=\theta_2=\theta}.
\label{eqappA8}
\end{equation}

%
% -------------------------------------------------------------------------------------------------------------------------------------------------------------------------------------
%

\section{\label{secB}Proof of Eq.~(\ref{eq2C4})}
In this section, we prove that the quantum fidelity of the environment can be reformulated in terms of the generalized reduced density matrix in Eq.~(\ref{eq2B3}), following Ref.~\cite{Yang2023}. Suppose that the dimension of a Hilbert space is $D$. Let $\theta$ denote a parameter to be estimated in the following. The key idea is that the joint system-environment state in Eq.~(\ref{eq2A2}) can be written in the form of the Schmidt decomposition as follows:
\begin{equation}
|\Psi_\theta(t)\rangle=\sum_{d=1}^Ds_d(\theta,t)|{\rm S}_{\theta,d}(t)\rangle\otimes|{\rm E}_{\theta,d}(t)\rangle,
\label{eqappB1}
\end{equation}
where
\begin{equation}
\sum_{d=1}^D\left[s_d(\theta,t)\right]^2=1,
\label{eqappB2}
\end{equation}
and $|{\rm S}_{\theta,d}(t)\rangle$ and $|{\rm E}_{\theta,d}(t)\rangle$ are orthonormal bases of the system and the environment, respectively. We then obtain the reduced density matrix of the environment by
\begin{equation}
\hat{\rho}_\theta^{\rm E}(t)=\sum_{d=1}^D\left[s_d(\theta,t)\right]^2|{\rm E}_{\theta,d}(t)\rangle\langle{\rm E}_{\theta,d}(t)|.
\label{eqappB3}
\end{equation}
Importantly, this representation allows us to rewrite Eq.~(\ref{eq2C3}) as follows:
\begin{eqnarray}
&&{\cal F}_{\rm E}(\theta_1,\theta_2,t) \nonumber\\
&&={\rm Tr}_{\rm Env}\left[\sqrt{\sum_{d,d^{\prime\prime}}\left[V(\theta_1,\theta_2)\right]_{dd^{\prime\prime}}|{\rm E}_{\theta_1,d}(t)\rangle\langle{\rm E}_{\theta_1,d^{\prime\prime}}(t)|}\right], \nonumber\\
\label{eqappB4}
\end{eqnarray}
where $V(\theta_1,\theta_2)$ is the $D\times D$ matrix defined as
\begin{eqnarray}
&&\left[V(\theta_1,\theta_2)\right]_{d,d^{\prime\prime}}=s_d(\theta_1,t)s_{d^{\prime\prime}}(\theta_1,t)\sum_{d^\prime}\{\left[s_{d^\prime}(\theta_2,t)\right]^2 \nonumber\\
&&\times\langle{\rm E}_{\theta_1,d}(t)|{\rm E}_{\theta_2,d^\prime}(t)\rangle\langle{\rm E}_{\theta_2,d^\prime}(t)|{\rm E}_{\theta_1,d^{\prime\prime}}(t)\rangle\}.
\label{eqappB5}
\end{eqnarray}

We here remark that Eq.~(\ref{eqappB4}) can be related to the generalized reduced density matrix written in terms of the Schmidt decomposition as
\begin{eqnarray}
&&\hat{\mu}_{\theta_1,\theta_2}(t)=\sum_{d,d^\prime}s_d(\theta_1,t)s_{d^\prime}(\theta_2,t)\langle{\rm E}_{\theta_2,d^\prime}(t)|{\rm E}_{\theta_1,d}(t)\rangle \nonumber\\
&&\times|{\rm S}_{\theta_1,d}(t)\rangle\langle{\rm S}_{\theta_2,d^\prime}(t)|.
\label{eqappB6}
\end{eqnarray}
Indeed, we have
\begin{equation}
\hat{\mu}_{\theta_1,\theta_2}(t)\hat{\mu}_{\theta_1,\theta_2}^\dag(t)=\sum_{d,d^{\prime\prime}}\left[V(\theta_1,\theta_2)\right]_{dd^{\prime\prime}}|{\rm S}_{\theta_1,d^{\prime\prime}}(t)\rangle\langle{\rm S}_{\theta_1,d}(t)|.
\label{eqappB7}
\end{equation}
Since the trace is invariant under the transpose, we obtain
\begin{equation}
{\cal F}_{\rm E}(\theta_1,\theta_2,t)={\rm Tr}_{\rm Sys}\left[\sqrt{\hat{\mu}_{\theta_1,\theta_2}(t)\hat{\mu}_{\theta_1,\theta_2}^\dag(t)}\right],
\label{eqappB8}
\end{equation}
which proves Eq.~(\ref{eq2C4}).

%
% -------------------------------------------------------------------------------------------------------------------------------------------------------------------------------------
%

\section{\label{secC}Proof of Eq.~(\ref{eq4C2})}
In this section, we consider the case of dissipative couplings where the system evolves towards a unique steady state and prove the bounds (\ref{eq4C2}) on the optimized QFIs (see Eqs.~(\ref{eq3E13}), (\ref{eq4B1}), and (\ref{eq4B4})) in terms of the steady-state value of bosonic excitations (\ref{eq4C1}). For the sake of notational simplicity, we express the optimized QFI rate and the number of bosonic excitations per mode in the long-time regime by $x$ and $y$, respectively (cf. Eqs.~(\ref{eq4B1}) and (\ref{eq4C1})):
\begin{align}
x&=8M\max_{|{\bm n}|=1}{\bm n}^{\rm T}(-X^{-1}\Gamma_{\rm st}){\bm n}, \label{eqappC1}\\
y&=\frac{1}{2M}{\rm Tr}[\Gamma_{\rm st}]. \label{eqappC2}
\end{align}
Here, we recall that $\Gamma_{\rm st}$ is defined in Eq.~(\ref{eq4A4}), and it is a real, symmetric, positive-semidefinite matrix. The assumptions on the matrix $X$ are summarized in Sec.~\ref{sec4A}.

We start from considering the upper bound
\begin{align}
x&\leq8M\left\|\frac{-X^{-1}\Gamma_{\rm st}-(X^{-1}\Gamma_{\rm st})^{\rm T}}{2}\right\| \nonumber\\
&\leq8M\|X^{-1}\Gamma_{\rm st}\| \nonumber\\
&\leq8M\|X^{-1}\|\|\Gamma_{\rm st}\|.
\label{eqappC3}
\end{align}
Since we have ${\rm Tr}[\Gamma_{\rm st}]\geq\|\Gamma_{\rm st}\|$ and
\begin{equation}
\|X^{-1}\|=\frac{\kappa(X)}{\|X\|}\leq\frac{c_3}{\|X\|}\leq\frac{c_3}{c_1},
\label{eqappC4}
\end{equation}
the resulting upper bound is
\begin{equation}
x\leq\frac{16c_3}{c_1}M^2y.
\label{eqappC5}
\end{equation}

To obtain the lower bound, we have
\begin{eqnarray}
&&\Lambda_{\rm max}\left(\frac{-X^{-1}\Gamma_{\rm st}-(X^{-1}\Gamma_{\rm st})^{\rm T}}{2}\right) \nonumber\\
&&\geq\frac{1}{2M}{\rm Tr}[-X^{-1}\Gamma_{\rm st}] \nonumber\\
&&=-\frac{1}{2M}{\rm Tr}\left[\frac{X^{-1}+(X^{-1})^{\rm T}}{2}\Gamma_{\rm st}\right] \nonumber\\
&&\geq-\frac{1}{2M}\Lambda_{\rm max}\left(\frac{X^{-1}+(X^{-1})^{\rm T}}{2}\right){\rm Tr}[\Gamma_{\rm st}],
\label{eqappC6}
\end{eqnarray}
where $\Lambda_{\rm max}(\cdot)$ represents the largest eigenvalue. Using the assumption (\ref{eq4A2}), we have
\begin{align}
X^{-1}+(X^{-1})^{\rm T}&=(X^{\rm T})^{-1}(X^{\rm T}+X)X^{-1} \nonumber\\
&\leq(X^{\rm T})^{-1}(-c_41_{2M})X^{-1} \nonumber\\
&=-c_4(XX^{\rm T})^{-1}.
\label{eqappC7}
\end{align}
We note that $XX^{\rm T}$ is positive definite, resulting in
\begin{align}
\Lambda_{\rm max}\left(\frac{X^{-1}+(X^{-1})^{\rm T}}{2}\right)&\leq-\frac{c_4}{2}\Lambda_{\rm min}\left((XX^{\rm T})^{-1}\right) \nonumber\\
&=-\frac{c_4}{2\|X\|^2}.
\label{eqappC8}
\end{align}
By combining $\|X\|=\kappa(X)/\|X^{-1}\|\leq c_2c_3$, we obtain the lower bound as
\begin{equation}
x\geq\frac{4c_4}{(c_2c_3)^2}My.
\label{eqappC9}
\end{equation}

%
% -------------------------------------------------------------------------------------------------------------------------------------------------------------------------------------
%

\section{\label{secD}Derivation of Eq.~(\ref{eq5A2})}
As discussed in Sec.~\ref{sec3B}, the time evolution of the covariance matrix is described by Eq.~(\ref{eq3D1}). In the case of zero-damping couplings leading to oscillatory dynamics, the spectral decomposition of the matrix $X$ is given by Eq.~(\ref{eq5A1}). In this section, we explain how to derive the general solution (\ref{eq5A2}). Let $\Gamma_\phi(0)$ denote an initial condition. The key idea is that one can solve the time-evolution equations of $\Gamma_{\alpha\alpha^\prime}\equiv P_\alpha\Gamma_\phi P_{\alpha^\prime}^\dag$ and $\bar{\Gamma}_{\alpha\alpha^\prime}\equiv P_\alpha\Gamma_\phi P_{\alpha^\prime}^{\rm T}$. First, from Eq.~(\ref{eq3D1}), the time-evolution equation of $\Gamma_{\alpha\alpha^\prime}$ is given by
\begin{equation}
\frac{d\Gamma_{\alpha\alpha^\prime}}{dt}=i(\lambda_\alpha-\lambda_{\alpha^\prime})\Gamma_{\alpha\alpha^\prime}+P_\alpha YP_{\alpha^\prime}^\dag.
\label{eqappD1}
\end{equation}
When $\alpha=\alpha^\prime$, the solution of Eq.~(\ref{eqappD1}) is written as
\begin{equation}
\Gamma_{\alpha\alpha}=P_\alpha\Gamma_\phi(0)P_\alpha^\dag+tP_\alpha YP_\alpha^\dag.
\label{eqappD2}
\end{equation}
Meanwhile, when $\alpha\neq\alpha^\prime$, the solution of Eq.~(\ref{eqappD1}) is written as
\begin{eqnarray}
&&\Gamma_{\alpha\alpha^\prime}=e^{i(\lambda_\alpha-\lambda_{\alpha^\prime})t}P_\alpha\Gamma_\phi(0)P_{\alpha^\prime}^\dag \nonumber\\
&&+\frac{P_\alpha YP_{\alpha^\prime}^\dag}{i(\lambda_\alpha-\lambda_{\alpha^\prime})}\left[e^{i(\lambda_\alpha-\lambda_{\alpha^\prime})t}-1\right].
\label{eqappD3}
\end{eqnarray}
Next, the time-evolution equation of $\bar{\Gamma}_{\alpha\alpha^\prime}$ is given by
\begin{equation}
\frac{d\bar{\Gamma}_{\alpha\alpha^\prime}}{dt}=i(\lambda_\alpha+\lambda_{\alpha^\prime})\bar{\Gamma}_{\alpha\alpha^\prime}+P_\alpha YP_{\alpha^\prime}^{\rm T},
\label{eqappD4}
\end{equation}
and the solution of Eq.~(\ref{eqappD4}) is written as
\begin{eqnarray}
&&\bar{\Gamma}_{\alpha\alpha^\prime}=e^{i(\lambda_\alpha+\lambda_{\alpha^\prime})t}P_\alpha\Gamma_\phi(0)P_{\alpha^\prime}^{\rm T} \nonumber\\
&&+\frac{P_\alpha YP_{\alpha^\prime}^{\rm T}}{i(\lambda_\alpha+\lambda_{\alpha^\prime})}\left[e^{i(\lambda_\alpha+\lambda_{\alpha^\prime})t}-1\right].
\label{eqappD5}
\end{eqnarray}
We remark that the solution of the time-evolution equations for $P_\alpha^\ast\Gamma_\phi P_{\alpha^\prime}^{\rm T}$ and $P_\alpha^\ast\Gamma_\phi P_{\alpha^\prime}^\dag$ can be solved in a similar manner. A general solution of Eq.~(\ref{eq3D1}) in the oscillatory dynamics is then constructed by the sum of the solutions of the time-evolution equation for $P_\alpha\Gamma_\phi P_{\alpha^\prime}^\dag$, $P_\alpha^\ast\Gamma_\phi P_{\alpha^\prime}^{\rm T}$, $P_\alpha\Gamma_\phi P_{\alpha^\prime}^{\rm T}$, and $P_\alpha^\ast\Gamma_\phi P_{\alpha^\prime}^\dag$, leading to Eq.~(\ref{eq5A2}).

%
% -------------------------------------------------------------------------------------------------------------------------------------------------------------------------------------
%

\section{\label{secE}Derivation of Eq.~(\ref{eq5D1})}
In this section, we describe how to derive Eq.~(\ref{eq5D1}). We use the time derivative of the global QFI (\ref{eq3B12}), the spectral decomposition of $X=\sigma{\mathbb H}_{\rm eff}$ (\ref{eq5A1}), and the general expression of the covariance matrix (\ref{eq5A2}) in the oscillatory dynamics. We need to evaluate the integrals of $e^{\sigma{\mathbb H}_{\rm eff}(t-\tau)}P_\alpha$ and $\tau e^{\sigma{\mathbb H}_{\rm eff}(t-\tau)}P_\alpha$ with respect to the variable $\tau$. Specifically, we have
\begin{eqnarray}
&&\int_0^td\tau\,e^{\sigma{\mathbb H}_{\rm eff}(t-\tau)}P_\alpha=\frac{i}{\lambda_\alpha}(1-e^{i\lambda_\alpha t})P_\alpha, \label{eqappE1}\\
&&\int_0^td\tau\,\tau e^{\sigma{\mathbb H}_{\rm eff}(t-\tau)}P_\alpha=\left(-\frac{it}{\lambda_\alpha}+\frac{1}{\lambda_\alpha^2}-\frac{e^{i\lambda_\alpha t}}{\lambda_\alpha^2}\right)P_\alpha. \label{eqappE2}\nonumber\\
\end{eqnarray}
Therefore, by substituting Eq.~(\ref{eq5A2}) into Eq.~(\ref{eq3B12}), we get
\begin{widetext}
\begin{eqnarray}
&&\frac{dI_{\rm G}}{dt}=16{\bm a}^{\rm T}{\rm Re}\left[-\sum_{\alpha\neq\alpha^\prime}\frac{P_\alpha YP_{\alpha^\prime}^\dag}{\lambda_\alpha(\lambda_\alpha-\lambda_{\alpha^\prime})}-\sum_{\alpha,\alpha^\prime}\frac{P_\alpha YP_{\alpha^\prime}^{\rm T}}{\lambda_\alpha(\lambda_\alpha+\lambda_{\alpha^\prime})}+\sum_\alpha\frac{P_\alpha YP_\alpha^\dag}{\lambda_\alpha^2}\right]{\bm a} \nonumber\\
&&+16{\bm a}^{\rm T}{\rm Im}\left[\sum_\alpha\frac{P_\alpha}{\lambda_\alpha}(-\Gamma_\phi(0)+tY)P_\alpha^\dag\right]{\bm a}+({\rm Oscillatory~terms}).
\label{eqappE3}
\end{eqnarray}
\end{widetext}

Next, we focus on the continuous-position-measurement case discussed in the main text. We express $P_\alpha YP_{\alpha^\prime}^\dag$ and $P_\alpha YP_{\alpha^\prime}^{\rm T}$ by
\begin{eqnarray}
&&P_\alpha YP_{\alpha^\prime}^\dag=\frac{1}{4}\left( \begin{array}{cc}
\Omega/\sqrt{\tilde{\lambda}_\alpha\tilde{\lambda}_{\alpha^\prime}} & -i\sqrt{\Omega/\tilde{\lambda}_\alpha} \vspace{5pt}\\
i\sqrt{\Omega/\tilde{\lambda}_{\alpha^\prime}}                      & 1
\end{array}\right)\otimes\tilde{P}_\alpha m_R\tilde{P}_{\alpha^\prime}^{\rm T}, \label{eqappE4}\nonumber\\
\\
&&P_\alpha YP_{\alpha^\prime}^{\rm T}=\frac{1}{4}\left( \begin{array}{cc}
-\Omega/\sqrt{\tilde{\lambda}_\alpha\tilde{\lambda}_{\alpha^\prime}} & -i\sqrt{\Omega/\tilde{\lambda}_\alpha} \vspace{5pt}\\
-i\sqrt{\Omega/\tilde{\lambda}_{\alpha^\prime}}                      & 1
\end{array}\right)\otimes\tilde{P}_\alpha m_R\tilde{P}_{\alpha^\prime}^{\rm T}. \label{eqappE5}\nonumber\\
\end{eqnarray}
We recall that the vector ${\bm a}$ is given by Eq.~(\ref{eq5B1}), and we have $\lambda_\alpha=\sqrt{\Omega\tilde{\lambda}_\alpha}$. We note that the second term of the right-hand side in Eq.~(\ref{eqappE3}) does not contribute to the time derivative $\dot{I}_G$, since the imaginary part of $P_\alpha YP_\alpha^\dag$ is an off-diagonal block matrix. Therefore, we get
\begin{eqnarray}
&&\lim_{t\rightarrow\infty}\frac{I_{\rm G}(t)}{t} \nonumber\\
&&=4{\bm b}^{\rm T}\left(-\sum_{\alpha\neq\alpha^\prime}\frac{2\tilde{P}_\alpha m_R\tilde{P}_{\alpha^\prime}^{\rm T}}{\tilde{\lambda}_\alpha(\tilde{\lambda}_\alpha-\tilde{\lambda}_{\alpha^\prime})}+\sum_{\alpha}\frac{3\tilde{P}_\alpha m_R\tilde{P}_\alpha^{\rm T}}{2\tilde{\lambda}_\alpha^2}\right){\bm b}. \nonumber\\
\label{eqappE6}
\end{eqnarray}
Since only the symmetric part of the matrix included in Eq.~(\ref{eqappE6}) contributes to the asymptotic rate of the global QFI (\ref{eq3E11}), we obtain Eq.~(\ref{eq5D1}) as the final expression of Eq.~(\ref{eqappE6}).

%
% -------------------------------------------------------------------------------------------------------------------------------------------------------------------------------------
%

\section{\label{secF}Proof of Eq.~(\ref{eq5D2})}
We here consider the case of zero-damping couplings and derive the bounds (\ref{eq5D2}) on the asymptotic growth rate of the optimized global QFI (cf. Eqs.~(\ref{eq3E13}) and (\ref{eq5D1})) in terms of the asymptotic generation rate of bosonic excitations (\ref{eq5C3}). For the sake of notational simplicity, we express these quantities by $x$ and $y$ as follows:
\begin{eqnarray}
&&x=4M \nonumber\\
&&\times\max_{|{\bm n}|=1}{\bm n}^{\rm T}\left(\sum_{\alpha,\alpha^\prime}\frac{\tilde{P}_\alpha m_R\tilde{P}_{\alpha^\prime}^{\rm T}}{\tilde{\lambda}_\alpha\tilde{\lambda}_{\alpha^\prime}}+\sum_\alpha\frac{\tilde{P}_\alpha m_R\tilde{P}_\alpha^{\rm T}}{2\tilde{\lambda}_\alpha^2}\right){\bm n}, \nonumber\\
\label{eqappF1}\\
&&y=\frac{1}{4M}{\rm Tr}\left[\sum_{\alpha=1}^M\left(\frac{1}{\tilde{\lambda}_\alpha}+1\right)\tilde{P}_\alpha m_R\tilde{P}_\alpha^{\rm T}\right], \label{appF2}
\end{eqnarray}
where we recall that $m_R$ is a real, symmetric, positive-semidefinite matrix. The quantities $\tilde{\lambda}_\alpha$ and $\tilde{P}_\alpha$ are the eigenvalues and the corresponding eigenprojectors of the effective Hamiltonian matrix $h_{\rm eff}=h+m_I$ (cf. Eq.~(\ref{eq5B3})). This matrix $h_{\rm eff}$ is a real, generally nonsymmetric $M\times M$ matrix, and as such, its right and left eigenvectors are distinct and form a biorthogonal system. The operators $\tilde{P}_\alpha$ project onto the right eigenspaces of $h_{\rm eff}$. The assumptions on the matrices are summarized in Sec.~\ref{sec5B}.

Our derivation begins by noting that the maximization over the normalized vector ${\bm n}$ in $x$ is equivalent to finding the largest eigenvalue of the matrix within the parentheses. Specifically, we have
\begin{equation}
x=4M\Lambda_{\rm max}(S),
\label{eqappF3}
\end{equation}
where $\Lambda_{\rm max}(\cdot)$ denotes the maximum eigenvalue, and $S$ is the symmetric matrix given by
\begin{equation}
S=h_{\rm eff}^{-1}m_R(h_{\rm eff}^{\rm T})^{-1}+\sum_{\alpha=1}^M\frac{\tilde{P}_\alpha m_R\tilde{P}_\alpha^{\rm T}}{2\tilde{\lambda}_\alpha^2}.
\label{eqappF4}
\end{equation}

To establish an upper bound on $x$, we use the Weyl's inequality, $\Lambda_{\rm max}(X+Y)\leq\Lambda_{\rm max}(X)+\Lambda_{\rm max}(Y)$, to obtain
\begin{align}
\frac{x}{4M}&\leq\Lambda_{\rm max}(h_{\rm eff}^{-1}m_R(h_{\rm eff}^{\rm T})^{-1})+\Lambda_{{\rm max}}\left(\sum_{\alpha=1}^M\frac{\tilde{P}_{\alpha}m_R\tilde{P}_{\alpha}^{T}}{2\tilde{\lambda}_{\alpha}^{2}}\right) \nonumber\\
&\leq\|h_{\rm eff}^{-1}\|^2\|m_R\|+{\rm Tr}\left[\sum_{\alpha=1}^M\frac{\tilde{P}_\alpha m_R\tilde{P}_\alpha^{\rm T}}{2\tilde{\lambda}_\alpha^2}\right] \nonumber\\
&\leq\frac{1}{c_3^2}\|m_R\|+\frac{1}{2c_1^2}{\rm Tr}\left[\sum_{\alpha=1}^M\tilde{P}_{\alpha}m_R\tilde{P}_{\alpha}^{T}\right] \nonumber\\
&\leq\frac{c_5}{c_3^2c_6}{\rm Tr}\left[\sum_{\alpha=1}^M\tilde{P}_\alpha m_R\tilde{P}_\alpha^{\rm T}\right]+\frac{2yM}{c_1^2(1+1/c_2)} \nonumber\\
&\leq\frac{c_5}{c_3^2c_6}\cdot\frac{4yM}{c_1^2(1+1/c_2)}+\frac{2yM}{c_1^2(1+1/c_2)}.
\label{eqappF5}
\end{align}
In the second inequality, we used
\begin{align}
\Lambda_{\rm max}(h_{\rm eff}^{-1}m_R(h_{\rm eff}^{\rm T})^{-1})&=\|h_{\rm eff}^{-1}m_R(h_{\rm eff}^{\rm T})^{-1}\| \nonumber\\
&\leq\|h_{\rm eff}^{-1}\|^2\|m_R\|,
\label{eqappF6}
\end{align}
and also used fact that the maximum eigenvalue of a positive semidefinite matrix is bounded by its trace. In the third inequality, we used
\begin{equation}
\|h_{\rm eff}^{-1}\|=\frac{1}{\sigma_{\rm min}(h_{\rm eff})}\leq\frac{1}{c_3},
\label{eqappF7}
\end{equation}
where $\sigma_{\rm min}(\cdot)$ denotes the minimum singular value. The resulting upper bound is
\begin{equation}
x\leq\frac{8(2c_1^2c_5+c_3^2c_6)}{c_1^2c_3^2c_6(1+1/c_2)}yM^2.
\label{eqappF8}
\end{equation}

To obtain a lower bound, we again start from the Weyl's inequality in the form $\Lambda_{\rm max}(X+Y)\geq\Lambda_{\rm min}(X)+\Lambda_{\rm max}(Y)$, resulting in
\begin{align}
\frac{x}{4M}&\geq\Lambda_{\rm min}(h_{{\rm eff}}^{-1}m_R(h_{{\rm eff}}^{T})^{-1})+\Lambda_{\rm max}\left(\sum_{\alpha=1}^M\frac{\tilde{P}_\alpha m_R\tilde{P}_\alpha^{\rm T}}{2\tilde{\lambda}_\alpha^2}\right) \nonumber\\
&\geq\Lambda_{\rm max}\left(\sum_{\alpha=1}^M\frac{\tilde{P}_\alpha m_R\tilde{P}_\alpha^{\rm T}}{2\tilde{\lambda}_\alpha^2}\right) \nonumber\\
&\geq\frac{1}{M}{\rm Tr}\left[\sum_{\alpha=1}^M\frac{\tilde{P}_\alpha m_R\tilde{P}_\alpha^{\rm T}}{2\tilde{\lambda}_\alpha^2}\right] \nonumber\\
&\geq\frac{1}{2c_2^2M}{\rm Tr}\left[\sum_{\alpha=1}^M\tilde{P}_\alpha m_R\tilde{P}_\alpha^{\rm T}\right] \nonumber\\
&\geq\frac{2y}{c_2^2(1+1/c_1)}.
\label{eqappF9}
\end{align}
In the second inequality, we used the fact that $h_{\rm eff}^{-1}m_R(h_{\rm eff}^{\rm T})^{-1}$ is positive semidefinite and thus $\Lambda_{\rm min}(h_{\rm eff}^{-1}m_R(h_{\rm eff}^{\rm T})^{-1})\geq0$. In the third inequality, we used the fact that the maximum eigenvalue is bounded below by the average eigenvalue. We thus get the lower bound
\begin{equation}
x\geq\frac{8yM}{c_2^2(1+1/c_1)}.
\label{eqappF10}
\end{equation}
Equations~(\ref{eqappF8}) and (\ref{eqappF10}) prove the bounds (\ref{eq5D2}) in the main text.

%
% -------------------------------------------------------------------------------------------------------------------------------------------------------------------------------------
%

\section{\label{secG}Derivation of Eq.~(\ref{eq5E1}) and proof of Eq.~(\ref{eq5E2})}
We first note that, following a line of reasoning analogous to the one used for the global QFI, the asymptotic growth rate of the difference term is given by (cf. Eq.~(\ref{eq3C7}))
\begin{equation}
\delta\dot{I}_{\rm st}\equiv\lim_{t\rightarrow\infty}\frac{\delta I(t)}{t}=4{\bm b}^{\rm T}(\tilde{m}_R^\prime\tilde{m}_R^+\tilde{m}_R^\prime){\bm b},
\label{eqapppG1}
\end{equation}
where
\begin{equation}
\tilde{m}_R=\sum_{\alpha=1}^M\tilde{P}_\alpha m_R\tilde{P}_\alpha^{\rm T},~\tilde{m}_R^\prime=\sum_{\alpha=1}^M\frac{\tilde{P}_\alpha m_R\tilde{P}_\alpha^{\rm T}}{\sqrt{2}\tilde{\lambda}_\alpha},
\label{eqappG2}
\end{equation}
and $\tilde{m}_R^+$ is the Moore-Penrose pseudoinverse being the inverse restricted to the support of $\tilde{m}_R$. From this result and Eq.~(\ref{eq5D1}), one can see that Eq.~(\ref{eq5E1}) holds true if
\begin{equation}
A=\sum_{\alpha=1}^M\frac{\tilde{P}_\alpha m_R\tilde{P}_\alpha^{\rm T}}{2\tilde{\lambda}_\alpha^2}
\label{eqappG3}
\end{equation}
and
\begin{equation}
B=\tilde{m}_R^\prime\tilde{m}_R^{-1}\tilde{m}_R^\prime
\label{eqappG4}
\end{equation}
are equal to each other; this means that the second term in Eq.~(\ref{eq5D1}) corresponds to the growth rate of the information difference, and thus the first term in Eq.~(\ref{eq5D1}) characterizes the asymptotic rate of the environmental QFI. The key point is that the projections $\tilde{P}_\alpha$ are written by the right and left eigenvectors of the effective non-Hermitian matrix, given in Eq.~(\ref{eq5B5}). By using this expression, we have
\begin{align}
\sum_{\alpha=1}^M\tilde{P}_\alpha m_R\tilde{P}_\alpha^{\rm T}&=\sum_{\alpha=1}^M{\bm r}_\alpha({\bm l}_\alpha^{\rm T}m_R{\bm l}_\alpha){\bm r}_\alpha^{\rm T} \nonumber\\
&=({\bm r}_1,\cdots,{\bm r}_M)\left( \begin{array}{ccc}
d_1 &        &     \vspace{5pt}\\
    & \ddots &     \vspace{5pt}\\
    &        & d_M
\end{array}\right)\left( \begin{array}{c}
{\bm r}_1^{\rm T} \vspace{5pt}\\
\vdots            \vspace{5pt}\\
{\bm r}_M^{\rm T}
\end{array}\right) \nonumber\\
&\equiv VDV^{\rm T},
\label{eqappG5}
\end{align}
where $d_\alpha\equiv{\bm l}_\alpha^{\rm T}m_R{\bm l}_\alpha$. We then have $\tilde{m}_R^+=(V^{\rm T})^{-1}D^+V^{-1}$, where $D^+=(d_1^+,\dots,d_M^+)$ with
\begin{eqnarray}
d_\alpha^+=\left\{ \begin{array}{ll}
1/d_\alpha, & (d_\alpha>0), \vspace{5pt}\\
0,          & (d_\alpha=0).
\end{array}\right.
\label{eqappG6}
\end{eqnarray}
Thus, the inverse is taken only over the subspace associated with nonzero $d_\alpha$, while $d_\alpha=0$ are projected out. Since ${\bm r}_\alpha^{\rm T}\tilde{m}_R^+{\bm r}_{\alpha^\prime}=\delta_{\alpha\alpha^\prime}d_\alpha^{-1}$, we obtain
\begin{align}
B&=\sum_{\alpha,\alpha^\prime}\frac{1}{2\tilde{\lambda}_\alpha\tilde{\lambda}_{\alpha^\prime}}(d_\alpha{\bm r}_\alpha{\bm r}_\alpha^{\rm T})\tilde{m}_R^{-1}(d_{\alpha^\prime}{\bm r}_{\alpha^\prime}{\bm r}_{\alpha^\prime}^{\rm T}) \nonumber\\
&=\sum_{\alpha=1}^M\frac{d_\alpha{\bm r}_\alpha{\bm r}_\alpha^{\rm T}}{2\tilde{\lambda}_\alpha^2},
\label{eqappG7}
\end{align}
which leads to $A=B$.

The inequality (\ref{eq5E2}) for the optimized environmental QFI (cf. Eqs.~(\ref{eq3E14}) and (\ref{eq5E1})) can be readily obtained by considering the operator norm of the first term in Eq.~(\ref{eq5D1}). As explained in Appendix~\ref{secF}, its upper bound can be given by $\|h_{\rm eff}^{-1}m_R(h_{\rm eff}^{\rm T})^{-1}\|\leq4c_5yM/[c_1^2c_3^2c_6(1+1/c_2)]$; we can also obtain another upper bound directly from the assumed bound $\|m_R\|\leq c_5M$, leading to $\|h_{\rm eff}^{-1}m_R(h_{\rm eff}^{\rm T})^{-1}\|\leq4c_5M/c_3^2$. Similarly, the lower bound is obtained as
\begin{equation}
\|h_{\rm eff}^{-1}m_R(h_{\rm eff}^{\rm T})^{-1}\|\geq\frac{\|m_R\|}{\|h_{\rm eff}\|^2}\geq\frac{c_4}{c_2^{\prime2}}.
\label{eqappG8}
\end{equation}
These relations give Eq.~(\ref{eq5E2}) in the main text.

%
% -------------------------------------------------------------------------------------------------------------------------------------------------------------------------------------
%

\section{\label{secH}Non-Hermitian skin effect}
The non-Hermitian skin effect is a phenomenon which induces the accumulation of numerous eigenstates at the edge~\cite{Yao2018,Okuma2020,Zhang2020NH}. In this section, we review a basic framework to investigate the non-Hermitian skin effect. We first introduce a simple model which exhibits the non-Hermitian skin effect and describe distinct behaviors for open and periodic boundary conditions. We then explain how to obtain the eigenvalue and the corresponding eigenstate of the model with the open boundary condition, based on a non-Hermitian spectrum theory~\cite{Yao2018,Yokomizo2019,Yokomizo2020}.

%
% -------------------------------------------------------------------------------------------------------------------------------------------------------------------------------------
%

\subsection{\label{secH1}Nonreciprocal model}
We start with a non-Hermitian system described by the eigenvalue equation,
\begin{equation}
\hat{H}|\psi\rangle=E|\psi\rangle,
\label{eqappH11}
\end{equation}
where $|\psi\rangle=(\psi_1,\dots,\psi_L)$ represents an eigenstate. We consider a real-space non-Hermitian Hamiltonian with nonreciprocal hopping given by
\begin{equation}
\hat{H}=\sum_{j=1}^{L-1}(t_R|j+1\rangle\langle j|+t_L|j\rangle\langle j+1|)+\sum_{j=1}^Lw|j\rangle\langle j|,
\label{eqappH12}
\end{equation}
where $t_R$ and $t_l$ take positive real values.

\begin{figure}[]
\includegraphics[width=8.5cm]{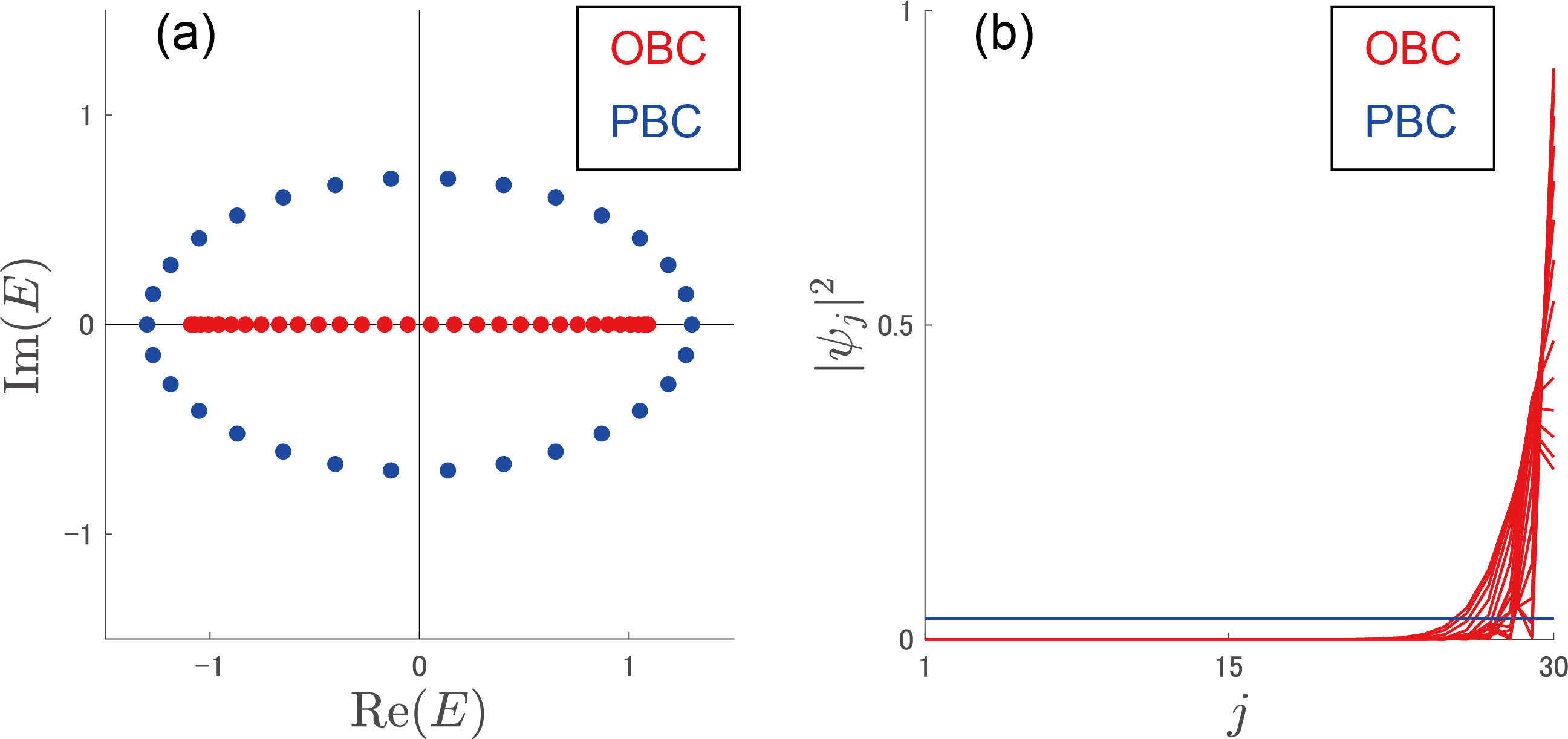}
\caption{\label{figappH}Distinct behaviors of the nonreciprocal model for a periodic boundary condition (PBC) and an open boundary condition (OBC). (a) Eigenvalues of the Hamiltonian given by Eq.~(\ref{eqappH12}). (b) Spatial profile of the eigenstate. The parameters are set to be $t_R=1,t_L=0.3,w=0$, and $L=30$.}
\end{figure}

In Fig.~\ref{figappH}, we show the numerical diagonalization results of the Hamiltonian for periodic and open boundary conditions. We point out that the distribution of the eigenvalue depends on boundary conditions as shown in Fig.~\ref{figappH}(a); the eigenvalues take real values under the open boundary condition, whereas they become complex numbers under the periodic boundary condition. Correspondingly, the behaviors of the eigenstate are distinct for the two boundary conditions as shown in Fig.~\ref{figappH}(b). In particular, all the eigenstates are localized at the right end of the system under the open boundary condition. Physically, the unidirectional flow induced by the nonreciprocal hopping leads to the accumulation of the eigenstates at the edge. Thus, the non-Hermitian skin effect exhibits the distinct eigenvalue distributions for the boundary conditions and the localization of the bulk eigenstate.

%
% -------------------------------------------------------------------------------------------------------------------------------------------------------------------------------------
%

\subsection{\label{secH2}Eigenvalue and eigenstate}
To investigate the system with the open boundary condition, we begin by showing an eigenvalue equation obtained from Eq.~(\ref{eqappH11}),
\begin{equation}
t_R\psi_{j-1}+w\psi_j+t_L\psi_{j+1}=E\psi_j.
\label{eqappH21}
\end{equation}
The open boundary condition is defined by
\begin{equation}
\psi_0=\psi_{L+1}=0.
\label{eqappH22}
\end{equation}
From a general theory for linear differential equations, the solution of Eq.~(\ref{eqappH21}) is written as a linear combination form, given by
\begin{equation}
\psi_j=\sum_{l=1}^2(\beta_l)^j\phi^{(l)},
\label{eqappH23}
\end{equation}
where $\phi^{(l)}$ is a combination coefficient, and $\beta_l$ is a solution of the eigenvalue equation,
\begin{equation}
t_R\beta^{-1}+w+t_L\beta=E.
\label{eqappH24}
\end{equation}

We can rewrite Eq.~(\ref{eqappH22}) as
\begin{eqnarray}
\left( \begin{array}{cc}
1               & 1               \vspace{5pt}\\
(\beta_1)^{L+1} & (\beta_2)^{L+1}
\end{array}\right)\left( \begin{array}{c}
\phi^{(1)} \vspace{5pt}\\
\phi^{(2)}
\end{array}\right)=0.
\label{eqappH25}
\end{eqnarray}
We then obtain $(\beta_1/\beta_2)^{L+1}=1$, which ensures that the combination coefficients $\phi^{(1)}$ and $\phi^{(2)}$ take nonzero values. This condition is rewritten as
\begin{equation}
\frac{\beta_1}{\beta_2}=e^{2i\theta_n},
\label{eqappH26}
\end{equation}
where $\theta_n=n\pi/(L+1)~(n=1,\dots,L)$. By combining the Vieta's theorem and the fact that the absolute values of $\beta_1$ and $\beta_2$ are equal, we obtain
\begin{equation}
|\beta_1|=|\beta_2|=\sqrt{\frac{t_R}{t_L}}.
\label{eqappH27}
\end{equation}
Hence, the values of $\beta_1$ and $\beta_2$ are determined as
\begin{equation}
\beta_1^{(n)}=\sqrt{\frac{t_R}{t_L}}e^{i\theta_n},~\beta_2^{(n)}=\sqrt{\frac{t_R}{t_L}}e^{-i\theta_n}.
\label{eqappH28}
\end{equation}

From Eqs.~(\ref{eqappH24}) and (\ref{eqappH28}), we can obtain the analytical form of the eigenvalue as follows:
\begin{equation}
E^{(n)}=w+2\sqrt{t_Rt_L}\cos\theta_n,
\label{eqappH29}
\end{equation}
which is consistent with the numerical result in Fig.~\ref{figappH}(a). Furthermore, Eq.~(\ref{eqappH23}) can be represented by
\begin{equation}
\psi_j^{(n)}\propto\left(\sqrt{\frac{t_R}{t_L}}\right)^j\sin\theta_n.
\label{eqappH210}
\end{equation}
One can infer from Eq.~(\ref{eqappH210}) that the eigenstate exhibits the exponential localization at the edge of the system, which is consistent with the numerical result in Fig.~\ref{figappH}(b). Importantly, all the eigenstates are localized with the common localization length $\xi$, and the analytical form of $\xi$ is given by
\begin{equation}
\frac{1}{\xi}=\log\sqrt{\frac{t_R}{t_L}}.
\label{eqappH211}
\end{equation}

%
% -------------------------------------------------------------------------------------------------------------------------------------------------------------------------------------
%

%\nocite{*}
%\bibliographystyle{apsrev4-2}
\bibliography{ASPLCGQM}% Produces the bibliography via BibTeX.
\end{document}